\documentclass[12pt,amsmath,amssymb,aps]{revtex4-1}
\usepackage{graphicx,epsfig,dcolumn,bm,epic,eepic,float}
\usepackage{amsmath}
\usepackage[T1]{fontenc}
\usepackage[utf8]{inputenc}
\usepackage{mathtools} 
\usepackage{pifont}
\usepackage{amsmath}
\usepackage{latexsym}
\usepackage{color}
\usepackage{cancel}
\usepackage{scrextend}
\usepackage{amsmath}
\usepackage{amsthm}
\usepackage{eucal}
\usepackage{amssymb}
\usepackage{mathrsfs}
\usepackage[bottom]{footmisc}
\usepackage{makeidx,shortvrb,latexsym}
\usepackage{epstopdf}
\usepackage{mathtools}
\usepackage{ragged2e}
\usepackage{autobreak}
\usepackage[T1]{fontenc}
\newcommand {\be}{\begin{equation}}
\newcommand {\ee}{\end{equation}}
\newcommand {\ba}{\begin{eqnarray}}
\newcommand {\ea}{\end{eqnarray}}
\newcommand {\bea}{\begin{eqnarray}}
\newcommand {\eea}{\end{eqnarray}}

\newcommand{\tr}{\mathrm{Tr}}
\pdfoutput=1

\numberwithin{equation}{section}
\begin{document}

\title{{\Large Maximally Symmetric Three Higgs Doublet Model}\vspace*{5mm}}

\author{\sc Neda Darvishi$^{\,a,b}$} 
\author{\sc M.R. Masouminia$^{\,c}$}
\author{\sc Apostolos Pilaftsis$^{\,b}$\vspace*{3mm}}
\affiliation{$^a$Faculty of Physics, University of Warsaw, Pasteura 5, 02-093 Warsaw, Poland}
\affiliation{$^b$School of Physics and Astronomy, The University of Manchester, Manchester M13 9PL, United Kingdom}
\affiliation{$^c$Institute for Particle Physics Phenomenology, Department of Physics, Durham University, Durham DH1 3LE, United Kingdom}
    
\begin{abstract}
${}$

\centerline{\bf ABSTRACT} \medskip

\noindent
We consider the general Three-Higgs Doublet Model (3HDM) and identify all limits that lead to exact SM alignment. After discussing the underlying symmetries that can naturally enforce such an alignment, we focus on the most economic setting,\- called here the Maximally Symmetric Three-Higgs Doublet Model (MS-3HDM). The potential\- of the MS-3HDM obeys an $\mathrm{Sp(6)}$ symmetry, softly broken by bilinear masses and explicitly by hypercharge and Yukawa couplings through renormalisation-group effects, whilst the theory allows for quartic coupling unification up to the Planck~scale.  Besides the two ratios of vacuum expectation values, $\tan\beta_{1,2}$, the MS-3HDM is predominantly governed by only three input parameters: the masses of the two charged Higgs bosons, $M_{h_{1,2}^{\pm}}$, and their mixing angle~$\sigma$.  Most remarkably, with these input parameters, we obtain definite predictions for the entire scalar mass spectrum of the theory, as well as for the SM-like Higgs-boson couplings to the gauge bosons and fermions. The predicted deviations of these couplings from their SM values might be probed at future precision high-energy colliders. The new phenomenological aspects of the MS-3HDM with respect to the earlier studied MS-2HDM are discussed.

\end{abstract}
\maketitle

\section{Introduction}
\label{sec:intro}
 
The expedition for physics Beyond the Standard Model (BSM), such as the exploration of non-standard scenarios with an extended Higgs sector, has strong~theoretical and experimental~motivations. The data collected from the CERN Large Hadron Collider (LHC) impose constraints on the coupling strengths of the Higgs boson, primarily on its interaction with the electro\-weak~(EW) gauge bosons, namely the $W^{\pm}$ and $Z$ bosons. These LHC data show that the coupling strengths of the observed 125-GeV scalar particle must be very close to those predicted by the Standard Model (SM)~\cite{ATLAS1,CMS1,Darvishi:2019uzp}. This simple fact severely restricts the form of possible scalar-sector extensions of the SM~\cite{Choi:2021nql}.

An interesting class of Higgs-sector extensions is the $n$-Higgs Doublet Model~($n$HDM) which enlarges the SM with $n\geq 2$ Higgs doublets~\cite{Lee1,Weinberg:1976hu, Pilaftsis:1999qt,Branco1}. In view of the aforementioned constraints on the SM-like Higgs boson couplings to the EW gauge bosons, the only viable scenarios of interest are those that allow for the so-called SM alignment limit~\cite{Ginzburg:1999fb,Chankowski:2000an, Lee:2012jn, Delgado:2013zfa,Dev:2014yca,Bernon:2015qea,Pilaftsis:2016erj,Darvishi:2017bhf,Benakli:2018vqz,Lane:2018ycs,Darvishi:2020teg,Eichten:2021qbm}.

The simplest $n$HDM is the Two-Higgs Doublet Model (2HDM). In the 2HDM, two SM-like Higgs scenarios accommodating a scalar resonance of mass $\sim \mathrm{125\, GeV}$ as observed at the LHC can be realised. In these scenarios, the interaction strength of the observed scalar particle to the $W^\pm$ and $Z$ bosons can approach its SM prediction.  In such a SM-aligned 2HDM, the other CP-even states do not couple to the EW gauge bosons at the tree level. In~a~general 2HDM without the imposition of any symmetry, exact SM alignment limit can be achieved in two different ways: (i)~either by postulating for all new-physics mass scales to be sufficiently large, or (ii)~by resorting to a fine-tuning among the parameters of the model~\cite{Georgi:1978ri,Gunion:2002zf,CPMK, Haber:2006ue,Carena:2013ooa,Haber:2015pua, Grzadkowski:2018ohf}. This generic feature persists even within more extended frameworks, such as the Three-Higgs Doublet (3HDM)~\cite{Ivanov:2012ry,Keus:2013hya,Darvishi:2019dbh,Aranda:2019vda,Das:2019yad,Kuncinas:2020wrn,Ivanov:2021pnr,Chakraborti:2021bpy}.

In this paper we analyse the general 3HDM, for which we find three distinct SM-like Higgs scenarios that possess an exact SM alignment limit. We then discuss the underlying symmetries that can naturally enforce such an alignment without the need to decouple all new-physics mass scales or to resort to {\em ad-hoc} arrangements among the parameters~\cite{Pilaftsis:2011ed,Dev:2014yca,Pilaftsis:2016erj,Darvishi:2020teg}. In this context, earlier studies~\cite{Darvishi:2019dbh,Darvishi:2020teg} have shown that the potential of $n$HDMs contains a large number of $\mathrm{SU(2)}_L$-preserving accidental symmetries as subgroups of the symplectic group $\mathrm{Sp(2}n)$. This maximal symmetry group plays an instrumental role in classifying accidental symmetries that may occur in the scalar potentials of $n$HDMs and $n$HDM-Effective Field Theories with higher-order operators~\cite{Pilaftsis:2016erj,Darvishi:2019dbh,Darvishi:2020teg,Birch-Sykes:2020btk}. Thus far, this classification has been done for: (i) the 2HDM~\cite{Pilaftsis:2011ed}, (ii) the 2HDM Effective Field Theory for higher-order operators up to dimension-6 and dimension-8~\cite{Birch-Sykes:2020btk}, and (iii)~the 3HDM~\cite{Ivanov:2012ry,Darvishi:2019dbh}. Interestingly enough, there are three continuous symmetries~\cite{Pilaftsis:2016erj} which, when imposed on the $n$HDM scalar potential, lead to SM alignment. In the present study, we will consider the most economic class of such scenarios, i.e.~the Maximally Symmetric $n$HDM (MS-$n$HDM). In particular, we will focus on the MS-3HDM and show how one can have successful quartic coupling unification up to the Planck scale.

The present study of the MS-3HDM extends a previous work on the MS-2HDM~\cite{Dev:2014yca,Dev:2017org,Hanson:2018uhf,Darvishi:2019ltl,Darvishi:2020paz}. Like in the MS-2HDM, the Sp(6) symmetry of the MS-3HDM potential gets violated by two sources: (i) softly by bilinear scalar mass terms $m^2_{ij}$ (with $i,j=1,2,3$), and (ii)~explicitly by renormalization-group~(RG) effects involving the hypercharge and Yukawa couplings.  The MS-3HDM is a very predictive scenario, as it only depends on fewer theoretical parameters than those in the general 3HDM. In addition to the two ratios $\tan\beta_{1,2}$ of vacuum expectation values${}$~(VEVs) involving the three Higgs doublets, the model is mainly governed by only three input parameters: the masses of the two charged Higgs bosons, $M_{h_{1,2}^{\pm}}$, and their mixing angle~$\sigma$.  Most notably, with these input parameters, we obtain sharp predictions for the entire scalar mass spectrum of the theory, including the interactions of all Higgs particles to the SM fields and all scalar self-interactions.

In analogy to the findings in the MS-2HDM~\cite{Darvishi:2019ltl}, we show how all quartic couplings in the MS-3HDM can unify at high-energy scales~$\mu_X$ and vanish identically at two distinct conformal points, called here~$\mu^{(1,2)}_X$, for which $\mu_X^{(1)} \lesssim 10^{13}\,\text{GeV}$ and $\mu_X^{(2)} \gtrsim 10^{21}\,\text{GeV}$. Assuming that all quartic couplings unify at~$\mu^{(1,2)}_X$,  we present definite predictions for the full scalar mass spectrum, as well as for the SM-like Higgs-boson interactions with the $W^\pm$ and $Z$ bosons, the top- and bottom-quarks, and tau-leptons. The deviations found for all these couplings from their SM values might be testable at future high-energy $e^+e^-$ colliders.

The layout of this paper is as follows. In Section~\ref{3HDM}, we discuss the basic features of the general 3HDM and identify all possible SM alignment limits that can take place in this model. For definiteness, we focus on the canonical SM-like Higgs scenario in the Type-V 3HDM and derive the conditions for achieving exact SM alignment. In the same section, we present analytic expressions that describe the misalignment predictions for the SM-like Higgs boson couplings to the EW gauge bosons and SM fermions. In Section~\ref{sym}, we define the 3HDM in the bilinear scalar-field formalism. Given that $\mathrm{Sp(6)}$ is the maximal symmetry of the 3HDM potential, we review the complete set of continuous maximal symmetries for the SM alignment limit that may take place in the 3HDM potential. Subsequently, we concentrate on the most minimal setting, the MS-3HDM, and clarify the origin of {\em natural} SM alignment in this model. In Section \ref{MS-3HDM}, we describe the Higgs-mass spectrum of the model and determine the breaking pattern of the $\mathrm{Sp(6)}$ symmetry due to RG effects and the soft-breaking mass terms $m^2_{ij}$. In~Section~\ref{QCU}, we evaluate two-loop RG effects on all relevant couplings in the 3HDM. Specifically, we show results obtained by a RG evolution of these couplings from the quartic coupling unification points $\mu^{(1,2)}_X$ down to the threshold scale $\mu_{\rm thr} = M_{h_1^{\pm}}$. In this analysis, we consider illustrative benchmark scenarios for the VEV ratios, $\tan\beta_1$ and $\tan\beta_2$, and the kinematic parameters of the charged Higgs sector. We~also present misalignment predictions for Higgs-boson couplings to the $W^\pm$ and $Z$ bosons, $\tau$-leptons, and $t$- and $b$-quarks. Section~\ref{conc} summarises our results and discusses the new phenomenological aspects of the MS-3HDM with respect to the MS-2HDM. Finally, technical details, as well as the one- and two-loop RG equations pertinent to the 3HDM, are presented in Appendices~\ref{ap:pot}, \ref{AppMassMat} and~\ref{RGEs}.

\vfill\eject

\section{The General 3HDM}\label{3HDM}

The general 3HDM augments the SM Higgs-sector with three 
Higgs doublets, denoted here as 
\begin{equation} {\Phi}_1\ =\ \left(
 \begin{matrix}
\Phi_1^+\\
 \Phi_1^0
 \end{matrix}\right)\;,\qquad 
{\Phi}_2\ =\ \left(
 \begin{matrix}
\Phi_2^+\\
 \Phi_2^0
 \end{matrix}\right)\;,\qquad 
{\Phi}_3\ =\ \left(
 \begin{matrix}
\Phi_3^+\\
 \Phi_3^0
 \end{matrix}\right)\;,
\end{equation} 
where the scalar doublets $\Phi_i$ (with $i=1,2,3$) carry the same $\mathrm{U(1)}_Y$-hypercharge quantum number: $Y_{\Phi} = {1/2}$. The general 3HDM potential invariant under $\mathrm{SU(2)}_L\otimes \mathrm{U(1})_Y$ may be succinctly expressed as follows~\cite{Botella:1994cs}:
\begin{equation}
  V=\ -\displaystyle\sum_{i,j=1}^{3}\, m_{ij}^2\, ( \Phi_i^{\dagger} \Phi_j)+ \ {1\over 2} \displaystyle\sum_{i,j\, k,l=1}^{3}\, 
  \lambda_{ijkl}\, ( \Phi_i^{\dagger} \Phi_j)( \Phi_k^{\dagger} \Phi_l),
  \label{pot-3hdm}
\end{equation}
with $\lambda_{ijkl}=\lambda_{klij}$. Thus, the $\mathrm{SU(2)}_L\otimes \mathrm{U(1})_Y$ invariant 3HDM potential contains $9$ real bilinear mass terms, $m^2_{ij}$, along with $45$ real quartic couplings.  Moreover, the scalar potential $V$ may realise up to 19 accidental symmetries, and this number can further increase by another 21 {\em custodial} symmetries for which only the $\text{SU(2)}_L$ group is preserved~\cite{Darvishi:2019dbh}. The explicit form of the 3HDM potential is given in Appendix~\ref{ap:pot}.

Moreover, extending the Higgs-sector of the SM with additional Higgs doublets leads generically to Flavor Changing Neutral Currents (FCNCs) at tree level, which are severely constrained by experiment. One simple way to suppress such tree-level FCNCs will be to enforce that fermions belonging to a specific family do not couple simultaneously to two or more different Higgs doublets in the Yukawa sector~\cite{Glashow:1976nt}. In the 3HDM, this condition allows for five independent types of Yukawa interactions as exhibited in Table~\ref{tab:1}.  More explicitly, in Type-I all fermions couple to the first doublet~$\Phi_1$, and none to the other two doublets~$\Phi_{2,3}$. In Type-II, the down-type quarks, $d^i_{L,R}$, and the charged leptons, $e^i_{L,R}$, couple to~$\Phi_1$, and the up-type quarks, $u^i_{L,R}$, couple to~$\Phi_2$, where the superscript $i=1,2,3$ labels the three generations of fermions. In Type-III, the down-type quarks couple to $\Phi_1$ and the up-type quarks and the charged leptons couple to $\Phi_2$. In Type-IV, all quarks couple to $\Phi_1$ and the charged leptons to $\Phi_2$. In Type-V, the down-type quarks couple to $\Phi_1$,  the up-type quarks to $\Phi_2$, and the charged leptons to $\Phi_3$. 

\begin{table}
\centering
\begin{tabular}{|c|ccc|}
\hline
\hline
Yukawa Types  \qquad &  \qquad$y_u$  &  \qquad   $ y_d$       \qquad  & \qquad    $y_e~$    \qquad  \\ \hline
Type-I   \qquad &   \qquad$\Phi_1$      & \qquad  $\Phi_1$     \qquad  & \qquad   $ \Phi_1$     \qquad   \\ \hline
Type-II  \qquad &  \qquad $\Phi_2$        & \qquad  $ \Phi_1$    \qquad  & \qquad   $ \Phi_1$     \qquad  \\ \hline
Type-III  \qquad &   \qquad$\Phi_2$       & \qquad  $ \Phi_1$    \qquad  & \qquad  $ \Phi_2$     \qquad   \\ \hline
Type-IV  \qquad &  \qquad $\Phi_1$       & \qquad  $ \Phi_1$    \qquad  & \qquad  $ \Phi_2$     \qquad   \\ \hline
Type-V   \qquad &  \qquad $\Phi_2$       & \qquad $ \Phi_1$    \qquad  & \qquad  $ \Phi_3$    \qquad  \\ \hline
\end{tabular}
\caption{The five independent types of Yukawa interaction for $n$HDMs, with $n\geq 3$ scalar doublets. In the 2HDM, only the top four interaction types can be realised.}
\label{tab:1}
\end{table}

In this paper, we consider the Type-V realisation of the 3HDM, where each of the Higgs doublets $\Phi_{1,2,3}$ couples to one fermionic family only. Specifically, the Yukawa Lagrangian of the Type-V 3HDM reads
\begin{align}
\begin{autobreak}
  -\mathcal{L}_Y\ =\ y_d^{ij} \overline{Q}_L^{i} \Phi_1 d^j_R\: +\: y_u^{ij} \overline{Q}_L^i \widetilde{\Phi}_2 u_R^j\: +\: y_e^{ij} {\overline{L}_L^{i}} {\Phi_3}\,e^j_R\
 +\ \mathrm{H.c.\,},
\end{autobreak}
\end{align}
where $Q^i_L = \big( u^i_L\,, d^i_L\big)^{\sf T}$, $L^i_L = \big( \nu^i_L\,, e^i_L\big)^{\sf T}$ (with $i=1,2,3$), $\widetilde{\Phi}_2 = i\sigma^2 \Phi^*_2$ is the hypercharge-conjugate of $\Phi_2$, and $\sigma^2$ is the second Pauli matrix.  To endow all SM fermions with a mass at tree level, all three scalar doublets must receive a non-zero VEV, i.e.  $\langle \Phi_1^0\rangle= v_1/\sqrt{2}$, $\langle \Phi_2^0 \rangle = v_2/\sqrt{2}$ and $\langle \Phi_3^0 \rangle = v_3/\sqrt{2}$. Here, we will restrict our attention to CP conservation and to CP-conserving vacua.

Performing the usual linear expansion of the scalar doublets $\Phi_j$ (with $j =1,2,3$) about their VEVs, we may re-express them as
\begin{equation}
\Phi_j \  =\ \left(
 \begin{matrix}
 \phi_j^+ \\
 {1 \over \sqrt{2}} (v_j + \phi_j + i\chi_j)
 \end{matrix}
 \right)\;.
\end{equation}
Taking into account minimization conditions on the CP-conserving 3HDM potential in~\eqref{pot-3hdm} gives rise to the following relation:
\begin{align}
m_{ii}^2\ = \frac{1}{2 v_i} \Big[ 
			&-2 m_{ij}^2 v_j-2 m_{ik}^2 v_k
			+2 \lambda_{ii} v_i^3
			+ (\lambda_{iikk}+\lambda_{ikik}+\lambda_{ikki})  v_i v_k^2
			\nonumber \\ &
			+ (\lambda_{iijj}+\lambda_{ijij} +\lambda_{ijji}) v_i v_j^2	
			+2 (\lambda_{iijk}+\lambda_{ijik}+\lambda_{jiik}) v_i v_j v_k
			\nonumber \\ &
			+(\lambda_{ijjk} +\lambda_{jijk}+\lambda_{jjik}) v_j^2 v_k	
			+ (\lambda_{ikjk}+\lambda_{ikkj}+\lambda_{kkij}) v_j v_k^2
			\nonumber \\ &	
			+\lambda_{jjij} v_j^3
			+3 \lambda_{iiij} v_i^2 v_j
			+3 \lambda_{iiik} v_i^2 v_k		
			+\lambda_{kkik} v_k^3 \Big],
			\label{min-ijk}
\end{align}   
where $i,j,k=1,2,3$ and  $i\neq j \neq k$. In~\eqref{min-ijk}, the VEVs $v_{1,2,3}$ of the three scalar doublets are given by  
\begin{equation}
v_1 = v \cos \beta_1 \cos \beta_2, \quad 
v_2 = v \sin \beta_1 \cos \beta_2, \quad 
v_3 = v \sin \beta_2.
\end{equation}
Note that the VEVs may equivalently be determined by the two ratios, $\tan\beta_1 = v_2 / v_1$ and $ \tan\beta_2 = v_3/\sqrt{v_1^2+ v_2^2}$,  given the constraint that $v \equiv \sqrt{v_1^2 + v_2^2 + v_3^2}$ is the VEV of the SM Higgs doublet.

For later convenience, we now define the three-dimensional rotational matrices about the individual axes $z,\, y$ and $x$ as follows:
\begin{align}
  &R_{12}(\alpha)= \left( \begin{array}{ccccc}
  \cos\alpha & \sin\alpha&0\\ 
  -\sin\alpha&\cos\alpha& 0\\
  0&0& 1
     \end{array}
     \right),\quad
R_{13}(\beta) = \left(
\begin{array}{ccccc} 
\cos\beta& 0& \sin\beta\\
0&1 & 0 \\
-\sin\beta & 0&\cos\beta\end{array}
     \right),\quad
    \nonumber \\
&R_{23}(\gamma) = \left(
\begin{array}{ccccc}  
 1 & 0 & 0\\
 0& \cos\gamma&  \sin\gamma\\
 0& -\sin\gamma& \cos\gamma
  \end{array}
\right),
\end{align}
as well as the two-dimensional rotational matrix
\begin{align}
R(\theta)&= \left(
\begin{array}{ccccc}
\cos \theta& \sin \theta\\ 
-\sin \theta&\cos \theta
     \end{array}
     \right).
\end{align}

Our first step is to transform all scalar fields from a weak basis with generic choice of vacua to the so-called Higgs basis~\cite{Georgi:1978ri,Haber:2006ue}, where only one Higgs doublet acquires the
SM VEV $v$. This can be achieved by virtue of a common orthogonal transformation that involves
the two mixing angles $\beta_1$ and $\beta_2$, i.e.
\begin{equation}
\begin{pmatrix}
H_{1}\\ H_2 \\ H_3
\end{pmatrix}
=
\mathcal{O}_\beta
\begin{pmatrix}
\phi_1 \\ \phi_2 \\ \phi_3
\end{pmatrix},
\qquad 
\begin{pmatrix}
G^0 \\ \eta_1 \\ \eta_2
\end{pmatrix}
=
\mathcal{O}_\beta
\begin{pmatrix}
\chi_1 \\ \chi_2 \\ \chi_3
\end{pmatrix},
\qquad
\begin{pmatrix}
G^{\pm} \\ \eta_1^{\pm} \\ \eta_2^{\pm}
\end{pmatrix}
=
\mathcal{O}_\beta
\begin{pmatrix}
\phi_1^{\pm} \\ \phi_2^{\pm} \\ \phi_3^{\pm}
\end{pmatrix},
\end{equation}
with $\mathcal{O}_\beta\equiv R_{13}(\beta_2)R_{12}(\beta_1)$. After spontaneous symmetry breaking (SSB), the $Z$ and $W^{\pm}$ gauge bosons become massive after absorbing in the unitary gauge the three would-be Goldstone bosons $G^0$ and $G^{\pm}$, respectively~\cite{Goldstone1}. Consequently, the model has nine scalar mass eigenstates: (i)~three CP-even scalars ($H,\,h_{1,2}$), (ii) two CP-odd scalars $(a_{1,2})$, and (iii) four charged scalars~($h_{1,2}^{\pm}$).

Our next step is therefore to determine the composition of the above scalar mass eigenstates in terms of their respective weak fields in the Higgs basis.  This can be done by the orthogonal transformations
\begin{equation}
  \begin{pmatrix} H \\ h_1 \\ h_2 \end{pmatrix} = \mathcal{O}^h
  \begin{pmatrix} H_1 \\ H_2 \\ H_3 \end{pmatrix}, \quad \begin{pmatrix}
 a_1 \\ a_2
\end{pmatrix}
=
R(\rho )
\begin{pmatrix}
\eta_1 \\ \eta_2
\end{pmatrix},
\quad
\begin{pmatrix}
h_1^{\pm} \\ h_2^{\pm}
\end{pmatrix}
=
R(\sigma)
\begin{pmatrix}
\eta_1^{\pm} \\ \eta_2^{\pm}
\end{pmatrix},
\end{equation}
where $\mathcal{O}^h = \mathcal{O}\,\mathcal{O}^{\sf T}_\beta$ and
\begin{align}
   \label{oab}
  \mathcal{O}&=R_{23}(\alpha) R_{13}(\alpha_2)R_{12}(\alpha_1)\,.
\end{align}
In the Higgs basis, spanned by $\{\eta_{1,2}\}$ and $\{\eta^\pm_{1,2}\}$, the CP-odd and charged scalar mass matrices reduce to the $2\times 2$ matrices given by 
\begin{align}
    \label{eq:MatrixPcharged}
  \mathcal{M}^2_{\text{P}}=
  \left(
  \begin{array}{ccc}
  M^2_{\text{P},22} &  M^2_{\text{P},23}
\\
  M^2_{\text{P},32} &  M^2_{\text{P},33}
\\
\end{array}
\right), \qquad
\mathcal{M}^2_{\pm}=
\left(
\begin{array}{ccc}
  M^2_{{\pm},22} &  M^2_{{\pm},23}
\\
  M^2_{{\pm},32} &  M^2_{{\pm},33}
  \\
\end{array}
  \right).
\end{align}
The elements of the above matrices are given explicitly in Appendices~\ref{AppOddMass} and~\ref{AppChargedMass}. 
Upon diagonalisation of the $2\times 2$ mass matrices given in~\eqref{eq:MatrixPcharged}, the masses of the two CP-odd scalars $a_{1,2}$ and the four charged Higgs bosons $h_{1,2}^\pm$ may be computed as 
\begin{align}
  M_{a_1,a_2}^2 &=  {1 \over 2}\bigg[\,M_{{\rm P}, 22}^2 + M_{{\rm P},33}^2 \mp \sqrt{(M_{{\rm P}, 22}^2- M_{{\rm P},33}^2)^2+4 M_{{\rm P},23}^4}\,\bigg]\,,\\[2mm]
  M^2_{h_1^{\pm},h_2^{\pm}} &=  {1\over 2}\bigg[\,M_{\pm, 22}^2 + M_{\pm,33}^2 \mp \sqrt{(M_{\pm,22}^2- M_{\pm,33}^2)^2+4 M_{\pm,23}^4}\,\bigg]\,.
\end{align}
In addition, the mixing angles $\rho$ and $\sigma$ may be determined by
\begin{align}
  \tan{2\rho}={2 M_{{\rm P},23} ^2\over M_{{\rm P},22} ^2-M_{{\rm P},33}^2}, \quad \tan{2\sigma}={2 M_{\pm,23} ^2 \over M_{\pm,22} ^2-M_{\pm,33}^2}.
\end{align}

Finally, the masses for the three CP-even scalars, $H$, $h_1$ and $h_2$ can be evaluated by diagonalising the squared mass matrix ${\mathcal{M}}^2_{\rm S}$, expressed in the general weak basis $\{\phi_{1,2,3}\}$, by employing the orthogonal matrix ${\mathcal{O}}$ defined in~\eqref{oab}.
In this general weak basis, ${\mathcal{M}}^2_{\rm S}$ reads
\begin{align}
    \label{eq:MS}
{\mathcal{M}}_{\rm S}^2=\left(
\begin{array}{ccc}
A & {C}_{1} & {C}_{2}
\\
{C}_{1} & B_1& {C}_{3}
\\
{C}_{2} & {C}_{3} &B_2
\\
\end{array}
\right),
\end{align}
where the analytic form of all its entries is given in Appendix~\ref{AppevenMass}. 
Therefore, the diagonal mass-basis matrix $\overline{\mathcal{M}}^2_{\rm S}$ for the physical CP-even scalars takes on the form
\begin{align}
\overline{\mathcal{M}}_{\rm S}^2 =
\left(
\begin{array}{ccc}
{M}_{H}^2 & 0 & 0
\\
0 & {M}_{h_1} ^2& 0
\\
0 & 0 &{M}_{h_2}^2
\\
\end{array}
\right)=\ \mathcal{O}\,
{\mathcal{M}}_{\rm S}^2 \,\mathcal{O}^{\sf T},
\label{mass-wtom}
\end{align}
with the convention of mass ordering: $M_H \le M_{h_1} \le M_{h_2}$.

 In order to determine all possible limits of SM alignment, it proves convenient to perform a diagonalisation of the CP-even mass matrix ${\mathcal{M}}^2_{\rm S}$ in two steps.  In the first step, we use two angles, e.g.~$\alpha_{1,2}$, to project out the mass eigenvalue of the SM-like Higgs state. For instance, if $H$ is identified with the observed SM-like Higgs boson, then ${\mathcal{M}}^2_{\rm S}$ may be block-diagonalised as follows:
\begin{align}
    \label{mass-wtomH}
\widetilde{\mathcal{M}}_{\rm S}^{2}=
\left(
\begin{array}{ccc}
{M}_{H}^2 & 0 & 0
\\
0 & \widetilde{B}_1& \widetilde{C}_3
\\
0 & \widetilde{C}_3 & \widetilde{B}_2
\\
\end{array}
\right)= \mathcal{O}_\alpha\,
{\mathcal{M}}_{\rm S}^2 \, \mathcal{O}_\alpha^{\sf T},
\end{align}
with $\mathcal{O}_\alpha = R_{13}(\alpha_2)R_{12}(\alpha_1)$.  Here, the mixing angles $\alpha_{1,2}$ are given in terms of the mass-matrix elements in~\eqref{eq:MS} by
 \begin{align}
   \label{alpha1}
\tan 2\alpha_{1}\ &=\ \frac{2C_1}{A-B_1},\; 
\\
 \label{alpha2}
\tan 2\alpha_{2}\ &=\ \frac{4C_3/\sin\alpha_1}{A+B_1-2 B_2+ \big[(A-B_1)/\cos 2\alpha_1\big]}\;,
\end{align}
where $C_3/C_2 =\tan \alpha_1$.
The second step of diagonalisation consists in bringing the matrix~$\widetilde{\mathcal{M}}_{\rm S}^{2}$ in a fully diagonal form.
This can be done by using the orthogonal transformation $R_{23}(\alpha)$,
 \begin{align}
\overline{\mathcal{M}}_{\rm S}^2 =
\left(
\begin{array}{ccc}
{M}_{H}^2 & 0 & 0
\\
0 & {M}_{h_1} ^2& 0
\\
0 & 0 &{M}_{h_2}^2
\\
\end{array}
\right)\, =\ R_{23}(\alpha) \,
\widetilde{\mathcal{M}}_{\rm S}^2 \,R^{\sf T}_{23}(\alpha)\,,
\label{mass-mHtom}
\end{align}
where the squared masses of the two heavy CP-even scalars, $h_{1,2}$, are
\begin{align}
    \label{h1,h2}
  M^2_{h_1,h_2} &=  {1 \over 2}\bigg[\,\widetilde{B}_1 + \widetilde{B}_2 \mp
                  \sqrt{(\widetilde{B}_1- \widetilde{B}_2 )^2+4 {\widetilde{C}_3}^2}\,\bigg].
\end{align}
Thus, the mixing angle $\alpha$ can be evaluated in terms of the above mass parameters
as 
\begin{align}
 \label{alpha3}
\tan{2\alpha}\ = \ {2 \widetilde{C}_3\over \widetilde{B}_1- \widetilde{B}_2}\,.
\end{align}

In this CP-conserving 3HDM, the SM-normalised couplings of the SM-like Higgs boson to
the EW gauge bosons ($V = Z,W^{\pm}$) are calculated to be:
\begin{align}
    \label{eq:gHVV}
  g_{HVV}\ =&\ \cos{\alpha_2} \cos{\beta_2} \cos({\beta_1-\alpha_1 }) \,+\, \sin{\alpha_2} \sin{\beta_2},\\[3mm]
     \label{eq:gh1VV}
g_{h_1VV}\ =&\cos\alpha \, \cos\beta_2 \sin( \beta_1-\alpha_1)\, + \, \sin\alpha \, \Big[ \cos\alpha_2 \sin\beta_2 
\nonumber \\ &  -\cos\beta_2 \sin\alpha_2\cos(\beta_1-\alpha_1)\,\Big], 
  \\[3mm]
  \label{eq:gh2VV}
  g_{h_2VV}\ =&\ \cos\alpha\, \Big[ \cos\alpha_2 \sin\beta_2-\cos\beta_2\sin\alpha_2\cos(\beta_1-\alpha_1)\,\Big]
                \nonumber \\ &-\sin\alpha \, \cos\beta_2 \sin(\beta_1-\alpha_1),
\end{align}
which obey the sum rule: $g_{HVV}^2+ g^2_{h_1VV}+ g^2_{h_2VV} = 1$.
Evidently, there are three possible scenarios for which the 125-GeV resonance
can identified with the SM-like Higgs boson. The first one is the so-called
{\em canonical} SM-like Higgs scenario, where $M_H \approx 125$~GeV with coupling
strength $g_{HVV} = 1$, but $g_{h_{1,2}VV} = 0$. In this case, there are two possible arrangements for
the mixing angles:
\begin{equation}
   \label{eq:HSM}
   \text{(i)}\ \beta_1\, =\, \alpha_1, \quad\beta_2\, =\, \alpha_2\,,\qquad
   \text{(ii)}\ \beta_2\, =\, \alpha_2\, =\, \pm{\pi }/2\,.
 \end{equation}
 The second possibility is that the $h_1$ boson represents the observed scalar resonance at 125~GeV, with $g_{h_1VV}=1$. In this case, the $H$ scalar is lighter than the $h_1$,
 whereas the $h_2$ is heavier. For this SM-like $h_1$ scenario, we have four possibilities
 to arrange the mixing angles:
 \begin{align}
    \label{eq:h1SM}
  \text{(i)}\quad \beta_1 =&\ \alpha_1, \qquad \beta_2\, =\, \alpha_2\pm{\pi }/2,
                             \quad\alpha=\pm{\pi }/2\,,\nonumber\\
   \text{(ii)}\quad \beta_1=&\ \alpha_1+{\pi }/2, \qquad \beta_2\, =\, \alpha, \qquad\alpha_2\,=\,0\,,\nonumber\\
   \text{(iii)}\quad \beta_1=&\ \alpha_1+{\pi }/2, \qquad \beta_2\, =\, \alpha=0\,,\nonumber\\
   \text{(iv)}\quad \beta_2=&\ \alpha\, =\, \pm{\pi }/2, \qquad \alpha_2\, =\, 0\,.
 \end{align}
 The last possibility is the option that $h_2$ represents the observed SM-like
 Higgs boson, with $g_{h_2VV}=1$. Here, the $H$ and $h_1$ scalars are lighter than
 the SM-like $h_2$ boson, with vanishing couplings to the EW gauge bosons. This scenario may be achieved for the following four choices
 of mixing angles:
 \begin{align}
   \label{eq:h2SM}
\text{(i)}\quad \beta_1 =&\ \alpha_1, \quad\beta_2\, =\, \alpha_2+{\pi }/2, \quad\alpha=0,\nonumber\\
\text{(ii)}\quad \beta_1 =&\ \alpha_1+{\pi }/2, \quad\beta_2\, =\, \alpha+{\pi }/2, \quad\alpha_2\, =\, 0,\nonumber\\
\text{(iii)}\quad \beta_1 =&\ \alpha_1\pm{\pi }/2, \quad\beta_2\, =\, 0,
                                                           \quad \alpha\, =\, \mp{\pi}/2,\nonumber\\
\text{(iv)}\quad \beta_2 =&\ {\pi }/2, \quad\alpha_2\, =\, \alpha=0\,.
\end{align}
In this paper, we consider the canonical SM-like $H$ scenario, in which 
SM alignment is obtained when $\beta_1=\alpha_1$ and $\beta_2=\alpha_2$,
according to the option~(i) in~\eqref{eq:HSM}.

In the Higgs basis $\{ H_{1,2,3}\}$, the $3 \times 3$ mass matrix of the CP-even scalars
is given by
 \begin{align}
\widehat{\mathcal{M}}_{\rm S}^2=\left(
\begin{array}{ccc}
    \label{mass-wtoH}
\widehat{A} & \widehat{C}_{1} & \widehat{C}_{2}
\\
\widehat{C}_{1} & \widehat{B}_{1}& \widehat{C}_{3}
\\
\widehat{C}_{2} & \widehat{C}_{3} &\widehat{B}_{2}
\\
\end{array}
\right)=\mathcal{O}_\beta\,
\mathcal{M}_{\rm S}^{2} \,  \mathcal{O}^{\sf T}_\beta\,,
\end{align}
with
\begin{align}
\widehat{A}\ =&\ 
 \frac{1}{4} \Big[\,A 
   + B_{1} 
   + 2 B_{2}
   +2 (A-B_{1}) c_{2\beta_1} c_{\beta_2}^{2} 
   + (A+B_{1}-2 B_{2})c_{2\beta_2} 
 \nonumber \\
 &+ 8 c_{\beta_2} (C_{1} s_{\beta_1} c_{\beta_1} c_{\beta_2}
  + C_{2} c_{\beta_1} s_{\beta_2}
  + C_{3} s_{\beta_1} s_{\beta_2})
  \Big]
, \nonumber \\[3mm]
\widehat{B}_{1} =&\ 
 \frac{1}{2} \Big[A
 + B_{1}
+ (B_{1}-A) c_{2\beta_1}
 - 2 C_{1} s_{2\beta_1}
 \Big]
, \nonumber \\[3mm]
\widehat{B}_{2} =&\ 
 \frac{1}{4} \Big\{
A+B_{1}+2 B_{2}+ 2 s_{\beta_2}^{2} \Big[ (A-B_{1}) c_{2\beta_1}+2 C_{1} s_{2\beta_1}\Big]
\nonumber \\
 & 
 - c_{2\beta_2} (A+B_{1}-2 B_{2})
 -4 s_{2\beta_2} (C_{2} c_{\beta_1}+C_{3} s_{\beta_1})
 \Big\}
, \nonumber \\[3mm]
\widehat{C}_{1} =&\
 c_{\beta_1} \Big[ (B_{1}-A) s_{\beta_1} c_{\beta_2}+C_{3} s_{\beta_2}\Big]\: +\: C_{1} c_{2\beta_1} c_{\beta_2}-C_{2} s_{\beta_1} s_{\beta_2}\,, \nonumber\\[3mm]
\widehat{C}_{2} =&\ -\frac{1}{4} s_{2\beta_2} \big((A-B_{1}) c_{2\beta_1}+A+B_{1}-2 B_{2}+2 C_{1} s_{2\beta_1}\big)+c_{2\beta_2} (C_{2} c_{\beta_1}+C_{3} s_{\beta_1}), \nonumber \\[3mm]
\widehat{C}_{3} =&\ 
                   c_{\beta_1} \Big[ (A-B_{1}) s_{\beta_1} s_{\beta_2}+C_{3} c_{\beta_2}\Big] -
                   C_{1} c_{2\beta_1} s_{\beta_2}-C_{2} s_{\beta_1} c_{\beta_2}\,, 
\end{align} 
where we employed the shorthand notation: $c_x\equiv \cos x$ and $s_x\equiv \sin x$.
In the SM alignment limit $\beta_1=\alpha_1$ and $\beta_2=\alpha_2$ under consideration,
the mass parameter $\widehat{A}$ becomes equal to~$M_H^2$, while the parameters $\widehat{C}_1$ and $\widehat{C}_2$ vanish. Thus, taking the limit $\widehat{C}_{1,2} \to 0$, the following relationships between the quartic couplings may be derived:
 \begin{align}
  \label{Ci=0}
&\lambda_{11}=\lambda_{22}=\lambda_{33},\nonumber \\
 &\lambda_{1122}=\lambda_{1133}=\lambda_{2233}, \nonumber \\ &\lambda_{1221}=\lambda_{1331}=\lambda_{2332},\nonumber \\
&\lambda_{1212}=\lambda_{1313}=\lambda_{2323}=2\lambda_{11}-\lambda_{1122}-\lambda_{1221},
\end{align}
while the remaining quartic couplings are zero. These conditions may be obtained independently of any values for the mixing angles $\beta_1$ and $\beta_2$. Moreover, for a small but calculable value of misalignment, the deviations of $\beta_1-\alpha_1$ and $\beta_2-\alpha_2$ from zero can be parameterised as follows:
 \begin{align}
   \label{alpha-hat-1}
\tan 2(\beta_1-\alpha_1)\ &=\ {2 \widehat{C}_{1}  \over \widehat{A}- \widehat{B}_1},\; 
\\
  \label{alpha-hat-2}
\tan 2(\beta_2-\alpha_2) \ &=\ {4 \widehat{C}_3/\sin(\beta_1-\alpha_1) \over  \widehat{A}+ \widehat{B}_1-2  \widehat{B}_2+ \big[(\widehat{A}- \widehat{B}_1)/\cos 2(\beta_1-\alpha_1)\big]}\;,
\end{align}
where relations analogous to~\eqref{alpha1} and \eqref{alpha2} have been used for
the hat quantities occurring in the CP-even mass matrix of \eqref{mass-wtoH}.

Having determined the mixing angles in terms of the matrix elements of $\widehat{\mathcal{M}}_{\rm S}^2$, we may now calculate the reduced $H$-boson couplings to the EW gauge bosons in a power expansion of $\widehat{C}_{1,2}/\widehat{B}_{1,2}$. To order $\widehat{C}_{1,2}^2/\widehat{B}^2_{1,2}$, these are given by the following approximate analytic expressions:
\begin{align}
    \label{ex-gc}
g_{HVV}\ &\simeq\ 1\, -\, { \widehat{C}_1^2 \over 4 (\widehat{A}- \widehat{B}_1)^2}\;,\\
    g_{h_1VV}\ &\simeq\ c_{\alpha} c_{\beta_2}\, {\widehat{C}_1 \over \widehat{A}- \widehat{B}_1} + s_{\alpha}\, {\widehat{C}_2 \over 4 (\widehat{A}- \widehat{B}_2)} \,\bigg(1+{ \widehat{C}_1 \over \widehat{A}- \widehat{B}_1}\bigg)\;,\\                                                                                                                                                g_{h_2VV}\ &\simeq\  c_{\alpha} {\widehat{C}_2 \over 4 (\widehat{A}- \widehat{B}_2)} \,\bigg(1+{ \widehat{C}_1 \over \widehat{A}- \widehat{B}_1}\bigg) - s_{\alpha}c_{\beta_2}\, {\, \widehat{C}_1 \over \widehat{A}- \widehat{B}_1}\; .
\end{align}
Likewise, the SM-normalised couplings of the CP-even scalars to up-type, down-type quarks and charged leptons are
\begin{align}
    \label{hqq}
 g_{(H/h_1/h_2) d \bar{d}}\ &=\ {v \over v_1} \mathcal{O}_{11/21/31}\,,\\
g_{(H/h_1/h_2) u \bar{u}}\ &=\ {v \over v_2} \mathcal{O}_{12/22/32}\,,\\
g_{(H/h_1/h_2) l \bar{l}}\ &=\ {v \over v_3} \mathcal{O}_{13/23/33}\,.  
\end{align}
To order $\widehat{C}_{1,2}^2/\widehat{B}^2_{1,2}$, the SM-normalised couplings of the $H$ boson
to fermions are dictated by the following approximate analytic formulae:
\begin{align}
     \label{ex-d}
 g_{H d \bar{d}}\ &\simeq\ 1- {(t_{\beta_1} + t_{\beta_2})\, \widehat{C}_1 \over   \widehat{A}- \widehat{B}_1}-{(1-t_{\beta_1}  t_{\beta_2})\, \widehat{C}_1^2 \over   (\widehat{A}- \widehat{B}_1)^2}\; ,\\
     \label{ex-u}
 g_{H u \bar{u}}\ &\simeq\ 1 - {t_{\beta_1}^{-1} \widehat{C}_1  \over  (\widehat{A}-\widehat{B}_1)}+{t_{\beta_2} (\widehat{A}-\widehat{B}_1-\widehat{C}_1 t_{\beta_1}^{-1})  \widehat{C}_2 \over  2(\widehat{A}-\widehat{B}_1)(\widehat{A}-\widehat{B}_2)}
-{ \widehat{C}_1^2 \over  2 (\widehat{A}- \widehat{B}_1)^2} 
-{ \widehat{C}_2^2 \over 8 (\widehat{A}- \widehat{B}_2)^2} \;, \\
     \label{ex-l}
 g_{H l \bar{l}}\ &\simeq\ 1-{2 t_{\beta_2}^{-1}(\widehat{A}-\widehat{B}_1) \widehat{C}_2 \over 4 (\widehat{A}-\widehat{B}_1)(\widehat{A}-\widehat{B}_2) +  \widehat{C}_1^2}
-{2(\widehat{A}-\widehat{B}_1)^2 \widehat{C}_2^2\over \big[4 (\widehat{A}-\widehat{B}_1)(\widehat{A}-\widehat{B}_2) +  \widehat{C}_1^2\big]^2}\ .
\end{align}
We note that the analytic expressions of the reduced $H$-boson couplings in~\eqref{ex-gc} and in~\eqref{ex-d}--\eqref{ex-l}  go to the SM value 1, when the exact SM alignment limit~$\widehat{C}_{1,2} \to 0$ is considered, or when the new-physics mass scales $\widehat{B}_{1,2} \sim M^2_{h_{1,2}}$ are taken to infinity.  Since we are interested in the former possibility which in turn implies a richer collider phenomenology, we will study scenarios in which SM alignment is accomplished by virtue of symmetries.

In~the next section, we will discuss restrictions on the model parameters of the 3HDM that emanate from SM alignment, and identify possible maximal symmetries that can be imposed on the 3HDM potential to fulfill these restrictions.

\vfill\eject

\section{Symmetries for SM alignment in 3HDM}\label{sym}

After outlining the basic features of the 3HDM in the previous section, we now review the key symmetries for SM alignment for this class of models~\cite{Pilaftsis:2016erj}. To this end, we introduce the $12$-dimensional $\mathrm{SU(2)}_L$-covariant ${\bm{\Phi}}$-multiplet~\cite{Battye:2011jj,Darvishi:2019dbh},
\begin{align}
\label{eq:bfPhi}
&{\bm{\Phi}}^{\mathsf{T}}  =\begin{pmatrix}
\Phi_1, \,
\Phi_2, \,
\Phi_3, \,
\widetilde{\Phi}_1, \,
\widetilde{\Phi}_2,  \,
\widetilde{\Phi}_3 \,
\end{pmatrix}^{\mathsf{T}},
\end{align}
where $\widetilde{\Phi}_i=i \sigma^{2} \Phi_i^{*}$ (with $i=1,2,3$) are the U(1)$_Y$ hypercharge-conjugates of $\Phi_i$ and $\sigma^{1,2,3}$ are the Pauli matrices. Moreover, the ${\bm{\Phi}}$-multiplet satisfies the Majorana-type property~\cite{Battye:2011jj},
\begin{align}
{\bm{\Phi}}\, =\, {\rm C}\,{{\bm \Phi}^*}, 
\end{align}
where ${\rm C}=\sigma^{2} \otimes {\bf{1}}_3 \,\otimes \, \sigma^{2}$ (with ${\rm C}={\rm C}^{-1}={\rm C}^*$) is the charge conjugation operator and ${\bf{1}}_3$ is the $3 \times 3$ identity matrix. With the help of the ${\bm{\Phi}}$-multiplet, one may now define the bilinear field vector~\cite{Darvishi:2019dbh},
\begin{equation}
R^{A} \, \equiv\,  {\bm{\Phi}}^{\dagger} {\Sigma}^{A} {\bm{\Phi}}\,,
\end{equation}
with ${A}=0,1,2,\dots,14$. The ${\Sigma}^{A}$ matrices have $12 \times 12$ elements and
can be expressed in terms of double tensor products as follows:
\begin{align} {\Sigma}^{A}\, =\, \Big(\sigma^0 \otimes t_S^a \otimes \sigma^0\,,
 \ \sigma^i \otimes t_A^b \otimes \sigma^0 \Big)\,,
\end{align}
where $t_S^a$ and $ t_A^b$ stand for the symmetric and the anti-symmetric generators of the $\mathrm{SU(3)}$ group, respectively. Requiring that the canonical $\mathrm{SU(2)}_L$ gauge-kinetic terms of the scalar doublets $\Phi_{1}$, $\Phi_{2}$ and $\Phi_{3}$ remain invariant, then the maximal symmetry of the 3HDM is the symplectic group $\mathrm{Sp(6)}$ for vanishing hypercharge gauge coupling $g'$ and fermion Yukawa couplings. In an earlier study~\cite{Darvishi:2019dbh}, the full classification of all $40$ accidental symmetries for the 3HDM has been presented. This classification is obtained by working out all distinct subgroups of the maximal symmetry group $G^{\Phi}_\mathrm{3HDM}=\left[\mathrm{Sp(6)}/Z_2\right]\otimes\mathrm{SU(2)}_L$ in a bilinear field-space formalism. From these $40$ symmetries, only a few possess the desirable property of natural SM alignment.

In the 3HDM, the constraints required for SM alignment can be realised naturally by virtue of three symmetries~\cite{Pilaftsis:2016erj,Darvishi:2020teg}: (i)\, the maximal symmetry group $\mathrm{Sp(6)}$, (ii)\, $\mathrm{SU(3)_{\rm HF}}$, and (iii)\, $\mathrm{ SO(3)_{\rm HF} \times}$ CP symmetries. Here, the abbreviation HF indicates Higgs Family symmetries that only involve the elements of $\Phi=(\Phi_1,\Phi_2,\Phi_3)^{\mathsf{T}}$ and not their complex conjugates. The construction of the potential invariant under these symmetries is facilitated by an earlier developed technique based on prime invariants~\cite{Pilaftsis:2016erj,Darvishi:2020teg,Darvishi:2019dbh}. 

The MS-3HDM potential can be constructed by means of the Sp(6)-invariant expression,
\begin{equation}
     \label{eq:S}
    S\, =\, \mathbf{\Phi}^\dagger\mathbf{\Phi}\, =\, \Phi_1^\dagger\Phi_1 + \Phi_2^\dagger\Phi_2 + \Phi_3^\dagger  \Phi_3\, .
\end{equation}
In particular, it takes on the following minimal form:
\begin{equation}
V_{\rm MS-3HDM}\ =\ -\, m^2\, \Big( \left| \Phi_1 \right|^2+
  \left| \Phi_2 \right|^2 + \left| \Phi_3 \right|^2 \Big)\: 
 +\: \lambda\, \Big( \left| \Phi_1 \right|^2 + \left| \Phi_2 \right|^2+ \left| \Phi_3 \right|^2
   \Big)^2, 
   \label{ms-3hdm-v0}
\end{equation}
where
\begin{align}
&m^2\,=\,m_{11}^2\,=\,m_{22}^2\,=\,m_{33}^2\,,\nonumber \\
&\lambda\, =\, \lambda_{11}\, =\, \lambda_{22}\, =\, \lambda_{33}\, =\, 2\lambda_{1122}\, =\, 2\lambda_{1133}\, =\, 2\lambda_{2233}\,,
\nonumber
\end{align}
and all other potential parameters are set to zero.

The $\mathrm{SU(3)_{\rm HF}}$-invariant 3HDM potential may be expressed as a function of $S$, given in~\eqref{eq:S}, and the $\mathrm{SU(3)_{\rm HF}}$-invariant $D^2 \equiv D^aD^a$, where 
\begin{align}
    \label{eq:Da}
 D^a\,  =\, {{\Phi}}^\dagger \sigma^a {{\Phi}}\, =\, \Phi_1^{\dagger} \sigma^a \Phi_1 +
\Phi_2^\dagger \sigma^a \Phi_2 + \Phi_3^\dagger \sigma^a \Phi_3\, ,
\end{align}
with $a=1,2,3$ and $\Phi = \big(\Phi_1\,,\, \Phi_2\,,\,\Phi_3\big)^{\sf T}$. 
Hence, the model parameters of the $\mathrm{SU(3)_{\rm HF}}$-invariant 3HDM potential satisfy the following relations:
\begin{align}
&m_{11}^2\, =\, m_{22}^2\, =\, m_{33}^2\,,\nonumber \\
& \lambda_{11}\, =\, \lambda_{22}\, =\, \lambda_{33}\,,\nonumber \\
& \lambda_{1122}\, =\, \lambda_{1133}\, =\, \lambda_{2233}\,,\nonumber \\
&\lambda_{1221}\, =\, \lambda_{1331}\, =\, \lambda_{2332}\, =\, 2 \lambda_{11}-\lambda_{1122}\,.
 \end{align}
 
 By analogy, an $\mathrm{SO(3)_{\rm HF}}$-invariant 3HDM potential can be constructed
 by utilising the invariants $S$ [cf.~\eqref{eq:S}], $D^2$ [cf.~\eqref{eq:Da}], and
 $T^2\equiv {\rm Tr}(TT^*)$, where
 \begin{align}
     \label{eq:T}
T\ =\, \Phi\, \Phi^{\mathsf{T}}\, . 
\end{align}
In this case, the following relationships among the potential parameters may be derived:
\begin{align}
&m_{11}^2\,=\,m_{22}^2\,=\,m_{33}^2\,,\nonumber \\
& \lambda_{11}\,=\,\lambda_{22}\,=\,\lambda_{33}\,, \nonumber \\
&
\lambda_{1122}\,=\,\lambda_{1133}\,=\,\lambda_{2233}\,, \quad 
\nonumber \\
&
\lambda_{1221}\,=\,\lambda_{1331}\,=\,\lambda_{2332}\,,\nonumber \\
&
\lambda_{1212}\,=\,\lambda_{1313}\,=\,\lambda_{2323}\,=\,2\lambda_{11} - \lambda_{1122} - \lambda_{1221}\,,
\end{align}
and all other parameters not listed above vanish.

In summary, if a 3HDM potential is invariant under one of the three symmetries: Sp(6), SU(3)$_{\rm HF}$ and SO(3)$_{\rm HF}\times$CP, then its parameters will satisfy the alignment conditions stated in~\eqref{Ci=0}, for any value of the mixing angles $\beta_1$ and $\beta_2$. To render such constrained 3HDM potentials phenomenologically viable, we may have to include arbitrary soft symmetry-breaking bilinear masses, $m^2_{ij}$ (with $i,j=1,2,3$). However, such soft symmetry breakings do not spoil SM alignment~\cite{Pilaftsis:2016erj,Darvishi:2020teg}.  Obviously, the MS-3HDM is the most economic setting that realises naturally such an alignment from the general class of 3HDMs.
In particular, we find that in the MS-3HDM, the mixing angles that diagonalise
the heavy sectors of the CP-even, CP-odd and charged- scalar mass matrices are all equal, i.e.
\begin{equation}
  \label{eq:alpharhosigma}
\alpha\, =\, \rho\, =\, \sigma\, . 
\end{equation}
As will see in the next sections, this is a distinct and unique feature of the MS-3HDM under study with respect to other aligned 3HDMs.

\section{Breaking Pattern of the Type-V MS-3HDM}\label{MS-3HDM}

In this section, we will discuss the breaking pattern of the Type-V MS-3HDM, where each Higgs doublet $\Phi_{1,2,3}$ couples to one type of fermions only, as given in~Table~\ref{tab:1}. In particular, we notice that after SSB and in the Born approximation, the MS-3HDM predicts one CP-even scalar $H$ with non-zero squared mass $M^2_H= 2 \lambda_{22} v^2$, while all other scalars, $h_{1,2},\, a_{1,2}$ and $h_{1,2}^\pm$, are massless pseudo-Goldstone bosons with sizeable gauge and Yukawa interactions. Hence, as we will see below, we need to consider two dominant sources that violate the Sp(6) symmetry of the theory: (i)~the RG effects of the gauge and Yukawa couplings on the potential parameters, and (ii)~soft symmetry-breaking bilinear masses, so as to make all the pseudo-Goldstone fields sufficiently heavy in agreement with current LHC data and other low-energy experiments. 

To start with, let us first consider the effects of RG evolution, within a perturbative framework of the MS-3HDM. Given that physical observables, such as $S$-matrix elements, e.g.~$S(\mu,g(\mu),m(\mu))$, are invariant under changes of the RG scale~$\mu$, one 
derives the well-known fundamental relation:
\begin{align}
\mu { \partial S \over \partial \mu}\: +\: \beta\, {\partial S \over \partial g}\:  -\: \gamma_{m^2}\, m^2 {\partial S \over \partial m^2}\ =\ 0\,.
\end{align}
In the above, the generic coupling constant $g$ and mass parameter~$m^2$ satisfy typical RG equations:
\begin{align}
  \beta(g)\, =\, \mu {\partial g(\mu) \over \partial \mu},   \qquad \beta(m^2)\,\equiv\, \gamma_{m^2}(g)\, m^2\,=\, - \mu\,{ \partial m^2(\mu) \over \partial \mu}\, ,
\end{align}
where $\beta(g)$ and $\gamma_{m^2}(g)$ are the beta function and the anomalous dimension, respectively. As~explicitly indicated, $\beta(g)$ and $\gamma_{m^2}(g)$ depend only on $g$ in the Minimal-Subtraction~(MS) scheme. In Appendix~\ref{RGEs} we present the complete set of one- and two-loop beta functions of all gauge, Yukawa and quartic couplings, for the general 3HDM, including beta functions for potential mass parameters and the VEVs of the scalar doublets.

Although the aforementioned RG effects generate masses for the CP-even and charged pseudo-Goldstone bosons, $h_{1,2}$ and $h_{1,2}^\pm$, the corresponding CP-odd pseudo-Goldstone bosons, $a_{1,2}$, remain massless. The latter two states play the role of axions thanks to the presence of two Peccei--Quinn~(PQ) symmetries $\mathrm{U(1)}_{\mathrm{PQ}} \otimes \mathrm{U(1)}'_{\mathrm{PQ}}$~\cite{Peccei:1977hh,Weinberg:1977ma,Wilczek:1977pj}.  However, these visible EW-scale axions have strong couplings to gauge and SM fermions, and as such, they are ruled out experimentally. For this reason, we admit a second source for violating the Sp(6) symmetry in the MS-3HDM, through the introduction of soft Sp(6)-breaking bilinear mass terms~$m^2_{ij}$.  As a consequence, the complete MS-3HDM potential is given by
\begin{equation}
   \label{eq:V+DV}
   V\ =\ V_{{\rm MS}-{\rm 3HDM}}\: +\: \Delta V\,,
\end{equation}
where
\begin{equation}
  \label{eq:DV}
  \Delta V\ =\ \displaystyle\sum_{i,j=1}^3\, m_{ij}^2\, ( \Phi_i^{\dagger} \Phi_j)\, .
\end{equation}
After EW symmetry breaking, the diagonal mass terms $m^2_{ii}$ may be eliminated in favour of the VEVs $v_{1,2,3}$ and the off-diagonal terms $m^2_{i\neq j}$ (with $i,j=1,2,3$), according to~\eqref{min-ijk}. Hence, to a very good approximation, the scalar masses are found to be:
\begin{align}
    \label{eq:mij}
 M_H^2\  &\simeq\ 2\lambda_{22} v^2, \nonumber \\ 
 M_{h_1}^2\ &\simeq \, M_{a_1}^2\ \simeq\ M_{h_1^{\pm}}^2\ \simeq\  {1 \over 2}\bigg[\,S_1 + S_2- \sqrt{(S_1- S_2 )^2+4 S_3^2}\,\bigg]\;, \nonumber \\
  M_{h_2}^2\ &\simeq\ M_{a_2}^2\ \simeq \ M_{h_2^{\pm}}^2\ \simeq\  {1 \over 2}\bigg[\,S_1 + S_2+\sqrt{(S_1- S_2 )^2+4 S_3^2}\,\bigg]\;, 
\end{align}
with 
\begin{align}
  S_1\ &=\ \frac{1}{c_{\beta_1} s_{\beta_1}}\, \Big[ {m_{12}^2}\ +\  { t_{\beta_2}}\big({{m_{13}^2 \, s_{\beta_1}^3} + m_{23}^2 \, c_{\beta_1}^3 }\big)\Big]\; ,
\nonumber \\
S_2\ &=\ \frac{1}{c_{\beta_2} s_{\beta_2}}\, \Big( m_{13}^2 \, c_{\beta_1}\ +\  m_{23}^2 \, s_{\beta_1} \Big)\;,
\nonumber \\
S_3\ &=\ \  \frac{1}{c_{\beta_2}}\, \Big( m_{13}^2 \, s_{\beta_1}\ -\ m_{23}^2 \, c_{\beta_1} \Big)\;.
\end{align}
Notice that with the introduction of soft symmetry-breaking bilinears, all pseudo-Goldstone bosons, $h_{1,2}$, $a_{1,2}$ and $h_{1,2}^\pm$, receive appreciable masses at the tree level. 

To sum up, we consider an $\mathrm{Sp(6)}$ symmetry imposed on the MS-3HDM potential which is exact (up to soft symmetry-breaking masses) at some high-energy scale $\mu_X$, at which  quartic coupling unification occurs. Then, the following breaking pattern will generally take place for the Type-V MS-3HDM:
\begin{eqnarray}
  \mathrm{Sp(6)} \otimes \mathrm{SU(2)}_L &\xrightarrow[]{y_u,y_d,y_e \neq 0}&
 \mathrm{Sp(2)} \otimes \mathrm{Sp(2)} \otimes \mathrm{Sp(2)}
 \otimes \mathrm{SU(2)}_L 
 \nonumber \\ 
 &\xrightarrow[]{g' \neq 0}&
\mathrm{U(1)}_{\mathrm{PQ}} \otimes \mathrm{U(1)}'_{\mathrm{PQ}} \otimes \mathrm{U(1)}_{\mathrm{Y}} \otimes \mathrm{SU(2)}_L
  \nonumber \\ 
 &\xrightarrow[]{\left< \Phi_{1,2,3} \right> \neq 0}&
\mathrm{U(1)}_{\mathrm{PQ}} \otimes \mathrm{U(1)}'_{\mathrm{PQ}} \otimes  \mathrm{U(1)}_{\mathrm{em}}
  \nonumber \\ 
 &\xrightarrow[]{m_{ij}^2 \neq 0}&
\mathrm{U(1)}_{\mathrm{em}}\; .
\end{eqnarray}
In our analysis detailed in the next section, we take charged scalar masses ${M_{h_{1,2}^{\pm}}\stackrel{>}{{}_\sim} 500\,\text{GeV}}$ as input parameters, while maintaining agreement with $B$-meson constraints~\cite{Aad:2021xzu}. For the benchmark scenarios that we will be studying, we will have $M_{h_1^{\pm}} \sim M_{h_2^{\pm}}$, such that we can ignore any RG effects between the two charged scalar masses, and so perform a matching of the MS-3HDM to the SM at a single threshold.

\vfill\eject

\section{Quartic Coupling Unification}\label{QCU}

In this section we will study quartic coupling unification in the Type-V 3HDM. In analogy to a previous analysis for the MS-2HDM~\cite{Darvishi:2019ltl}, we find that in the MS-3HDM all quartic couplings can unify to a single value $\lambda$ at very high-energy scales~$\mu_X$. As the highest scale of unification, we take the values at which $\lambda (\mu_X) = 0$. Like in the MS-2HDM, we will see that there are two such conformally invariant unification points in the MS-3HDM which we distinguish them as~$\mu^{(1,2)}_X$, with $\mu^{(1)}_X \le \mu^{(2)}_X$. 

In our analysis we employ two-loop Renormalization Group Equations (RGEs) to evaluate the running of all relevant MS-3HDM parameters from the unification point~$\mu_X$ to the thres\-hold mass scale~$\mu_{\rm thr} = M_{h_1^{\pm}} \sim M_{h_2^\pm}$. Below $\mu_{\rm thr}$, we assume that the SM is a good effective field theory, and as such, we use the two-loop SM RGEs given in~\cite{Carena:2015uoe} to match the relevant MS-3HDM couplings to the corresponding SM quartic coupling $\lambda_{\rm SM}$, the Yukawa couplings, and the $\mathrm{SU(2)}_L$ and $\mathrm{U(1)}_Y$ gauge couplings: $g_2$ and $g'$. Unless stated explicitly otherwise,
we use the following baseline model for our analysis, where all input parameters
are given at the RG scale $\mu = M_{h_1^{\pm}}$:
\begin{align}
  \label{eq:benchmark}
&\tan\beta_1\, =\, 50\,,\qquad \tan\beta_2\, =\, 0.018\,,\quad \sigma~\text{[rad]}\, =\, 0.012\,,\nonumber\\[1mm]
&\big( M_{h^\pm_1}\,, M_{h^\pm_2}\big)~\text{[TeV]}\ =\ \big( 0.5\, , 0.525\big)\,,\,
\big( 1\,, 1.025\big)\,,\, \big( 10\,,  10.025\big)\, .
\end{align}
In addition, the values of the two-loop SM couplings at different threshold scales, $\mu_{\rm thr} = m_t,\, M_{h_1^{\pm}}$, are determined in a fashion similar to~\cite{Darvishi:2019ltl}.

\begin{figure}
\centering
\includegraphics[width=0.65\textwidth]{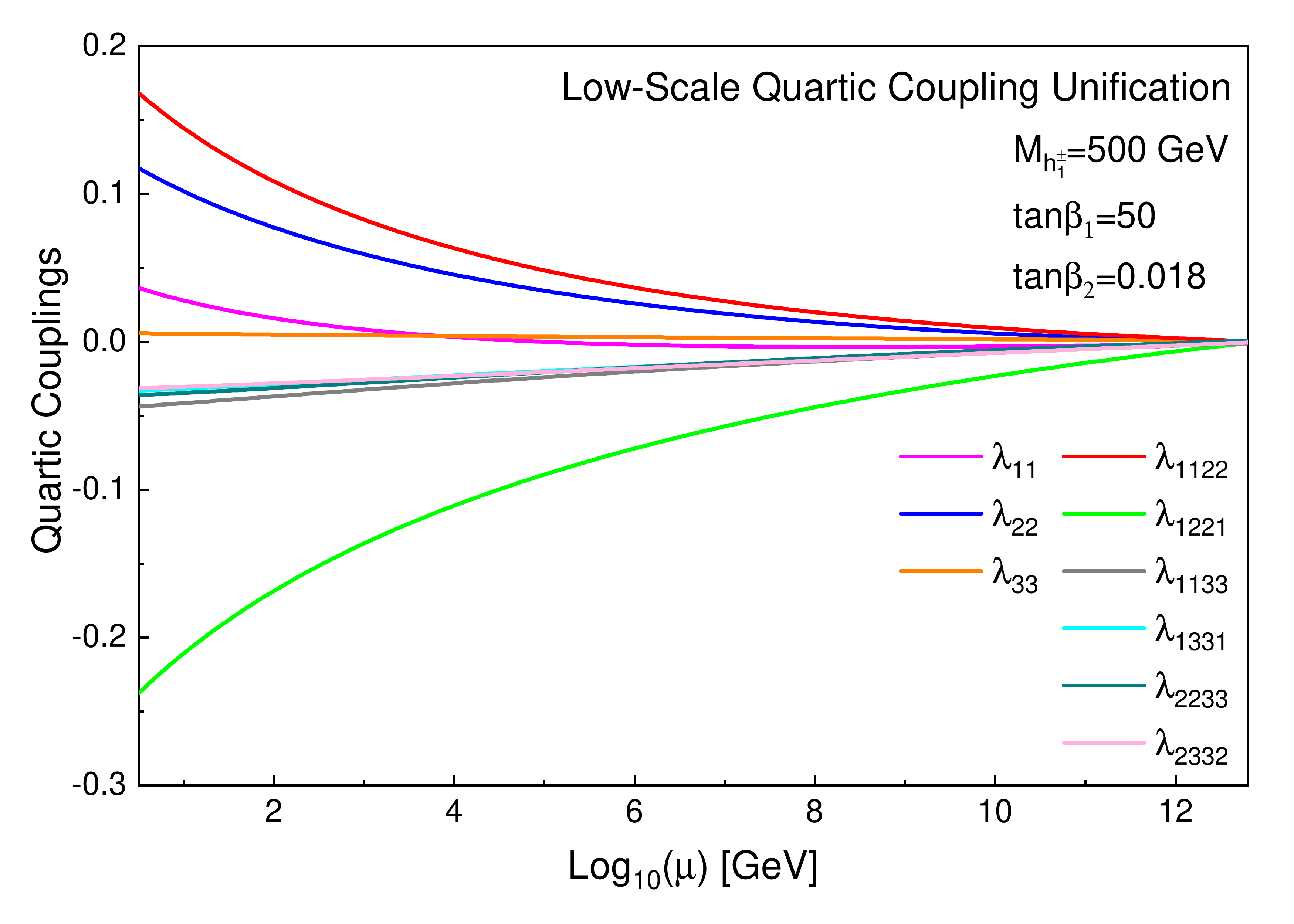}
\caption{The RG evolution of the quartic couplings from the threshold 
scale $M_{h_1^\pm} = 500\,\mathrm{GeV}$ up to their first quartic coupling unification scale~${\mu^{(1)}_X \sim 10^{13}}\,\mathrm{GeV}$ for $\tan\beta_1= 50$ and $\tan\beta_2 =0.018$.}
\label{first}
\end{figure}

In Figure~\ref{first}, we display the RG evolution of all quartic couplings for the benchmark model of~\eqref{eq:benchmark}, with a low charged Higgs mass $M_{h_1^\pm} = 500$~GeV.  We observe that the quartic coupling $\lambda_{22}$, which determines the SM-like Higgs-boson mass~$M_H$, decreases at high RG scales, due to the running of the top-Yukawa coupling~$y_t$. The coupling $\lambda_{22}$ turns negative just above the quartic coupling unification scale $\mu^{(1)}_X \sim 10^{13}$\,GeV, at which all quartic couplings vanish. Below $\mu^{(1)}_X$, the MS-3HDM quartic couplings exhibit different RG runnings, and especially the couplings $\lambda_{ijij}$ (with $i\neq j$ and $i,j=1,2,3$) take on non-zero values. Also, for energy scales above $\mu^{(1)}_X$, we understand that the MS-3HDM has to be embedded into a higher-scale UV-complete theory. Nevertheless, according to the results from \cite{Darvishi:2019ltl}, such a potential leads to a metastable but sufficiently long-lived EW vacuum, whose lifetime is many orders of magnitude longer than the age of our Universe. For this reason, we do not apply the usual, over-restrictive constraints derived from positivity conditions on the scalar potential that would imply an absolutely stable EW~vacuum.

\begin{figure}
\centering
\includegraphics[width=0.65\textwidth]{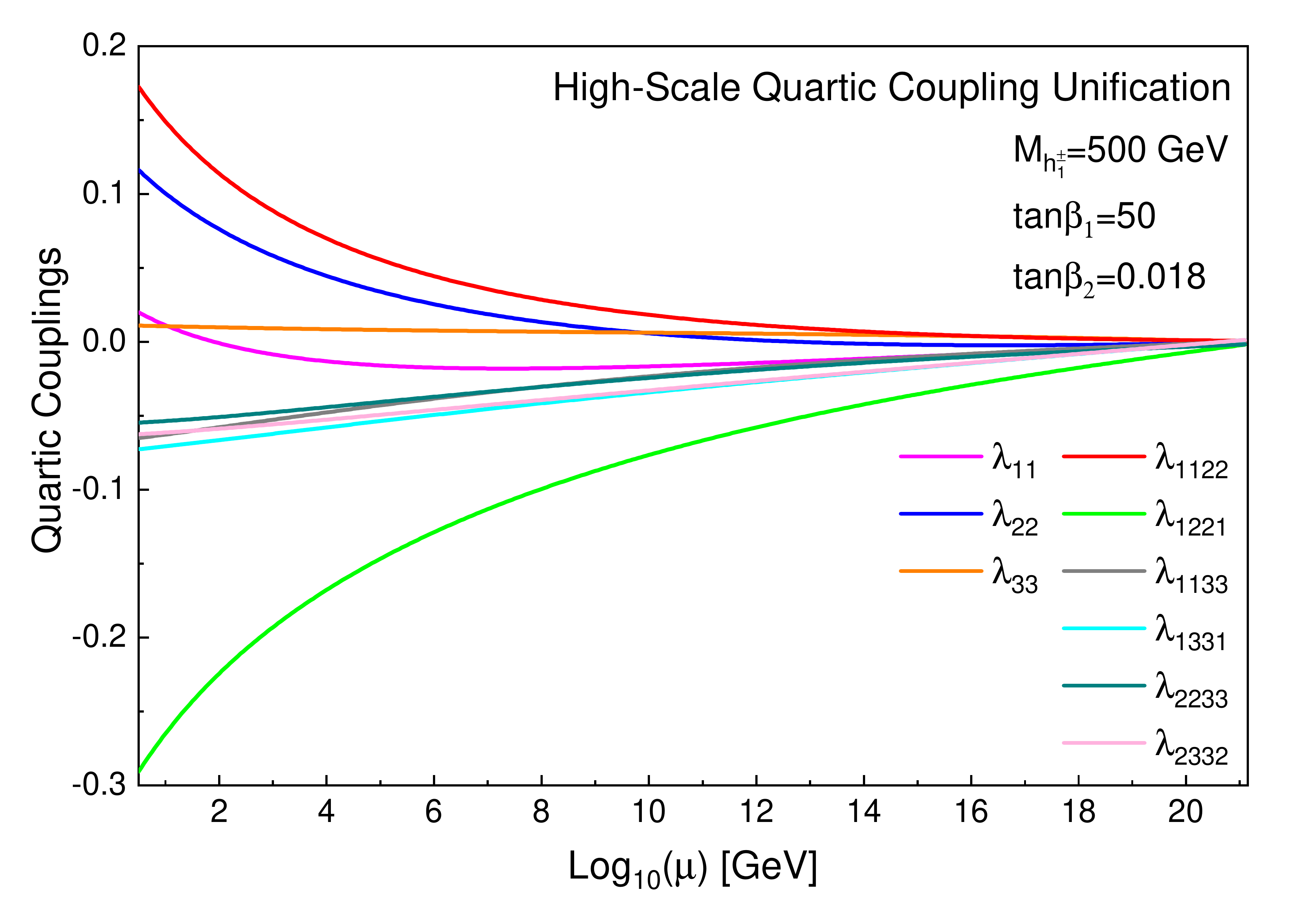}
\caption{The second conformal point of quartic coupling unification at $\mu^{(2)}_X \sim 10^{21}\,\mathrm{GeV}$ is shown, for the same input parameters as in~Fig.~\ref{first}.}
\label{second}
\end{figure}

\begin{figure}
\centering
\includegraphics[width=0.65\textwidth]{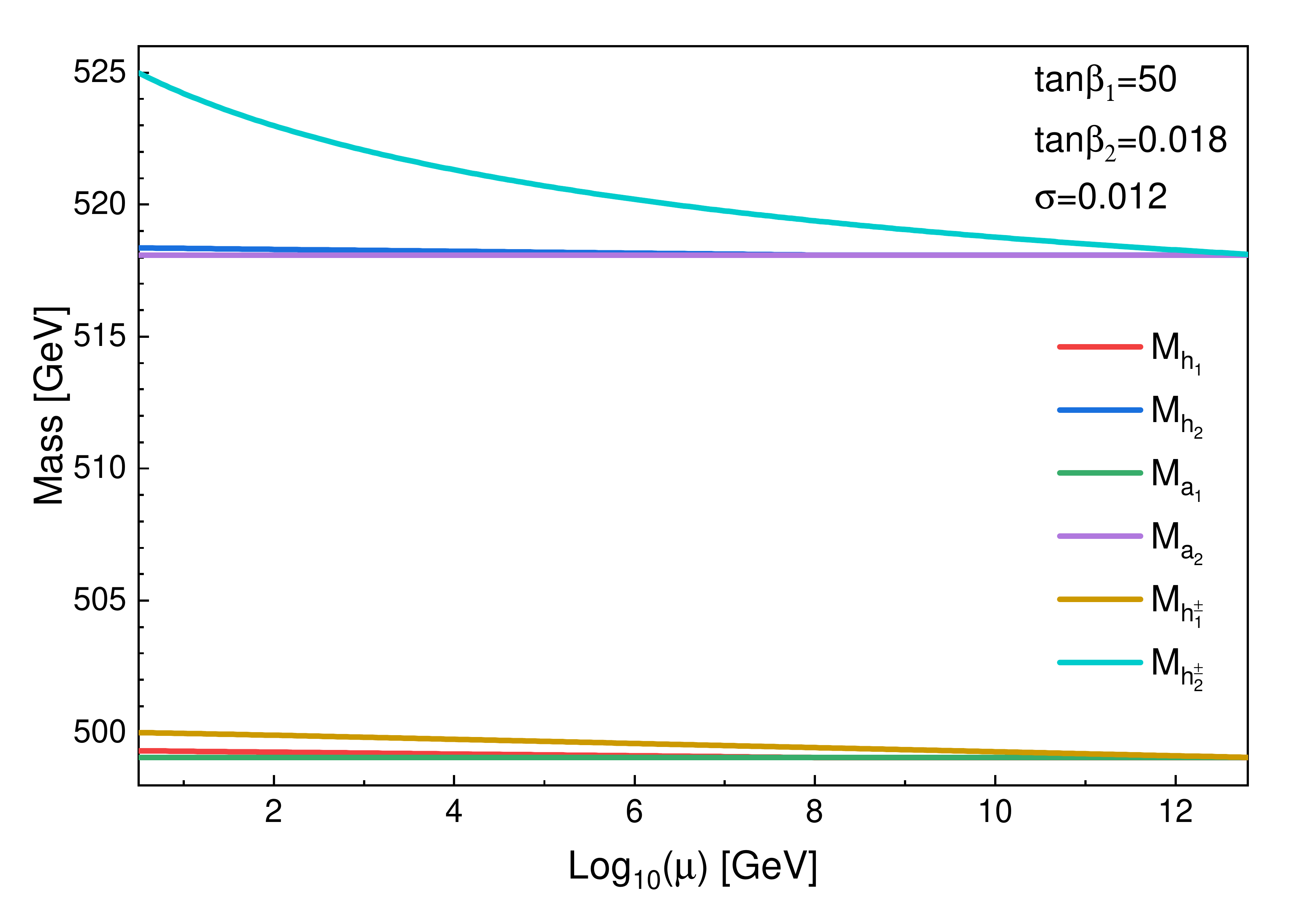}
\caption{The scalar mass spectrum of the MS-3HDM with the input parameters $\tan \beta_1=50$, $\tan \beta_2=0.018$, $M_{h_1^\pm} = 500$~GeV, $M_{h_2^\pm} = 525$~GeV and the mixing angle $\sigma= 0.012$. }
\label{masses}
\end{figure}

In addition to the conformal unification point $\mu^{(1)}_X$, there is in general a second and higher conformal point $\mu^{(2)}_X \sim 10^{21}\,\mathrm{GeV}$ in the MS-3HDM.  This is exemplified in Figure~\ref{second}. This higher conformal point occurs, when the quartic coupling~$\lambda_{22}$ increases at high RG scales and crosses zero for the second time. This happens when the running of the top-Yukawa coupling overtakes that of the gauge coupling. Obviously, in such a theoretical setting, any embedding of the MS-3HDM into a  candidate UV-complete theory must primarily include UV aspects of quantum gravity.

In Figure~\ref{masses}, we give numerical estimates of the mass spectrum of the MS-3HDM,
for the benchmark model in~\eqref{eq:benchmark}, for which a low charged Higgs mass $M_{h_1^\pm} = 500$~GeV and the first unification point $\mu_X = \mu^{(1)}_X$ are chosen. At this unification scale $\mu^{(1)}_X$, the masses of all heavy Higgs bosons, $h_{1,2}$, $a_{1,2}$ and $h_{1,2}^{\pm}$, are mainly determined by the soft-breaking mass terms, which are: $m^2_{12}\approx 75^2 \; \mathrm{GeV}^2$, $m^2_{23}\approx 70^2 \; \mathrm{GeV}^2$, $m^2_{13}\approx 20^2 \; \mathrm{GeV}^2$, and the VEV ratios $\tan \beta_1=50$ and $\tan \beta_2=0.018$. As a consequence of~\eqref{eq:mij}, the heavy mass spectrum of the MS-3HDM
becomes degenerate at the unification point~$\mu_X$, clustering about the two different 
charged Higgs masses, $M_{h_1^\pm} = 500$~GeV and $M_{h_2^\pm} = 525$~GeV, with a
mixing angle $\sigma\,\text{[rad]} = 0.012$ for the $h^\pm_1$-$h^\pm_2$ system. As shown in Figure~\ref{masses}, RG effects will break these mass degeneracies from a few MeV, for $M_{h_1}- M_{a_1}$ and $M_{h_2} - M_{a_2}$, up to about $30$~GeV, for~${M_{h^\pm_2} - M_{a_2}}$.

\begin{figure}[t]
\centering
\includegraphics[width=0.65\textwidth]{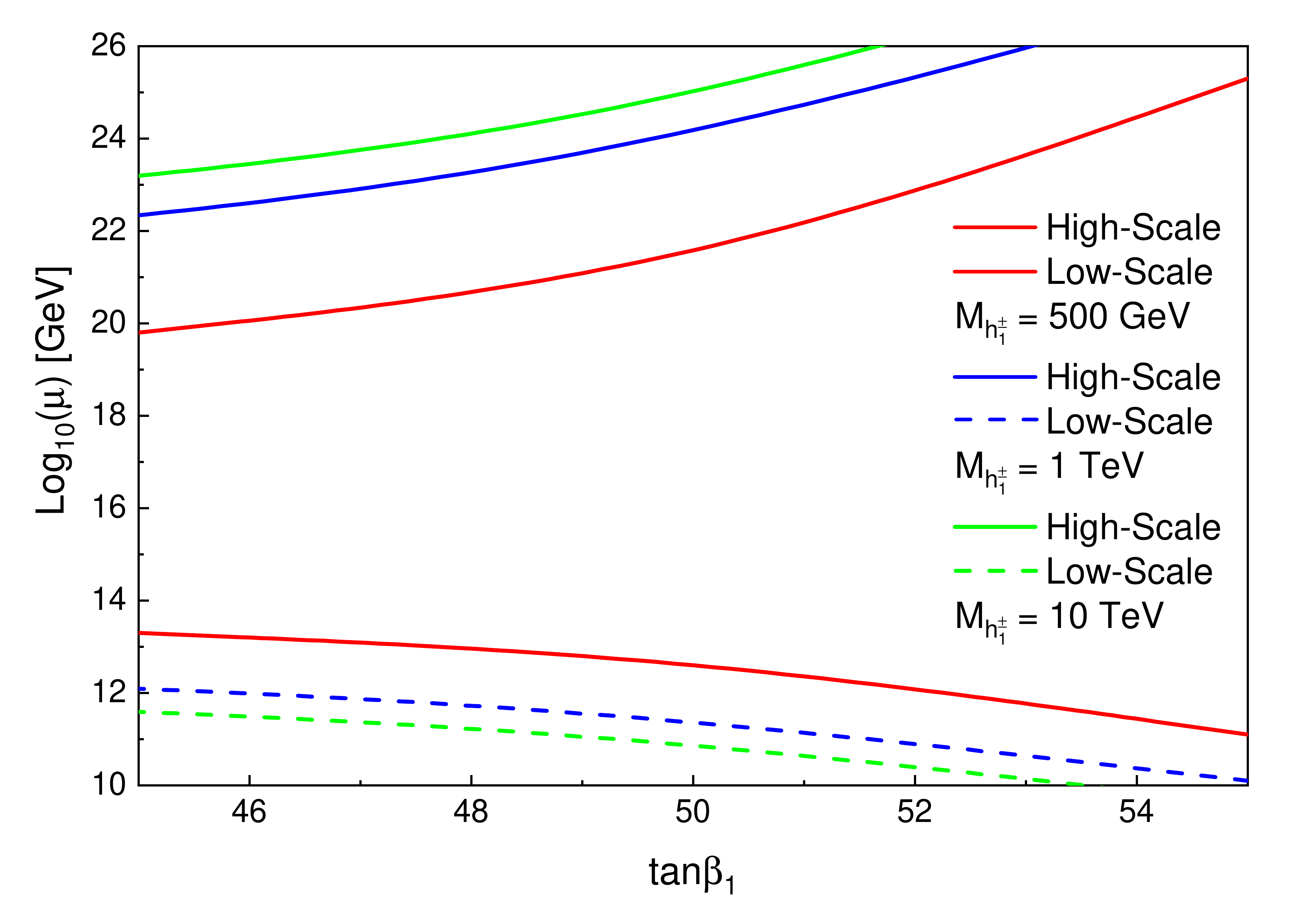}
\caption{Sets of quartic coupling unification points in the $(\tan\beta_{1},\, \log_{10}\mu )$
 plane, for charged Higgs-boson masses $M_{h_1^{\pm}}=500\,\mathrm{GeV},\,1\,\mathrm{TeV}\, \mathrm{and} \,\,10 \; \mathrm{TeV}$.  The dashed and solid curves show two sets of the low-scale and the high-scale quartic coupling unification points, $\mu^{(1)}_X$ and $\mu^{(2)}_X$, respectively.}
\label{tanb1}
\end{figure}

\begin{figure}
\centering
\includegraphics[width=0.65\textwidth]{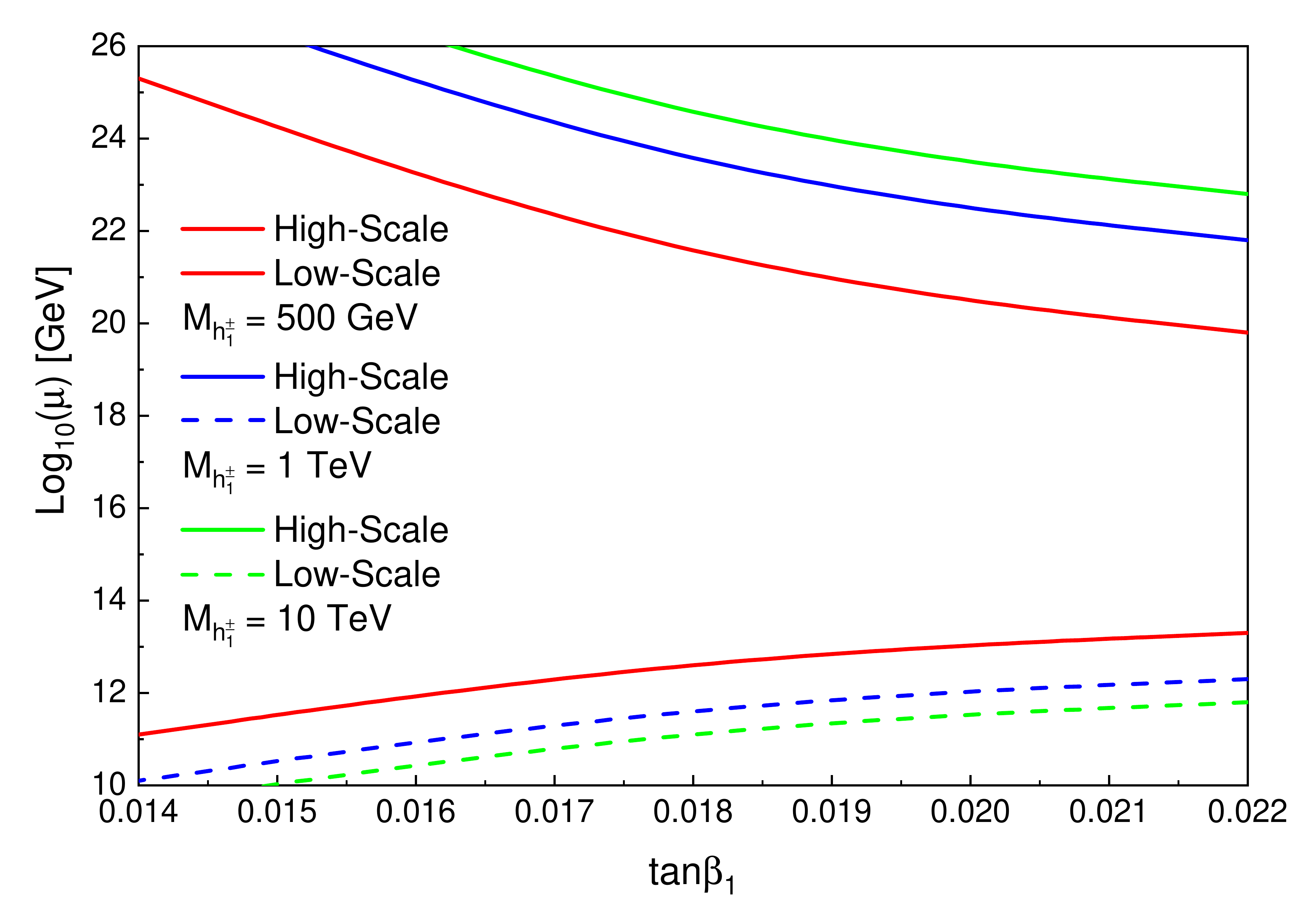}
\caption{The same as in Fig.~\ref{tanb1} but for sets of quartic coupling 
unification points in the $(\tan\beta_{2},\, \log_{10}\mu )$ plane.}
\label{tanb2}
\end{figure}

Figures~\ref{tanb1} and~\ref{tanb2} show all conformally-invariant quartic coupling unification points in the $(\tan\beta_1, \log_{10}\mu)$ and $(\tan\beta_2, \log_{10}\mu)$ planes, by considering different values of threshold scales~$\mu_{\rm thr}$, i.e.~for $\mu_{\rm thr} = M_{h_1^{\pm}}=500\,\mathrm{GeV},\,1\,\mathrm{TeV}\, \mathrm{and} \,\,10 \; \mathrm{TeV}$. In both figures, the lower curves (dashed curves) correspond to sets
of low-scale quartic coupling unification points, while the upper curves (solid curves) give the corresponding sets of high-scale unification points. From Figures~\ref{tanb1} and~\ref{tanb2}, we may also observe the domains in which the $\lambda_{22}$ coupling becomes negative. Evidently, as the threshold scale~$\mu_{\rm thr} = M_{h_1^\pm}$ increases, the size of the negative $\lambda_{22}$ domain increases. This becomes more pronounced for larger values of $\tan \beta_1$ and smaller values of $\tan \beta_2$. So, one may obtain lower and upper bounds on $\tan \beta_1$ and $\tan \beta_2$ for this unified theoretical\- framework at different threshold scales~$\mu_{\rm thr}$. For example, if $\mu_{\rm thr}=500\,\mathrm{GeV}$, we may deduce the bounds: $40 \lesssim \tan \beta_1\lesssim 55$ and $0.013 \lesssim \tan \beta_2 \lesssim 0.022$.

\begin{figure}[t]
\centering
\includegraphics[width=0.65\textwidth]{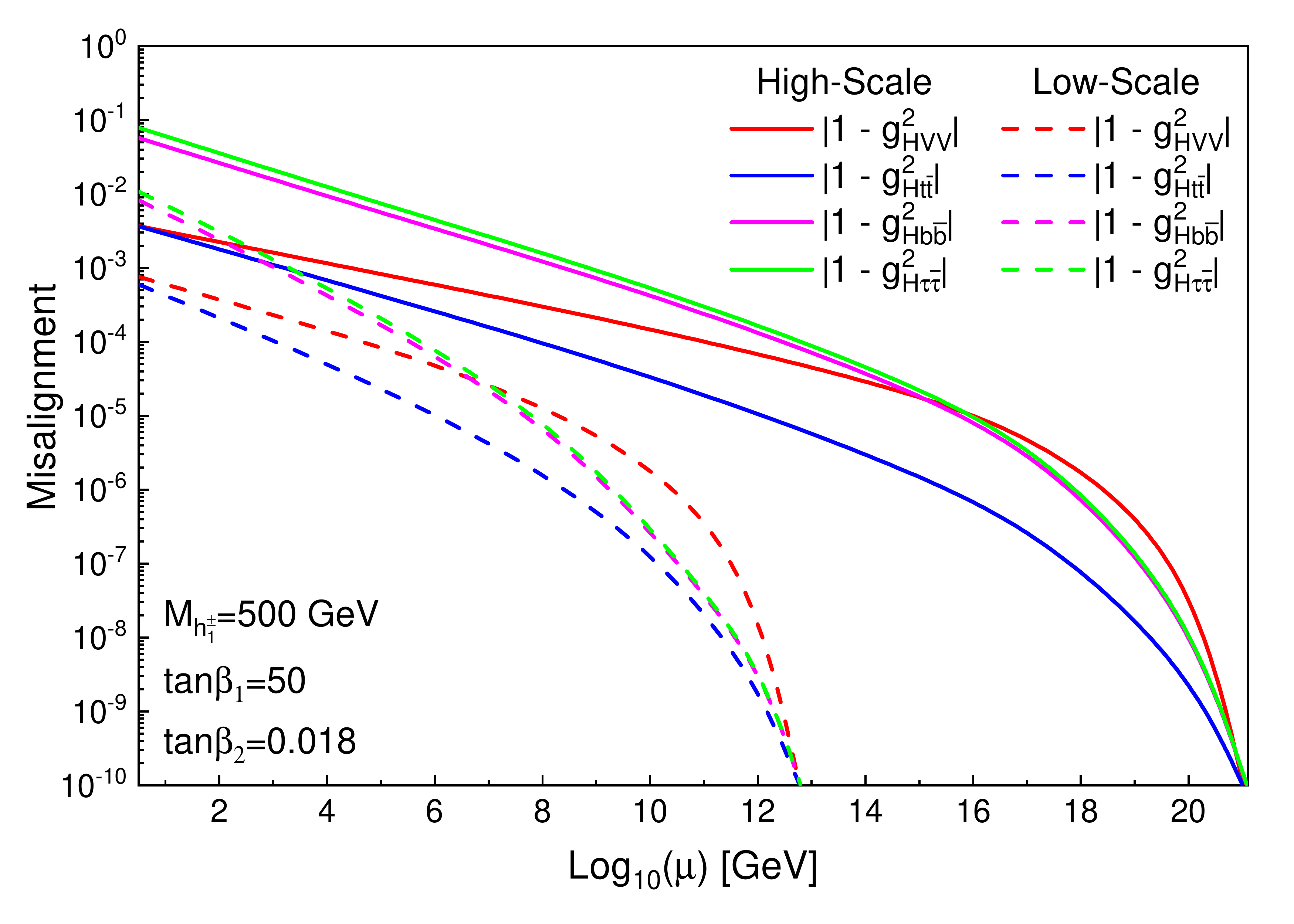}
\caption{Numerical estimates of the misalignment parameter $|1 - g^2_{HXX}|$ pertinent to the $HXX$-coupling (with $X=V,t,b,\tau$ and $V= W^\pm,Z$) as functions of the RG scale~$\mu$, for the low-scale and the high-scale quartic coupling unification scenarios, considering  $M_{h_1^\pm} = 500\,\mathrm{GeV}$, $\tan\beta_1=50$ and $\tan\beta_2=0.018$.}
\label{MisAlign}
\end{figure}

We have already seen how at the conformal points, $\mu^{(1)}_X$ and $\mu^{(2)}_X$, all quartic couplings vanish simultaneously, leading to an exact SM alignment. Nevertheless, for lower RG scales, the $\mathrm{Sp(6)}$ symmetry is broken, giving rise to calculable non-zero misalignment predictions. As was discussed in Section \ref{3HDM}, this misalignment can be derived using our analytic expressions given in~\eqref{ex-gc}. In Figure~\ref{MisAlign}, we present our numerical estimates of the predicted deviations for the SM-like Higgs-boson coupling $HXX$ (with $X= W^\pm,Z,t,b,\tau$) from its respective SM value. Specifically, Figure~\ref{MisAlign} exhibits the dependence of the misalignment parameter $|1 - g^2_{HXX}|$ (with $g_{H_{\mathrm{SM}}XX}=1$) as functions of the RG scale~$\mu$, for both the low- and high-scale quartic coupling unification scenarios. Evidently, the deviation of the normalised coupling $g_{HXX}$ from its SM value gets larger for the higher-scale unification scenario. Moreover, the degree of misalignment for~$g_{Hbb}$ and ~$g_{H\tau \tau}$ approach values larger than $10\%$. The ATLAS and CMS data~\cite{201606} for $g_{Hbb}=0.49^{+0.26}_{-0.19}$ and $g_{Hbb}=0.57^{+0.16}_{-0.16}$, which can be fitted to the SM at the $3\sigma$ level, reduces only to $2\sigma$ in the MS-3HDM in the high-scale unification scenario. We observe that the normalised couplings, $g_{HVV}$ and $g_{Htt}$, approach their SM values $g_{H_{\mathrm{SM}}V V} = g_{H_{\mathrm{SM}}tt}  = 1$ at the two quartic coupling unification points, $\mu^{(1)}_X$ and $\mu^{(2)}_X$.  

\section{Conclusions}\label{conc}

We have analysed the basic low-energy structure of the general 3HDM. We have found that this model can realise three distinct SM-like Higgs scenarios of SM alignment. Our study was focused on the canonical SM-like Higgs scenario of the Type-V, for which conditions on the model parameters for achieving exact SM alignment were derived. Interestingly enough, there are three continuous symmetries which, when imposed on the $n$HDM scalar potential, are sufficient to ensure SM alignment. These are: (i)~$\mathrm{Sp(6)}$, (ii)~$\mathrm{SU(3)_{\rm HF}}$,
and (iii)~$\mathrm{SO(3)_{\rm HF}}$. Amongst these symmetries, the most economic setting is the Maximally Symmetric Three-Higgs Doublet\- Model (MS-3HDM), whose potential obeys an $\mathrm{Sp(6)}$ symmetry.  The Sp(6) symmetry is softly broken by bilinear masses $m^2_{ij}$ (with $i,j=1,2,3$), as well as explicitly by hyper\-charge and Yukawa couplings through RG effects, whilst the theory allows for quartic coupling unification up to the Planck~scale.

The MS-3HDM is a remarkably predictive scenario, as it only depends on a few theoretical parameters when compared to the large number of independent parameters that are required in the general 3HDM. In fact, besides the ratios of the Higgs-doublet VEVs, $\tan\beta_{1,2}$, the model is mainly governed by only three input parameters: the masses of the two charged Higgs bosons, $M_{h_{1,2}^{\pm}}$, and their mixing angle~$\sigma$. Most notably, with the help of these input parameters, we have obtained misalignment predictions for the entire scalar mass spectrum of the theory, including the interactions of all Higgs particles to the SM fields.

We have presented the one- and two-loop RG equations pertinent to the general 3HDM.  We have used these to evaluate the two-loop RG effects on all relevant couplings in the MS-3HDM. In particular, we have shown that all quartic couplings in the MS-3HDM can unify at high-energy scales~$\mu_X$ and vanish simultaneously at two distinct conformal points, that are denoted by~$\mu^{(1,2)}_X$ with $\mu_X^{(1)} \lesssim 10^{13}\,\text{GeV}$ and $\mu_X^{(2)} \gtrsim 10^{21}\,\text{GeV}$. These limits have been obtained by considering the RG evolution of the quartic couplings from the unification points $\mu^{(1,2)}_X$ down to the threshold scale $\mu_{\rm thr} = M_{h_1^{\pm}}=500$~GeV, 1~TeV and 10~TeV. For our analysis, we considered a typical benchmark scenario for the VEV ratios, $\tan\beta_{1,2}$, and the kinematic parameters of the charged Higgs sector.  We have obtained misalignment predictions for the Higgs-boson couplings to the $W^\pm$, the $Z$ bosons, and $t$-quarks. These turned out to be very close to their SM value, $g_{HZZ}= g_{Htt} = g_{H\tau\tau} =1$, and so they are in excellent agreement with the current LHC observations. On the other hand, the LHC data measuring the strength of the $Hb\bar{b}$-coupling, $g^{\rm exp}_{Hbb}$, differs from its SM value by $3\sigma$, but this deviation reduces significantly to less than $2\sigma$ in the MS-3HDM.

The present results for the MS-3HDM, along with an earlier study of the MS-2HDM~\cite{Darvishi:2019ltl},  demonstrate the high predictive power of maximally symmetric settings in $n$HDMs. Such settings not only can naturally provide the experimentally favoured SM alignment, but also allow us to obtain sharp predictions for the entire scalar mass spectrum of the theory, including the interactions of all Higgs particles to the SM fields and all scalar self-interactions. This~fact opens up new interesting theoretical vistas that merit detailed exploration in the near future. We therefore plan to return to dedicated investigations of the Higgs self-interactions in this framework that may be probed via multi-Higgs and charged Higgs production events at the LHC and future high-energy colliders.

\section*{Acknowledgements}
\noindent
The work of AP is supported in part by the
Lancaster–Manchester–Sheffield Consortium\- for Fundamental Physics,
under STFC research grant ST/P000800/1. 
\\
MRM is supported by the UK Science and Technology Facilities Council (grant numbers ST/P001246/1). The work of MRM has also received funding from the European Union’s Horizon 2020 research and innovation program as part of the Marie Skłodowska-Curie Innovative Training Network MCnetITN3 (grant agreement no. 722104).
\\
The work of ND is supported in part by the National Science Centre (Poland) as a research project, decision no 2017/25/B/ST2/00191 and by the Polish National Science Centre HARMONIA grant under contract UMO-2015/20/M/ST2/00518 (2016-2021).

\vfill\eject

\appendix
\section{The 3HDM Potential} \label{ap:pot}

The most general 3HDM potential invariant under $\mathrm{SU(2)}_L\otimes \mathrm{U(1})_Y$ may be explicitly expressed in terms of three Higgs doublets $\Phi_i$ ($i=1,2,3$) as follows:

{\small
\begin{align}
V_{\mathrm{3HDM}} = 
&
- m_{11}^2 (\Phi_1^{\dagger} \Phi_1)
- m_{22}^2 (\Phi_2^{\dagger} \Phi_2)
- m_{33}^2 (\Phi_3^{\dagger} \Phi_3)
 - \bigg[ m_{12}^2 (\Phi_1^{\dagger} \Phi_2)
+m_{13}^2 (\Phi_1^{\dagger} \Phi_3)
     \nonumber\\ &
+m_{23}^2 (\Phi_2^{\dagger} \Phi_3)\: +\:
   {\mathrm{H.c.}} \bigg]
+\lambda_{11} (\Phi_1^{\dagger} \Phi_1)^2 
+ \lambda_{22} (\Phi_2^{\dagger} \Phi_2)^2 
+ \lambda_{33} (\Phi_3^{\dagger} \Phi_3)^2
     \nonumber \\ &
+ \lambda_{1122} (\Phi_1^{\dagger} \Phi_1) (\Phi_2^{\dagger} \Phi_2) 
+ \lambda_{1133} (\Phi_1^{\dagger} \Phi_1) (\Phi_3^{\dagger} \Phi_3) 
+ \lambda_{2233} (\Phi_2^{\dagger} \Phi_2) (\Phi_3^{\dagger} \Phi_3) 
    \nonumber \\ &
+ \lambda_{1221} (\Phi_1^{\dagger} \Phi_2) (\Phi_2^{\dagger} \Phi_1)
+ \lambda_{1331} (\Phi_1^{\dagger} \Phi_3) (\Phi_3^{\dagger} \Phi_1) 
+ \lambda_{2332} (\Phi_2^{\dagger} \Phi_3) (\Phi_3^{\dagger} \Phi_2)
     \nonumber \\ &
 + \bigg[\, {\lambda_{1212} \over 2} (\Phi_1^{\dagger} \Phi_2)^2 
+ {\lambda_{1313} \over 2} (\Phi_1^{\dagger} \Phi_3)^2 
+ {\lambda_{2323} \over 2}  (\Phi_2^{\dagger} \Phi_3)^2 
\nonumber \\ & 
+ \lambda_{1213}  (\Phi_1^{\dagger} \Phi_2) (\Phi_1^{\dagger} \Phi_3)
+ \lambda_{2113} (\Phi_2^{\dagger} \Phi_1) (\Phi_1^{\dagger} \Phi_3)
+ \lambda_{1323}    (\Phi_1^{\dagger} \Phi_3) (\Phi_2^{\dagger} \Phi_3)
\nonumber \\ & 
+ \lambda_{1332}   (\Phi_1^{\dagger} \Phi_3) (\Phi_3^{\dagger} \Phi_2)
+ \lambda_{2123}  (\Phi_2^{\dagger} \Phi_1) (\Phi_2^{\dagger} \Phi_3)
+ \lambda_{1223} (\Phi_1^{\dagger} \Phi_2) (\Phi_2^{\dagger} \Phi_3) 
     \nonumber \\ & 
+ \lambda_{1112}  (\Phi_1^{\dagger} \Phi_1) (\Phi_1^{\dagger} \Phi_2)
+ \lambda_{2212}  (\Phi_2^{\dagger} \Phi_2) (\Phi_1^{\dagger} \Phi_2)  
+ \lambda_{1113}  (\Phi_1^{\dagger} \Phi_1) (\Phi_1^{\dagger} \Phi_3) 
     \nonumber \\ & 
+ \lambda_{1123}  (\Phi_1^{\dagger} \Phi_1) (\Phi_2^{\dagger} \Phi_3)
+ \lambda_{2213}  (\Phi_2^{\dagger} \Phi_2) (\Phi_1^{\dagger} \Phi_3)  
+ \lambda_{2223}  (\Phi_2^{\dagger} \Phi_2) (\Phi_2^{\dagger} \Phi_3) 
     \nonumber \\ & 
+ \lambda_{3312}  (\Phi_3^{\dagger} \Phi_3) (\Phi_1^{\dagger} \Phi_2) 
+ \lambda_{3313}  (\Phi_3^{\dagger} \Phi_3)(\Phi_1^{\dagger} \Phi_3) 
+ \lambda_{3323}  (\Phi_3^{\dagger} \Phi_3) (\Phi_2^{\dagger} \Phi_3) 
+{\mathrm{H.c.}}\bigg]\;,
\label{eq:V3HDM}
\end{align}}
\hspace{-1.5mm}where $\lambda_{ii}\equiv \lambda_{iiii}/2$. Furthermore, assuming a CP-conserving 3HDM potential the following three minimisation conditions can be obtained: 
\begin{align}
m_{11}^2\ = \frac{1}{2 v_1} \Big[ 
			&-2 m_{12}^2 v_2-2 m_{13}^2 v_3
			+2 \lambda_{11} v_1^3
			+ (\lambda_{1133}+\lambda_{1313}+\lambda_{1331})  v_1 v_3^2
			\nonumber \\ &
			+ (\lambda_{1122}+\lambda_{1212} +\lambda_{1221}) v_1 v_2^2	
			+2 (\lambda_{1123}+\lambda_{1213}+\lambda_{2113}) v_1 v_2 v_3
			\nonumber \\ &
			+(\lambda_{1223} +\lambda_{2123}+\lambda_{2213}) v_2^2 v_3	
			+ (\lambda_{1323}+\lambda_{1332}+\lambda_{3312}) v_2 v_3^2
			\nonumber \\ &	
			+\lambda_{2212} v_2^3
			+3 \lambda_{1112} v_1^2 v_2
			+3 \lambda_{1113} v_1^2 v_3		
			+\lambda_{3313} v_3^3 \Big],
\end{align}  
\begin{align}
m_{22}^2\ = \frac{1}{2 v_{2}} \Big[ 
			&-2 m_{12}^2 v_{1}
			-2 m_{23}^2 v_{3}
			+2 \lambda_{22}v_{2}^3
			+(\lambda_{1122}+\lambda_{1212}+\lambda_{1221}) v_{1}^2v_{2}
			\nonumber \\ &
			+(\lambda_{2233}+\lambda_{2323}+\lambda_{2332}) v_{2} v_{3}^2
			+2 (\lambda_{2213}+\lambda_{1223}+\lambda_{2123}) v_{1} v_{2} v_{3}
			\nonumber \\ &
			+(\lambda_{1323}+\lambda_{1332}+\lambda_{3312}) v_{1} v_{3}^2
			+(\lambda_{1123}+\lambda_{1213}+\lambda_{2113}) v_{1}^2 v_{3}
			\nonumber \\ &
			+\lambda_{1112} v_{1}^3
			+3 \lambda_{2212} v_{1} v_{2}^2
			+3 \lambda_{2223} v_{2}^2 v_{3}
			+\lambda_{3323}v_{3}^3 
			\Big],
\end{align}    
\begin{align}
m_{33}^2\ = \frac{1}{2 v_{3}} \Big[
			 &-2 m_{13}^2 v_{1}
			 -2 m_{23}^2 v_{2}
			 +2\lambda_{33} v_{3}^3
			 +(\lambda_{1133}+\lambda_{1313}+\lambda_{1331}) v_{1}^2 v_{3}
			 \nonumber \\ &
			 +(\lambda_{2233}+\lambda_{2323}+\lambda_{2332}) v_{2}^2 v_{3}
			 +(\lambda_{1223}+\lambda_{2123}) v_{1}v_{2}^2
			 \nonumber \\ &
			  +(\lambda_{1123}+\lambda_{1213}+\lambda_{2113}) v_{1}^2v_{2}
			 +2 (\lambda_{1323}+\lambda_{1332}+\lambda_{3312}) v_{1} v_{2} v_{3}
			 \nonumber \\ &
			 +\lambda_{1113} v_{1}^3
			 +\lambda_{2223} v_{2}^3
			 +3 \lambda_{3313} v_{1}v_{3}^2		 
			 +3 \lambda_{3323} v_{2} v_{3}^2 
			 \Big].
\end{align}

\section{Mass Matrices}\label{AppMassMat}

As discussed in Section \ref{3HDM}, there are nine physical Higgs states in 3HDM: (i)~three CP-even scalars ($H,\,h_{1,2}$), (ii) two CP-odd scalars $(a_{1,2})$, and (iii) four charged scalars~($h_{1,2}^{\pm}$). The details for deriving the masses of these particles are given in Subsections~\ref{AppOddMass}, \ref{AppChargedMass} and \ref{AppevenMass}, respectively. 

\subsection{CP-odd Pseudo-Scalar Mass Matrix} \label{AppOddMass}

The mass matrix pertinent to the CP-odd scalars or pseudo-scalars in the general 3HDM is
given by
\begin{align}
\mathcal{M}^2_{\text{P}}=
\left(
\begin{array}{ccc}

 M^2_{\text{P},22} &  M^2_{\text{P},23}
\\
 M^2_{\text{P},32} &  M^2_{\text{P},33}

\end{array}
\right),
\end{align}
with
{\allowdisplaybreaks \small
\begin{align}
\begin{autobreak}
M^2_{\text{P},22} =
m_{12}^{2} s_{2\beta_1}
 +m_{12}^{2} c_{\beta_1}^{2} t_{\beta_1}^{-1}
 +s_{\beta_1} t_{\beta_1} (m_{12}^{2} s_{\beta_1}
 +m_{13}^{2} t_{\beta_2})
 +m_{23}^{2}
   c_{\beta_1} t_{\beta_1}^{-1} t_{\beta_2}
 +\frac{1}{4} t_{\beta_1}^{-1} \big(-s_{\beta_1} \big(v^2 s_{\beta_1} t_{\beta_1}^{2} c_{\beta_2}^{2} (c_{2\beta_1} (\lambda_{1112}
 +\lambda_{2212})
 +2 \lambda_{1212} s_{2\beta_1}
 +\lambda_{1112}
 +3 \lambda_{2212})
 +\frac{1}{2} v^2 t_{\beta_1}^{2} \big(s_{2\beta_2} (c_{2\beta_1} (\lambda_{1113}
 -\lambda_{1223}
 +3 \lambda_{2123}
 -\lambda_{2213})
 +4 \lambda_{1213} s_{2\beta_1}
 +\lambda_{1113}
 +\lambda_{1223}
 +5 \lambda_{2123}
 +\lambda_{2213})
 +4 \lambda_{3313} s_{\beta_2}^{2} t_{\beta_2}\big)
 +v^2 t_{\beta_1} c_{\beta_1}^{
 -1} s_{\beta_2}^{2} (2 \lambda_{1313} s_{2\beta_1}
 +c_{2\beta_1} (
 -3 \lambda_{1323}
 +\lambda_{1332}
 +\lambda_{3312})
 -\lambda_{1323}
 +3 (\lambda_{1332}
 +\lambda_{3312}))
 -\lambda_{2223} v^2 s_{2\beta_1} s_{2\beta_2}\big)
 +v^2 c_{\beta_1}^{2} \big(c_{\beta_2}^{2} (c_{2\beta_1} (\lambda_{1112}
 +\lambda_{2212})
 -3 \lambda_{1112}
 -\lambda_{2212})
 -8 \lambda_{2123} s_{\beta_1} s_{\beta_2} c_{\beta_2}
 -2 s_{\beta_2}^{2} (\lambda_{1323}
 +\lambda_{1332}
 +\lambda_{3312})\big)
 -2 v^2 c_{3\beta_1}^{3} c_{\beta_2} (2 \lambda_{1212} s_{\beta_1} c_{\beta_2}
 +s_{\beta_2} (\lambda_{1123}
 +\lambda_{1213}
 +\lambda_{2113}))
 -2 v^2 c_{\beta_1} \big(4 \lambda_{1212} s_{3\beta_1}^{3} c_{\beta_2}^{2}
 +s_{\beta_1}^{2} s_{\beta_2} c_{\beta_2} (4 \lambda_{1213}
 +3 \lambda_{2223})
 +s_{\beta_2}^{2} (2 \lambda_{2323} s_{\beta_1}
 +\lambda_{3323} t_{\beta_2})\big)\big)
,\end{autobreak}
\end{align}
\begin{align}
\begin{autobreak}
M^2_{\text{P},33} = 
s_{\beta_2}^{
 -1} c_{\beta_2}^{
 -1} (m_{13}^{2} c_{\beta_1}
 +m_{23}^{2} s_{\beta_1})
 +
\frac{1}{16} \big(
 -c_{\beta_1} s_{\beta_2}^{
 -1} c_{\beta_2}^{
 -1} \big(v^2 c_{2\beta_2} (3 \lambda_{1113}
 +\lambda_{1223}
 +\lambda_{2123}
 +\lambda_{2213}
 -4 \lambda_{3313})
 +v^2 (3 \lambda_{1113}
 +\lambda_{1223}
 +\lambda_{2123}
 +\lambda_{2213}
 +4 \lambda_{3313})\big)
 +2 v^2 c_{3\beta_1} t_{\beta_2}^{
 -1} (
 -\lambda_{1113}
 +\lambda_{1223}
 +\lambda_{2123}
 +\lambda_{2213})
 -s_{\beta_1} s_{\beta_2}^{
 -1} c_{\beta_2}^{
 -1} \big(v^2 c_{2\beta_2} (\lambda_{1123}
 +\lambda_{1213}
 +\lambda_{2113}
 +3 \lambda_{2223}
 -4 \lambda_{3323})
 +v^2 (\lambda_{1123} +\lambda_{1213} +\lambda_{2113}
 +3 \lambda_{2223}
 +4 \lambda_{3323})\big)
 -2 v^2 s_{3\beta_1} t_{\beta_2}^{
 -1} (\lambda_{1123}
 +\lambda_{1213}
 +\lambda_{2113}
 -\lambda_{2223})
 +8 v^2 c_{2\beta_1} (\lambda_{2323}
 -\lambda_{1313})
 -16 \lambda_{1323} v^2 s_{2\beta_1}
 -8 v^2 (\lambda_{1313}
 +\lambda_{2323})\big)
,\end{autobreak}
\end{align}
\begin{align}
\begin{autobreak}
M^2_{\text{P},23}  = 
c_{\beta_2}^{
 -1} (m_{13}^{2} s_{\beta_1}
 -m_{23}^{2} c_{\beta_1})
 +\frac{1}{8} \big(v^2 c_{\beta_2}(-s_{\beta_1} (\lambda_{1113}+3 \lambda_{1223}-5 \lambda_{2123}+3 \lambda_{2213}
 -4 \lambda_{3313})
 +s_{3\beta_1} (-\lambda_{1113}+\lambda_{1223}+\lambda_{2123}
 +\lambda_{2213})
 +c_{\beta_1} (3 \lambda_{1123}
 -5 \lambda_{1213}
 +3 \lambda_{2113}
 +\lambda_{2223}
 -4 \lambda_{3323})
 +c_{3\beta_1} (\lambda_{1123}
 +\lambda_{1213}
 +\lambda_{2113}
 -\lambda_{2223}))
 +4 v^2 s_{\beta_2} (s_{2\beta_1} (\lambda_{2323}
 -\lambda_{1313})
 +2 \lambda_{1323} c_{2\beta_1})
 +4 c_{\beta_2}^{
 -1} \big(\lambda_{3323} v^2 c_{\beta_1}
 -\lambda_{3313} v^2 s_{\beta_1}\big)\big).
\end{autobreak}
\end{align}
}
The above mass matrix $\mathcal{M}^2_{\text{P}}$ can be diagonalised with a rotational matrix $ R(\rho)$ in the form 
\begin{align}
\overline{\mathcal{M}}^2_{\text{P}}= R({\rho})\, \mathcal{M}^2_{\text{P}}\,R^{\sf T}(\rho) =
\left(
\begin{array}{ccc}
 {M}_{a_1}^2 & 0
\\
0  & {M}_{a_2}^2
\end{array}
\right),
\end{align}
with
\begin{align}
{M}_{a_1}^2 &=  M_{\text{P},22}^2 c_{\rho}^{2} +
                      M_{\text{P},23}^2 s_{2 \rho} + M_{\text{P},33}^2
                      s_{\rho}^{2} 
, \nonumber \\
{M}_{a_2}^2 &= {1 \over 2} (M_{\text{P},33}^2-M_{\text{P},22}^2) s_{2\rho} + M_{\text{P},23}^2 c_{2 \rho} 
,
\end{align}
where
\begin{align}
  \tan{2\rho}={2 M_{{\rm P},23} ^2\over M_{{\rm P},22} ^2-M_{{\rm P},33}^2}.
\end{align}

\subsection{Charged Scalar Mass Matrix} \label{AppChargedMass}

The mass matrix for the charged scalars in the general 3HDM is given by
\begin{align}
\mathcal{M}_{\pm}^2=
\left(
\begin{array}{ccc}
 M^2_{\pm,22} & M^2_{\pm,23} 
\\
 M^2_{\pm,23} & M^2_{\pm,33}
\end{array}
\right),
\end{align}
with
{\allowdisplaybreaks \small
\begin{align}
\begin{autobreak}
M^2_{\pm,22} = 
c_{\beta_1} t_{\beta_1}^{-1}\big(m_{12}^{2} c_{\beta_1}^{-3}
 +t_{\beta_2} \big(m_{13}^{2} t_{\beta_1}^{3}
 +m_{23}^{2} \big) \big)
 -\frac{1}{2} c_{\beta_1}^{3} t_{\beta_1}^{-1} 
 \big(\frac{1}{2} v^2 c_{\beta_1}^{
 -5} c_{\beta_2}^{2} (c_{2\beta_1} (\lambda_{1112}
 -\lambda_{2212})
 +s_{2\beta_1} (\lambda_{1212}
 +\lambda_{1221})
 +\lambda_{1112}
 +\lambda_{2212})
 +v^2 s_{\beta_2} c_{\beta_2} \big(\lambda_{1113} t_{\beta_1}^{3}
 +t_{\beta_1}^{5} (\lambda_{1223}
 +\lambda_{2123}
 +\lambda_{2213})
 +\lambda_{2223} t_{\beta_1}^{2}
 +\lambda_{1123}
 +\lambda_{1213}
 +\lambda_{2113}\big)
 +v^2 t_{\beta_1} c_{\beta_1}^{
 -2} s_{2\beta_2} (t_{\beta_1} (\lambda_{1213}
 +\lambda_{2113})
 +\lambda_{1223}
 +\lambda_{2123})
 +\frac{1}{8} v^2 c_{\beta_1}^{
 -5} s_{\beta_2}^{2} (2 s_{2\beta_1} (\lambda_{1313}
 +\lambda_{1331}
 +\lambda_{2323}
 +\lambda_{2332})
 +2 c_{4\beta_1} (\lambda_{1323}
 +\lambda_{1332})
 +s_{4\beta_2}(\lambda_{2323} 
 +\lambda_{2332}
 -\lambda_{1313}
 -\lambda_{1331} )
 +6 (\lambda_{1323}
 +\lambda_{1332})
 +8 \lambda_{3312})
 +v^2 c_{\beta_1}^{
 -2} s_{\beta_2}^{2} t_{\beta_2} \big(\lambda_{3313} t_{\beta_1}^{3} +\lambda_{3323}\big)\big)
,\end{autobreak}
\end{align}
\begin{align}
\begin{autobreak}
M^2_{\pm,33} = 
s_{\beta_2}^{-1} c_{\beta_2}^{-1} (m_{13}^{2} c_{\beta_1}
 +m_{23}^{2} s_{\beta_1})
 -\frac{1}{32} s_{\beta_2}^{-1} c_{\beta_2}^{-1} \big(16 \lambda_{1113} v^2 c_{3\beta_1}^{3} c_{\beta_2}^{2}
  +\lambda_{1123} v^2 \big(  s_{\beta_1 -2\beta_2}
 + s_{3\beta_1 -2\beta_2}
 + s_{\beta_1 +2\beta_2}
 + s_{3\beta_1 +2\beta_2} \big)
 +\lambda_{1213} v^2 \big( s_{\beta_1 -2\beta_2}
 + s_{3\beta_1 -2\beta_2}
 + s_{\beta_1 +2\beta_2}
 + s_{3\beta_1 +2\beta_2} \big)
 +\lambda_{1223} v^2 \big( c_{\beta_1 -2\beta_2}
 - c_{3\beta_1 -2\beta_2}
 +c_{\beta_1 +2\beta_2}
 -c_{3\beta_1 +2\beta_2} \big)
 -2 \lambda_{1313} v^2 \big( s_{2\beta_1 -2\beta_2}
 + s_{2\beta_1 +2\beta_2} \big)
 +4 \lambda_{1323} v^2 \big( c_{2\beta_1 -2\beta_2}
 - c_{2\beta_1 +2\beta_2} \big)
 -2 \lambda_{1331} v^2 \big (s_{2\beta_1 -2\beta_2}
 + s_{2\beta_1 +2\beta_2} \big)
 +4 \lambda_{1332} v^2 \big( c_{2\beta_1 -2\beta_2}
 - c_{2\beta_1 +2\beta_2} \big)
 +\lambda_{2113} v^2 \big( s_{\beta_1  -2\beta_2}
 + s_{3\beta_1 -2\beta_2}
 + s_{\beta_1 +2\beta_2}
 + s_{3\beta_1  +2\beta_2} \big)
 +\lambda_{2123} v^2 \big( c_{\beta_1 -2\beta_2}
 - c_{3\beta_1 -2\beta_2}
 + c_{\beta_1 +2\beta_2}
 - c_{3\beta_1 +2\beta_2} \big)
 +\lambda_{2213} v^2 \big( c_{\beta_1 -2\beta_2}
 - c_{3\beta_1 -2\beta_2}
 + c_{\beta_1 +2\beta_2}
 - c_{3\beta_1 +2\beta_2} \big)
 + \lambda_{2223} v^2 \big( 3 s_{\beta_1 -2\beta_2}
 -s_{3\beta_1  -2\beta_2}
 +3s_{\beta_1 +2\beta_2}
 -s_{3\beta_1  +2\beta_2} \big)
 +2 \lambda_{2323} v^2 \big( s_{2\beta_1 -2\beta_2}
 - s_{2\beta_1 +2\beta_2} \big)
 +2 \lambda_{2332} v^2 \big( s_{2\beta_1 -2\beta_2}
 - s_{2\beta_1 +2\beta_2} \big)
 -4 \lambda_{3313} v^2 \big( c_{\beta_1 -2\beta_2}
 - c_{\beta_1 +2\beta_2} \big)
 -4 \lambda_{3323} v^2 \big( s_{\beta_1 -2\beta_2}
 - s_{\beta_1 +2\beta_2} \big)
 +2 \lambda_{1123} v^2 \big( s_{\beta_1}
 + s_{3\beta_1} \big)
 +2 \lambda_{1213} v^2 \big( s_{\beta_1}
 + s_{3\beta_1} \big)
  +2 v^2 c_{\beta_1} (\lambda_{1223}
 +\lambda_{2123}
 +\lambda_{2213}
 +4 \lambda_{3313})
 -2 v^2 c_{3\beta_1} (\lambda_{1223}
 +\lambda_{2123}
 +\lambda_{2213})
 +2 v^2 s_{\beta_1} (\lambda_{2113} 
 +3 \lambda_{2223} 
 +4 \lambda_{3323})
 +2 v^2 s_{3\beta_1} (\lambda_{2113} 
 - \lambda_{2223} )
 
 +4 v^2 s_{2\beta_2}(\lambda_{1313}
 + \lambda_{1331}
 + \lambda_{2323} 
 + \lambda_{2332})\big) 
,\end{autobreak}
\end{align}
\begin{align}
\begin{autobreak}
M^2_{\pm,23} = 
c_{\beta_2}^{-1} (m_{13}^{2} s_{\beta_1}
 -m_{23}^{2} c_{\beta_1})
 +\frac{1}{16} c_{\beta_2} \big(8 v^2 c_{\beta_1}^{2} (s_{\beta_1} (
 -\lambda_{1113}
 +\lambda_{1223}
 +\lambda_{2123})
 +t_{\beta_2} (\lambda_{1323}
 +\lambda_{1332}))
 -4 v^2 s_{\beta_1} \big(2 s_{\beta_1} t_{\beta_2} (\lambda_{1323}
 +\lambda_{1332})
 +s_{2\beta_1} (\lambda_{1213}
 +\lambda_{2113}
 -\lambda_{2223})
 +2 \lambda_{2213} s_{\beta_1}^{2}
 +2 \lambda_{3313} t_{\beta_2}^{2}\big)
 -v^2 s_{2\beta_1} s_{3\beta_2} c_{\beta_2}^{
 -3} (\lambda_{1313}
 +\lambda_{1331}
 -\lambda_{2323}
 -\lambda_{2332})
 +v^2 s_{2\beta_1} t_{\beta_2} c_{\beta_2}^{
 -2} (
 -\lambda_{1313}
 -\lambda_{1331}
 +\lambda_{2323}
 +\lambda_{2332})
 +8 \lambda_{3323} v^2 c_{\beta_1} t_{\beta_2}^{2}
 +8 \lambda_{1123} v^2 c_{3\beta_1}^{3}\big)
\,.\end{autobreak}
\end{align}
}
The above mass matrix $\mathcal{M}_{\pm}^2$ can be diagonalised with
rotational matrix $R({\sigma})$ in the form 
\begin{align}
\overline{\mathcal{M}}^2_\pm= R({\sigma})\, \mathcal{M}^2_\pm \, R({\sigma})^{-1} =
\left(
\begin{array}{ccc}
M_{h_1^{\pm}}^2 & 0
\\
 0 & M_{h_2^{\pm}}^2

\end{array}
\right),
\end{align}
with 
\begin{align}
{M}_{h_1^{\pm}}^2 &=  M_{\pm, 22}^2 c_{\sigma}^{2} + M_{\pm,23}^2 s_{2 \sigma} + M_{\pm,33}^2 s_{\sigma}^{2}
, \nonumber \\
{M}_{h_2^{\pm}}^2 &= {1 \over 2} (M_{\pm,33}^2-M_{\pm,22}^2) s_{2\sigma} + M_{\pm,23}^2 c_{2 \sigma} 
\end{align}
where
\begin{align}
\tan{2\sigma}={2 M_{\pm,23} ^2 \over M_{\pm,22} ^2-M_{\pm,33}^2}.
\end{align}

\subsection{CP-even Scalar Mass Matrix} \label{AppevenMass}

The CP-even scalar mass matrix $\mathcal{M}_{\text{S}}^2$ in the generic basis $\phi_{1,2,3}$ may be expressed as follows:
\begin{align}
 \mathcal{M}_{\rm S}^2 =
\left(
\begin{array}{ccc}
A & {C}_{1} & {C}_{2}
\\
{C}_{1} & B_1& {C}_{3}
\\
{C}_{2} & {C}_{3} &B_2
\\
\end{array}
\right),
\end{align}
where the elements are
{\allowdisplaybreaks \small
\begin{align}
A & = m_{12}^2 t_{\beta_1}
  +m_{13}^2 c_{\beta_1}^{-1} t_{\beta_2}
  +2 \lambda_{11} v^2 c_{\beta_1}^{2} c_{\beta_2}^{2}
  - \frac{1}{2} v^2 s_{\beta_1} c_{\beta_1} c_{\beta_2}^{2} ( \lambda_{2212} t_{\beta_1}^{2}
  -3 \lambda_{1112})
  \nonumber \\ &
  -\frac{1}{2} v^2 c_{\beta_1} s_{\beta_2} c_{\beta_2} \big( t_{\beta_1}^{2} (\lambda_{2213}+\lambda_{1223}
  +\lambda_{2123})-3 \lambda_{1113}\big)
  \nonumber \\ &
  -\frac{1}{2} v^2 s_{\beta_2}^{2} \big( \lambda_{3313} c_{\beta_1}^{-1} t_{\beta_2}
  +t_{\beta_1} (\lambda_{1323}+\lambda_{1332}+\lambda_{3312}) \big),
 \nonumber \\[0.15in]
B_1 & = m_{12}^2 t_{\beta_1}^{-1}
  +m_{23}^2 s_{\beta_1}^{-1} t_{\beta_2}
  +2 \lambda_{22} v^2 s_{\beta_1}^{2} c_{\beta_2}^{2}
  -\frac{1}{2} v^2 s_{\beta_1} 
  c_{\beta_1} c_{\beta_2}^{2} ( \lambda_{1112} t_{\beta_1}^{-2} -3 \lambda_{2212} )
  \nonumber \\ &
  -\frac{1}{2} v^2 s_{\beta_1}  s_{\beta_2} c_{\beta_2} 
  \big( t_{\beta_1}^{-2} (\lambda_{1123} +\lambda_{1213}+\lambda_{2113})-3 \lambda_{2223} \big)
  \nonumber \\ &
  -\frac{1}{2} v^2 s_{\beta_2}^2 \big( \lambda_{3323} s_{\beta_1}^{-1} t_{\beta_2}
  +t_{\beta_1}^{-1} (\lambda_{1323} +\lambda_{1332}+\lambda_{3312}) \big),
 \nonumber \\[0.15in]
B_2 & = 2m_{13}^2 c_{\beta_1} t_{\beta_2}^{-1}
  +2 m_{23}^2 s_{\beta_1} t_{\beta_2}^{-1}
  +4 \lambda_{33} v^2 s_{\beta_2}^{2}+3 v^2 s_{\beta_2} c_{\beta_2} (\lambda_{3313} c_{\beta_1} +\lambda_{3323} s_{\beta_1})
  \nonumber \\ &
  - v^2 c_{\beta_2}^2 t_{\beta_2}^{-1} 
    \Big(
      \lambda_{1113} c_{\beta_1}^{3}
      +s_{\beta_1} c_{\beta_1}^{2} (\lambda_{1123}+\lambda_{1213}+\lambda_{2113})
        \nonumber \\ &
      +s_{\beta_1}^{2} c_{\beta_1} (\lambda_{1223}+\lambda_{2123}+\lambda_{2213})
      +\lambda_{2223} s_{\beta_1}^{3}
    \Big),
 \nonumber \\[0.15in]
C_1 & = - m_{12}^2 
  + \frac{1}{2} v^2 s_{2 \beta_1} c_{\beta_2}^{2} (\lambda_{1122}+\lambda_{1221})
  + \frac{1}{2} v^2 c_{\beta_2}^{2} ( 3 \lambda_{1112} c_{\beta_1}^{2}+\lambda_{1212} s_{2 \beta_1}+3 \lambda_{2212} s_{\beta_1}^{2} )
  \nonumber \\ &
  + \frac{1}{2} v^2 s_{2 \beta_2} 
    \big( 
    c_{\beta_1} (\lambda_{1123}+\lambda_{1213} +\lambda_{2113})
    +s_{\beta_1} (\lambda_{1223}+\lambda_{2123}+\lambda_{2213})
    \big)
     \nonumber \\ &
  + \frac{1}{2} v^2 s_{\beta_2}^{2} (\lambda_{1323}+\lambda_{1332}+\lambda_{3312}),
 \nonumber \\[0.15in]
C_2 & = -m_{13}^2 
  +\frac{1}{2} v^2 c_{\beta_1} s_{2 \beta_2} (\lambda_{1133}+\lambda_{1331})
  \nonumber \\ &
  + \frac{1}{2} v^2 c_{\beta_2}^{2} 
      \Big(
      3 \lambda_{1113} c_{\beta_1}^{2}
      +s_{2 \beta_1} (\lambda_{1123}+\lambda_{1213}+\lambda_{2113})
      +s_{\beta_1}^{2} (\lambda_{1223}+\lambda_{2123}+\lambda_{2213})
      \Big)
    \nonumber \\ &
  + \frac{1}{2} v^2 s_{2 \beta_2} \big(\lambda_{1313} c_{\beta_1}+s_{\beta_1} (\lambda_{1323}+\lambda_{1332}+\lambda_{3312})\big)
  + \frac{1}{2} v^2 3 \lambda_{3313} s_{\beta_2}^{2},
 \nonumber \\[0.15in]
C_3 & = -m_{23}^2
  +\frac{1}{2} v^2 s_{\beta_1} s_{2 \beta_2} (\lambda_{2233}+\lambda_{2332})
  \nonumber \\ &
  +\frac{1}{2} v^2 c_{\beta_2}^{2} 
      \Big(
      c_{\beta_1}^{2} (\lambda_{1123}+\lambda_{1213}+\lambda_{2113}+\lambda_{2213})+s_{2 \beta_1} (\lambda_{1223}+\lambda_{2123})+3 \lambda_{2223} s_{\beta_1}^{2}
      \Big)
    \nonumber \\ &
  + \frac{1}{2} v^2  s_{2 \beta_2} \big(c_{\beta_1} (\lambda_{1323}+\lambda_{1332}+\lambda_{3312})+\lambda_{2323} s_{\beta_1}\big)+3 \lambda_{3323} s_{\beta_2}^{2}.
\end{align}
}

\section{Renormalization Group Equations for 3HDM} \label{RGEs}

In this section, we present the complete set of one- and two-loop beta functions of all gauge, Yukawa and quartic couplings, for the Type-V 3HDM, including beta functions for potential mass parameters and the VEVs of the scalar doublets. These are presented in Subsections~\ref{RGE:a},~\ref{RGE:b},~\ref{RGE:c},~\ref{RGE:d}~and~\ref{RGE:e}. To this end, we use the following conventions~\cite{Lyonnet:2013dna}:
\begin{equation} 
\beta\left(g\right)\, \equiv\, \mu \frac{d g}{d \mu}\, \equiv\, \frac{1}{(4 \pi)^{2}}\beta^{(1)}(g)+\frac{1}{(4 \pi)^{4}}\beta^{(2)}(g)\,,
\end{equation}
where $\beta^{(1)}$ and $\beta^{(2)}$ refer to the one- and two-loop RGEs, respectively.

\subsection{One- and Two-Loop RGEs of Gauge Couplings}\label{RGE:a}

The one- and two-loop RGEs of the gauge couplings take on the following forms:
{\allowdisplaybreaks \small
\begin{align}
\begin{autobreak}
\beta^{(1)}(g_1) =\frac{43}{6} g_1^{3}
\,.\end{autobreak}
\end{align}
\begin{align}
\begin{autobreak}
\beta^{(2)}(g_1) =
  \frac{217}{18} g_1^{5}
+ \frac{15}{2} g_1^{3} g_2^{2}
+ \frac{44}{3} g_1^{3} g_3^{2}
-  \frac{5}{6} g_1^{3} \tr\left(y_d y_d^{\dagger} \right)
-  \frac{17}{6} g_1^{3} \tr\left(y_u y_u^{\dagger} \right)
-  \frac{5}{2} g_1^{3} \tr\left(y_e y_e^{\dagger} \right)
\,.\end{autobreak}
\end{align}
\begin{align}
\begin{autobreak}
\beta^{(1)}(g_2) =- \frac{17}{6} g_2^{3}
\,.\end{autobreak}
\end{align}
\begin{align}
\begin{autobreak}
\beta^{(2)}(g_2) =

 \frac{5}{2} g_1^{2} g_2^{3}

+ \frac{61}{6} g_2^{5}

+ 12 g_2^{3} g_3^{2}

-  \frac{3}{2} g_2^{3} \tr\left(y_d y_d^{\dagger} \right)

-  \frac{3}{2} g_2^{3} \tr\left(y_u y_u^{\dagger} \right)

-  \frac{1}{2} g_2^{3} \tr\left(y_e y_e^{\dagger} \right)
\,.\end{autobreak}
\end{align}
\begin{align}
\begin{autobreak}
\beta^{(1)}(g_3) =-7 g_3^{3}
\,.\end{autobreak}
\end{align}
\begin{align}
\begin{autobreak}
\beta^{(2)}(g_3) =

\frac{11}{6} g_1^{2} g_3^{3}

+ \frac{9}{2} g_2^{2} g_3^{3}

- 26 g_3^{5}

- 2 g_3^{3} \tr\left(y_d y_d^{\dagger} \right)

- 2 g_3^{3} \tr\left(y_u y_u^{\dagger} \right)
\,.\end{autobreak}
\end{align}
}

\subsection{One- and Two-Loop RGEs of Yukawa Couplings}\label{RGE:b}

The one- and two-loop RGEs of the Yukawa couplings can be given as follows:
{\allowdisplaybreaks \small
\begin{align}
\begin{autobreak}
\beta^{(1)}(y_d) =

 \frac{3}{2} y_d y_d^{\dagger} y_d

+ \frac{1}{2} y_u y_u^{\dagger} y_d

+ 3 \tr\left(y_d y_d^{\dagger} \right) y_d

-  \frac{5}{12} g_1^{2} y_d

-  \frac{9}{4} g_2^{2} y_d

- 8 g_3^{2} y_d
\,.\end{autobreak}
\end{align}
\begin{align}
\begin{autobreak}
\beta^{(2)}(y_d) =

\frac{3}{2} y_d y_d^{\dagger} y_d y_d^{\dagger} y_d

-  \frac{1}{4} y_d y_d^{\dagger} y_u y_u^{\dagger} y_d

-  \frac{1}{4} y_u y_u^{\dagger} y_u y_u^{\dagger} y_d

-  \frac{27}{4} \tr\left(y_d y_d^{\dagger} y_d y_d^{\dagger} \right) y_d

-  \frac{27}{4} \tr\left(y_d y_d^{\dagger} \right) y_d y_d^{\dagger} y_d

-  \frac{9}{4} \tr\left(y_d y_d^{\dagger} y_u y_u^{\dagger} \right) y_d

-  \frac{9}{4} \tr\left(y_u y_u^{\dagger} \right) y_u y_u^{\dagger} y_d

- 12 \lambda_{11} y_d y_d^{\dagger} y_d

- 2 \lambda_{1122} y_u y_u^{\dagger} y_d

+ 2 \lambda_{1221} y_u y_u^{\dagger} y_d

+ 6 \lambda_{11}^{2} y_d

+ \lambda_{1122}^{2} y_d

+ \lambda_{1122} \lambda_{1221} y_d

+ \lambda_{1221}^{2} y_d

+ \lambda_{1133}^{2} y_d

+ \lambda_{1133} \lambda_{1331} y_d

+ \lambda_{1331}^{2} y_d

+ \frac{3}{2} \left|{\lambda_{1212}}\right|^{2} y_d

+ \frac{3}{2} \left|{\lambda_{1313}}\right|^{2} y_d

+ \frac{187}{48} g_1^{2} y_d y_d^{\dagger} y_d

+ \frac{135}{16} g_2^{2} y_d y_d^{\dagger} y_d

+ 16 g_3^{2} y_d y_d^{\dagger} y_d

-  \frac{53}{144} g_1^{2} y_u y_u^{\dagger} y_d

+ \frac{33}{16} g_2^{2} y_u y_u^{\dagger} y_d

+ \frac{16}{3} g_3^{2} y_u y_u^{\dagger} y_d

+ \frac{25}{24} g_1^{2} \tr\left(y_d y_d^{\dagger} \right) y_d

+ \frac{45}{8} g_2^{2} \tr\left(y_d y_d^{\dagger} \right) y_d

+ 20 g_3^{2} \tr\left(y_d y_d^{\dagger} \right) y_d

-  \frac{11}{24} g_1^{4} y_d

-  \frac{9}{4} g_1^{2} g_2^{2} y_d

+ \frac{31}{9} g_1^{2} g_3^{2} y_d

-  \frac{19}{4} g_2^{4} y_d

+ 9 g_2^{2} g_3^{2} y_d

- 108 g_3^{4} y_d
\,.\end{autobreak}
\end{align}
\begin{align}
\begin{autobreak}
\beta^{(1)}(y_u) =

 \frac{1}{2} y_d y_d^{\dagger} y_u

+ \frac{3}{2} y_u y_u^{\dagger} y_u

+ 3 \tr\left(y_u y_u^{\dagger} \right) y_u

-  \frac{17}{12} g_1^{2} y_u

-  \frac{9}{4} g_2^{2} y_u

- 8 g_3^{2} y_u
\,.\end{autobreak}
\end{align}
\begin{align}
\begin{autobreak}
\beta^{(2)}(y_u) =

-  \frac{1}{4} y_d y_d^{\dagger} y_d y_d^{\dagger} y_u

-  \frac{1}{4} y_u y_u^{\dagger} y_d y_d^{\dagger} y_u

+ \frac{3}{2} y_u y_u^{\dagger} y_u y_u^{\dagger} y_u

-  \frac{9}{4} \tr\left(y_d y_d^{\dagger} \right) y_d y_d^{\dagger} y_u

-  \frac{9}{4} \tr\left(y_d y_d^{\dagger} y_u y_u^{\dagger} \right) y_u

-  \frac{27}{4} \tr\left(y_u y_u^{\dagger} y_u y_u^{\dagger} \right) y_u

-  \frac{27}{4} \tr\left(y_u y_u^{\dagger} \right) y_u y_u^{\dagger} y_u

- 2 \lambda_{1122} y_d y_d^{\dagger} y_u

+ 2 \lambda_{1221} y_d y_d^{\dagger} y_u

- 12 \lambda_{22} y_u y_u^{\dagger} y_u

+ 6 \lambda_{22}^{2} y_u

+ \lambda_{1122}^{2} y_u

+ \lambda_{1122} \lambda_{1221} y_u

+ \lambda_{1221}^{2} y_u

+ \lambda_{2233}^{2} y_u

+ \lambda_{2233} \lambda_{2332} y_u

+ \lambda_{2332}^{2} y_u

+ \frac{3}{2} \left|{\lambda_{1212}}\right|^{2} y_u

+ \frac{3}{2} \left|{\lambda_{2323}}\right|^{2} y_u

-  \frac{41}{144} g_1^{2} y_d y_d^{\dagger} y_u

+ \frac{33}{16} g_2^{2} y_d y_d^{\dagger} y_u

+ \frac{16}{3} g_3^{2} y_d y_d^{\dagger} y_u

+ \frac{223}{48} g_1^{2} y_u y_u^{\dagger} y_u

+ \frac{135}{16} g_2^{2} y_u y_u^{\dagger} y_u

+ 16 g_3^{2} y_u y_u^{\dagger} y_u

+ \frac{85}{24} g_1^{2} \tr\left(y_u y_u^{\dagger} \right) y_u

+ \frac{45}{8} g_2^{2} \tr\left(y_u y_u^{\dagger} \right) y_u

+ 20 g_3^{2} \tr\left(y_u y_u^{\dagger} \right) y_u

+ \frac{449}{72} g_1^{4} y_u

-  \frac{3}{4} g_1^{2} g_2^{2} y_u

+ \frac{19}{9} g_1^{2} g_3^{2} y_u

-  \frac{19}{4} g_2^{4} y_u

+ 9 g_2^{2} g_3^{2} y_u

- 108 g_3^{4} y_u
\,.\end{autobreak}
\end{align}
\begin{align}
\begin{autobreak}
\beta^{(1)}(\xi_{e}) =

 \frac{3}{2} y_e y_e^{\dagger} y_e

+ \tr\left(y_e y_e^{\dagger} \right) y_e

-  \frac{15}{4} g_1^{2} y_e

-  \frac{9}{4} g_2^{2} y_e
\,.\end{autobreak}
\end{align}
\begin{align}
\begin{autobreak}
\beta^{(2)}(y_e) =

 \frac{3}{2} y_e y_e^{\dagger} y_e y_e^{\dagger} y_e

-  \frac{9}{4} \tr\left(y_e y_e^{\dagger} y_e y_e^{\dagger} \right) y_e

-  \frac{9}{4} \tr\left(y_e y_e^{\dagger} \right) y_e y_e^{\dagger} y_e

- 12 \lambda_{33} y_e y_e^{\dagger} y_e

+ 6 \lambda_{33}^{2} y_e

+ \lambda_{1133}^{2} y_e

+ \lambda_{1133} \lambda_{1331} y_e

+ \lambda_{1331}^{2} y_e

+ \lambda_{2233}^{2} y_e

+ \lambda_{2233} \lambda_{2332} y_e

+ \lambda_{2332}^{2} y_e

+ \frac{3}{2} \left|{\lambda_{1313}}\right|^{2} y_e

+ \frac{3}{2} \left|{\lambda_{2323}}\right|^{2} y_e

+ \frac{129}{16} g_1^{2} y_e y_e^{\dagger} y_e

+ \frac{135}{16} g_2^{2} y_e y_e^{\dagger} y_e

+ \frac{25}{8} g_1^{2} \tr\left(y_e y_e^{\dagger} \right) y_e

+ \frac{15}{8} g_2^{2} \tr\left(y_e y_e^{\dagger} \right) y_e

+ \frac{509}{24} g_1^{4} y_e

+ \frac{9}{4} g_1^{2} g_2^{2} y_e

-  \frac{19}{4} g_2^{4} y_e
\,.\end{autobreak}
\end{align}
}

\subsection{One- and Two-Loop RGEs of Quartic Couplings}\label{RGE:c}

We obtain the following one- and two-loop RGEs of all quartic couplings for the 3HDM Type-V:
{\allowdisplaybreaks \small
\begin{align}
\begin{autobreak}
\beta^{(1)}(\lambda_{11}) =

24 \lambda_{11}^{2}

+ 2 \lambda_{1122}^{2}

+ 2 \lambda_{1122} \lambda_{1221}

+ \lambda_{1221}^{2}

+ 2 \lambda_{1133}^{2}

+ 2 \lambda_{1133} \lambda_{1331}

+ \lambda_{1331}^{2}

+ \left|{\lambda_{1212}}\right|^{2}

+ \left|{\lambda_{1313}}\right|^{2}

- 3 g_1^{2} \lambda_{11}

- 9 g_2^{2} \lambda_{11}

+ \frac{3}{8} g_1^{4}

+ \frac{3}{4} g_1^{2} g_2^{2}

+ \frac{9}{8} g_2^{4}

+ 12 \lambda_{11} \tr\left(y_d y_d^{\dagger} \right)

- 6 \tr\left(y_d y_d^{\dagger} y_d y_d^{\dagger} \right)
\,.\end{autobreak}
\end{align}
\begin{align}
\begin{autobreak}
\beta^{(2)}(\lambda_{11}) =

- 312 \lambda_{11}^{3}

- 20 \lambda_{11} \lambda_{1122}^{2}

- 20 \lambda_{11} \lambda_{1122} \lambda_{1221}

- 12 \lambda_{11} \lambda_{1221}^{2}

- 20 \lambda_{11} \lambda_{1133}^{2}

- 20 \lambda_{11} \lambda_{1133} \lambda_{1331}

- 12 \lambda_{11} \lambda_{1331}^{2}

- 14 \lambda_{11} \left|{\lambda_{1212}}\right|^{2}

- 14 \lambda_{11} \left|{\lambda_{1313}}\right|^{2}

- 8 \lambda_{1122}^{3}

- 16 \lambda_{1122} \lambda_{1221}^{2}

- 20 \lambda_{1122} \left|{\lambda_{1212}}\right|^{2}

- 12 \lambda_{1122}^{2} \lambda_{1221}

- 6 \lambda_{1221}^{3}

- 8 \lambda_{1133}^{3}

- 16 \lambda_{1133} \lambda_{1331}^{2}

- 20 \lambda_{1133} \left|{\lambda_{1313}}\right|^{2}

- 12 \lambda_{1133}^{2} \lambda_{1331}

- 6 \lambda_{1331}^{3}

- 22 \lambda_{1221} \left|{\lambda_{1212}}\right|^{2}

- 22 \lambda_{1331} \left|{\lambda_{1313}}\right|^{2}

+ 36 g_1^{2} \lambda_{11}^{2}

+ 108 g_2^{2} \lambda_{11}^{2}

+ 4 g_1^{2} \lambda_{1122}^{2}

+ 12 g_2^{2} \lambda_{1122}^{2}

+ 4 g_1^{2} \lambda_{1122} \lambda_{1221}

+ 12 g_2^{2} \lambda_{1122} \lambda_{1221}

+ 2 g_1^{2} \lambda_{1221}^{2}

+ 3 g_2^{2} \lambda_{1221}^{2}

+ 4 g_1^{2} \lambda_{1133}^{2}

+ 12 g_2^{2} \lambda_{1133}^{2}

+ 4 g_1^{2} \lambda_{1133} \lambda_{1331}

+ 12 g_2^{2} \lambda_{1133} \lambda_{1331}

+ 2 g_1^{2} \lambda_{1331}^{2}

+ 3 g_2^{2} \lambda_{1331}^{2}

-  g_1^{2} \left|{\lambda_{1212}}\right|^{2}

-  g_1^{2} \left|{\lambda_{1313}}\right|^{2}

+ \frac{673}{24} g_1^{4} \lambda_{11}

+ \frac{39}{4} g_1^{2} g_2^{2} \lambda_{11}

-  \frac{29}{8} g_2^{4} \lambda_{11}

+ \frac{5}{2} g_1^{4} \lambda_{1122}

+ \frac{15}{2} g_2^{4} \lambda_{1122}

+ \frac{5}{4} g_1^{4} \lambda_{1221}

+ \frac{5}{2} g_1^{2} g_2^{2} \lambda_{1221}

+ \frac{15}{4} g_2^{4} \lambda_{1221}

+ \frac{5}{2} g_1^{4} \lambda_{1133}

+ \frac{15}{2} g_2^{4} \lambda_{1133}

+ \frac{5}{4} g_1^{4} \lambda_{1331}

+ \frac{5}{2} g_1^{2} g_2^{2} \lambda_{1331}

+ \frac{15}{4} g_2^{4} \lambda_{1331}

-  \frac{407}{48} g_1^{6}

-  \frac{587}{48} g_1^{4} g_2^{2}

-  \frac{317}{48} g_1^{2} g_2^{4}

+ \frac{277}{16} g_2^{6}

- 144 \lambda_{11}^{2} \tr\left(y_d y_d^{\dagger} \right)

- 12 \lambda_{1122}^{2} \tr\left(y_u y_u^{\dagger} \right)

- 12 \lambda_{1122} \lambda_{1221} \tr\left(y_u y_u^{\dagger} \right)

- 6 \lambda_{1221}^{2} \tr\left(y_u y_u^{\dagger} \right)

- 4 \lambda_{1133}^{2} \tr\left(y_e y_e^{\dagger} \right)

- 4 \lambda_{1133} \lambda_{1331} \tr\left(y_e y_e^{\dagger} \right)

- 2 \lambda_{1331}^{2} \tr\left(y_e y_e^{\dagger} \right)

- 6 \left|{\lambda_{1212}}\right|^{2} \tr\left(y_u y_u^{\dagger} \right)

- 2 \left|{\lambda_{1313}}\right|^{2} \tr\left(y_e y_e^{\dagger} \right)

+ \frac{25}{6} g_1^{2} \lambda_{11} \tr\left(y_d y_d^{\dagger} \right)

+ \frac{45}{2} g_2^{2} \lambda_{11} \tr\left(y_d y_d^{\dagger} \right)

+ 80 g_3^{2} \lambda_{11} \tr\left(y_d y_d^{\dagger} \right)

+ \frac{5}{4} g_1^{4} \tr\left(y_d y_d^{\dagger} \right)

+ \frac{9}{2} g_1^{2} g_2^{2} \tr\left(y_d y_d^{\dagger} \right)

-  \frac{9}{4} g_2^{4} \tr\left(y_d y_d^{\dagger} \right)

- 3 \lambda_{11} \tr\left(y_d y_d^{\dagger} y_d y_d^{\dagger} \right)

- 9 \lambda_{11} \tr\left(y_d y_d^{\dagger} y_u y_u^{\dagger} \right)

+ \frac{4}{3} g_1^{2} \tr\left(y_d y_d^{\dagger} y_d y_d^{\dagger} \right)

- 32 g_3^{2} \tr\left(y_d y_d^{\dagger} y_d y_d^{\dagger} \right)

+ 30 \tr\left(y_d y_d^{\dagger} y_d y_d^{\dagger} y_d y_d^{\dagger} \right)

+ 6 \tr\left(y_d y_d^{\dagger} y_d y_d^{\dagger} y_u y_u^{\dagger} \right)
\,.\end{autobreak}
\end{align}
\begin{align}
\begin{autobreak}
\beta^{(1)}(\lambda_{22}) =

24 \lambda_{22}^{2}

+ 2 \lambda_{1122}^{2}

+ 2 \lambda_{1122} \lambda_{1221}

+ \lambda_{1221}^{2}

+ 2 \lambda_{2233}^{2}

+ 2 \lambda_{2233} \lambda_{2332}

+ \lambda_{2332}^{2}

+ \left|{\lambda_{1212}}\right|^{2}

+ \left|{\lambda_{2323}}\right|^{2}

- 3 g_1^{2} \lambda_{22}

- 9 g_2^{2} \lambda_{22}

+ \frac{3}{8} g_1^{4}

+ \frac{3}{4} g_1^{2} g_2^{2}

+ \frac{9}{8} g_2^{4}

+ 12 \lambda_{22} \tr\left(y_u y_u^{\dagger} \right)

- 6 \tr\left(y_u y_u^{\dagger} y_u y_u^{\dagger} \right)
\,.\end{autobreak}
\end{align}
\begin{align}
\begin{autobreak}
\beta^{(2)}(\lambda_{22}) =

- 312 \lambda_{22}^{3}

- 20 \lambda_{1122}^{2} \lambda_{22}

- 12 \lambda_{1221}^{2} \lambda_{22}

- 20 \lambda_{22} \lambda_{2233}^{2}

- 20 \lambda_{22} \lambda_{2233} \lambda_{2332}

- 12 \lambda_{22} \lambda_{2332}^{2}

- 14 \lambda_{22} \left|{\lambda_{2323}}\right|^{2}

- 8 \lambda_{1122}^{3}

- 20 \lambda_{1122} \lambda_{1221} \lambda_{22}

- 16 \lambda_{1122} \lambda_{1221}^{2}

- 20 \lambda_{1122} \left|{\lambda_{1212}}\right|^{2}

- 12 \lambda_{1122}^{2} \lambda_{1221}

- 6 \lambda_{1221}^{3}

- 8 \lambda_{2233}^{3}

- 16 \lambda_{2233} \lambda_{2332}^{2}

- 20 \lambda_{2233} \left|{\lambda_{2323}}\right|^{2}

- 12 \lambda_{2233}^{2} \lambda_{2332}

- 6 \lambda_{2332}^{3}

- 14 \lambda_{22} \left|{\lambda_{1212}}\right|^{2}

- 22 \lambda_{1221} \left|{\lambda_{1212}}\right|^{2}

- 22 \lambda_{2332} \left|{\lambda_{2323}}\right|^{2}

+ 36 g_1^{2} \lambda_{22}^{2}

+ 108 g_2^{2} \lambda_{22}^{2}

+ 4 g_1^{2} \lambda_{1122}^{2}

+ 12 g_2^{2} \lambda_{1122}^{2}

+ 4 g_1^{2} \lambda_{1122} \lambda_{1221}

+ 12 g_2^{2} \lambda_{1122} \lambda_{1221}

+ 2 g_1^{2} \lambda_{1221}^{2}

+ 3 g_2^{2} \lambda_{1221}^{2}

+ 4 g_1^{2} \lambda_{2233}^{2}

+ 12 g_2^{2} \lambda_{2233}^{2}

+ 4 g_1^{2} \lambda_{2233} \lambda_{2332}

+ 12 g_2^{2} \lambda_{2233} \lambda_{2332}

+ 2 g_1^{2} \lambda_{2332}^{2}

+ 3 g_2^{2} \lambda_{2332}^{2}

-  g_1^{2} \left|{\lambda_{1212}}\right|^{2}

-  g_1^{2} \left|{\lambda_{2323}}\right|^{2}

+ \frac{673}{24} g_1^{4} \lambda_{22}

+ \frac{39}{4} g_1^{2} g_2^{2} \lambda_{22}

-  \frac{29}{8} g_2^{4} \lambda_{22}

+ \frac{5}{2} g_1^{4} \lambda_{1122}

+ \frac{15}{2} g_2^{4} \lambda_{1122}

+ \frac{5}{4} g_1^{4} \lambda_{1221}

+ \frac{5}{2} g_1^{2} g_2^{2} \lambda_{1221}

+ \frac{15}{4} g_2^{4} \lambda_{1221}

+ \frac{5}{2} g_1^{4} \lambda_{2233}

+ \frac{15}{2} g_2^{4} \lambda_{2233}

+ \frac{5}{4} g_1^{4} \lambda_{2332}

+ \frac{5}{2} g_1^{2} g_2^{2} \lambda_{2332}

+ \frac{15}{4} g_2^{4} \lambda_{2332}

-  \frac{407}{48} g_1^{6}

-  \frac{587}{48} g_1^{4} g_2^{2}

-  \frac{317}{48} g_1^{2} g_2^{4}

+ \frac{277}{16} g_2^{6}

- 144 \lambda_{22}^{2} \tr\left(y_u y_u^{\dagger} \right)

- 12 \lambda_{1122}^{2} \tr\left(y_d y_d^{\dagger} \right)

- 12 \lambda_{1122} \lambda_{1221} \tr\left(y_d y_d^{\dagger} \right)

- 6 \lambda_{1221}^{2} \tr\left(y_d y_d^{\dagger} \right)

- 4 \lambda_{2233}^{2} \tr\left(y_e y_e^{\dagger} \right)

- 4 \lambda_{2233} \lambda_{2332} \tr\left(y_e y_e^{\dagger} \right)

- 2 \lambda_{2332}^{2} \tr\left(y_e y_e^{\dagger} \right)

- 6 \left|{\lambda_{1212}}\right|^{2} \tr\left(y_d y_d^{\dagger} \right)

- 2 \left|{\lambda_{2323}}\right|^{2} \tr\left(y_e y_e^{\dagger} \right)

+ \frac{85}{6} g_1^{2} \lambda_{22} \tr\left(y_u y_u^{\dagger} \right)

+ \frac{45}{2} g_2^{2} \lambda_{22} \tr\left(y_u y_u^{\dagger} \right)

+ 80 g_3^{2} \lambda_{22} \tr\left(y_u y_u^{\dagger} \right)

-  \frac{19}{4} g_1^{4} \tr\left(y_u y_u^{\dagger} \right)

+ \frac{21}{2} g_1^{2} g_2^{2} \tr\left(y_u y_u^{\dagger} \right)

-  \frac{9}{4} g_2^{4} \tr\left(y_u y_u^{\dagger} \right)

- 9 \lambda_{22} \tr\left(y_d y_d^{\dagger} y_u y_u^{\dagger} \right)

- 3 \lambda_{22} \tr\left(y_u y_u^{\dagger} y_u y_u^{\dagger} \right)

-  \frac{8}{3} g_1^{2} \tr\left(y_u y_u^{\dagger} y_u y_u^{\dagger} \right)

- 32 g_3^{2} \tr\left(y_u y_u^{\dagger} y_u y_u^{\dagger} \right)

+ 6 \tr\left(y_d y_d^{\dagger} y_u y_u^{\dagger} y_u y_u^{\dagger} \right)

+ 30 \tr\left(y_u y_u^{\dagger} y_u y_u^{\dagger} y_u y_u^{\dagger} \right)
\,.\end{autobreak}
\end{align}
\begin{align}
\begin{autobreak}
\beta^{(1)}(\lambda_{33}) =

 24 \lambda_{33}^{2}

+ 2 \lambda_{1133}^{2}

+ 2 \lambda_{1133} \lambda_{1331}

+ \lambda_{1331}^{2}

+ 2 \lambda_{2233}^{2}

+ 2 \lambda_{2233} \lambda_{2332}

+ \lambda_{2332}^{2}

+ \left|{\lambda_{1313}}\right|^{2}

+ \left|{\lambda_{2323}}\right|^{2}

- 3 g_1^{2} \lambda_{33}

- 9 g_2^{2} \lambda_{33}

+ \frac{3}{8} g_1^{4}

+ \frac{3}{4} g_1^{2} g_2^{2}

+ \frac{9}{8} g_2^{4}

+ 4 \lambda_{33} \tr\left(y_e y_e^{\dagger} \right)

- 2 \tr\left(y_e y_e^{\dagger} y_e y_e^{\dagger} \right)\,.
\end{autobreak}
\end{align}
\begin{align}
\begin{autobreak}
\beta^{(2)}(\lambda_{33}) =

- 312 \lambda_{33}^{3}

- 20 \lambda_{1133}^{2} \lambda_{33}

- 12 \lambda_{1331}^{2} \lambda_{33}

- 20 \lambda_{2233}^{2} \lambda_{33}

- 12 \lambda_{2332}^{2} \lambda_{33}

- 8 \lambda_{1133}^{3}

- 20 \lambda_{1133} \lambda_{1331} \lambda_{33}

- 16 \lambda_{1133} \lambda_{1331}^{2}

- 20 \lambda_{1133} \left|{\lambda_{1313}}\right|^{2}

- 12 \lambda_{1133}^{2} \lambda_{1331}

- 6 \lambda_{1331}^{3}

- 8 \lambda_{2233}^{3}

- 20 \lambda_{2233} \lambda_{2332} \lambda_{33}

- 16 \lambda_{2233} \lambda_{2332}^{2}

- 20 \lambda_{2233} \left|{\lambda_{2323}}\right|^{2}

- 12 \lambda_{2233}^{2} \lambda_{2332}

- 6 \lambda_{2332}^{3}

- 14 \lambda_{33} \left|{\lambda_{1313}}\right|^{2}

- 22 \lambda_{1331} \left|{\lambda_{1313}}\right|^{2}

- 14 \lambda_{33} \left|{\lambda_{2323}}\right|^{2}

- 22 \lambda_{2332} \left|{\lambda_{2323}}\right|^{2}

+ 36 g_1^{2} \lambda_{33}^{2}

+ 108 g_2^{2} \lambda_{33}^{2}

+ 4 g_1^{2} \lambda_{1133}^{2}

+ 12 g_2^{2} \lambda_{1133}^{2}

+ 4 g_1^{2} \lambda_{1133} \lambda_{1331}

+ 12 g_2^{2} \lambda_{1133} \lambda_{1331}

+ 2 g_1^{2} \lambda_{1331}^{2}

+ 3 g_2^{2} \lambda_{1331}^{2}

+ 4 g_1^{2} \lambda_{2233}^{2}

+ 12 g_2^{2} \lambda_{2233}^{2}

+ 4 g_1^{2} \lambda_{2233} \lambda_{2332}

+ 12 g_2^{2} \lambda_{2233} \lambda_{2332}

+ 2 g_1^{2} \lambda_{2332}^{2}

+ 3 g_2^{2} \lambda_{2332}^{2}

-  g_1^{2} \left|{\lambda_{1313}}\right|^{2}

-  g_1^{2} \left|{\lambda_{2323}}\right|^{2}

+ \frac{673}{24} g_1^{4} \lambda_{33}

+ \frac{39}{4} g_1^{2} g_2^{2} \lambda_{33}

-  \frac{29}{8} g_2^{4} \lambda_{33}

+ \frac{5}{2} g_1^{4} \lambda_{1133}

+ \frac{15}{2} g_2^{4} \lambda_{1133}

+ \frac{5}{4} g_1^{4} \lambda_{1331}

+ \frac{5}{2} g_1^{2} g_2^{2} \lambda_{1331}

+ \frac{15}{4} g_2^{4} \lambda_{1331}

+ \frac{5}{2} g_1^{4} \lambda_{2233}

+ \frac{15}{2} g_2^{4} \lambda_{2233}

+ \frac{5}{4} g_1^{4} \lambda_{2332}

+ \frac{5}{2} g_1^{2} g_2^{2} \lambda_{2332}

+ \frac{15}{4} g_2^{4} \lambda_{2332}

-  \frac{407}{48} g_1^{6}

-  \frac{587}{48} g_1^{4} g_2^{2}

-  \frac{317}{48} g_1^{2} g_2^{4}

+ \frac{277}{16} g_2^{6}

- 48 \lambda_{33}^{2} \tr\left(y_e y_e^{\dagger} \right)

- 12 \lambda_{1133}^{2} \tr\left(y_d y_d^{\dagger} \right)

- 12 \lambda_{1133} \lambda_{1331} \tr\left(y_d y_d^{\dagger} \right)

- 6 \lambda_{1331}^{2} \tr\left(y_d y_d^{\dagger} \right)

- 12 \lambda_{2233}^{2} \tr\left(y_u y_u^{\dagger} \right)

- 12 \lambda_{2233} \lambda_{2332} \tr\left(y_u y_u^{\dagger} \right)

- 6 \lambda_{2332}^{2} \tr\left(y_u y_u^{\dagger} \right)

- 6 \left|{\lambda_{1313}}\right|^{2} \tr\left(y_d y_d^{\dagger} \right)

- 6 \left|{\lambda_{2323}}\right|^{2} \tr\left(y_u y_u^{\dagger} \right)

+ \frac{25}{2} g_1^{2} \lambda_{33} \tr\left(y_e y_e^{\dagger} \right)

+ \frac{15}{2} g_2^{2} \lambda_{33} \tr\left(y_e y_e^{\dagger} \right)

-  \frac{25}{4} g_1^{4} \tr\left(y_e y_e^{\dagger} \right)

+ \frac{11}{2} g_1^{2} g_2^{2} \tr\left(y_e y_e^{\dagger} \right)

-  \frac{3}{4} g_2^{4} \tr\left(y_e y_e^{\dagger} \right)

-  \lambda_{33} \tr\left(y_e y_e^{\dagger} y_e y_e^{\dagger} \right)

- 4 g_1^{2} \tr\left(y_e y_e^{\dagger} y_e y_e^{\dagger} \right)

+ 10 \tr\left(y_e y_e^{\dagger} y_e y_e^{\dagger} y_e y_e^{\dagger} \right)
\,.\end{autobreak}
\end{align}
\begin{align}
\begin{autobreak}
\beta^{(1)}(\lambda_{1122}) =

12 \lambda_{11} \lambda_{1122}

+ 4 \lambda_{11} \lambda_{1221}

+ 12 \lambda_{1122} \lambda_{22}

+ 4 \lambda_{1122}^{2}

+ 4 \lambda_{1221} \lambda_{22}

+ 2 \lambda_{1221}^{2}

+ 4 \lambda_{1133} \lambda_{2233}

+ 2 \lambda_{1133} \lambda_{2332}

+ 2 \lambda_{1331} \lambda_{2233}

+ 2 \left|{\lambda_{1212}}\right|^{2}

- 3 g_1^{2} \lambda_{1122}

- 9 g_2^{2} \lambda_{1122}

+ \frac{3}{4} g_1^{4}

-  \frac{3}{2} g_1^{2} g_2^{2}

+ \frac{9}{4} g_2^{4}

+ 6 \lambda_{1122} \tr\left(y_d y_d^{\dagger} \right)

+ 6 \lambda_{1122} \tr\left(y_u y_u^{\dagger} \right)

- 12 \tr\left(y_d y_d^{\dagger} y_u y_u^{\dagger} \right)
\,.\end{autobreak}
\end{align}
\begin{align}
\begin{autobreak}
\beta^{(2)}(\lambda_{1122}) =

- 72 \lambda_{11} \lambda_{1122}^{2}

- 32 \lambda_{11} \lambda_{1122} \lambda_{1221}

- 28 \lambda_{11} \lambda_{1221}^{2}

- 36 \lambda_{11} \left|{\lambda_{1212}}\right|^{2}

- 72 \lambda_{1122}^{2} \lambda_{22}

- 28 \lambda_{1221}^{2} \lambda_{22}

- 60 \lambda_{11}^{2} \lambda_{1122}

- 60 \lambda_{1122} \lambda_{22}^{2}

- 12 \lambda_{1122}^{3}

- 32 \lambda_{1122} \lambda_{1221} \lambda_{22}

- 16 \lambda_{1122} \lambda_{1221}^{2}

- 2 \lambda_{1122} \lambda_{1133}^{2}

- 2 \lambda_{1122} \lambda_{1133} \lambda_{1331}

- 16 \lambda_{1122} \lambda_{1133} \lambda_{2233}

- 8 \lambda_{1122} \lambda_{1133} \lambda_{2332}

- 2 \lambda_{1122} \lambda_{1331}^{2}

- 8 \lambda_{1122} \lambda_{1331} \lambda_{2233}

- 2 \lambda_{1122} \lambda_{2233}^{2}

- 2 \lambda_{1122} \lambda_{2233} \lambda_{2332}

- 2 \lambda_{1122} \lambda_{2332}^{2}

- 18 \lambda_{1122} \left|{\lambda_{1212}}\right|^{2}

- 3 \lambda_{1122} \left|{\lambda_{1313}}\right|^{2}

- 3 \lambda_{1122} \left|{\lambda_{2323}}\right|^{2}

- 16 \lambda_{11}^{2} \lambda_{1221}

- 16 \lambda_{1221} \lambda_{22}^{2}

- 4 \lambda_{1122}^{2} \lambda_{1221}

- 12 \lambda_{1221}^{3}

- 2 \lambda_{1221} \lambda_{1331}^{2}

- 4 \lambda_{1221} \lambda_{1331} \lambda_{2332}

- 2 \lambda_{1221} \lambda_{2332}^{2}

- 2 \lambda_{1221} \left|{\lambda_{1313}}\right|^{2}

- 2 \lambda_{1221} \left|{\lambda_{2323}}\right|^{2}

- 8 \lambda_{1133} \lambda_{1331} \lambda_{2233}

- 8 \lambda_{1133} \lambda_{2233}^{2}

- 8 \lambda_{1133} \lambda_{2233} \lambda_{2332}

- 8 \lambda_{1133} \lambda_{2332}^{2}

- 12 \lambda_{1133} \left|{\lambda_{2323}}\right|^{2}

- 4 \lambda_{1331} \lambda_{2233}^{2}

- 2 \lambda_{1331} \lambda_{2332}^{2}

- 2 \lambda_{1331} \left|{\lambda_{2323}}\right|^{2}

- 8 \lambda_{1133}^{2} \lambda_{2233}

- 8 \lambda_{1331}^{2} \lambda_{2233}

- 4 \lambda_{1133}^{2} \lambda_{2332}

- 2 \lambda_{1331}^{2} \lambda_{2332}

- 36 \lambda_{22} \left|{\lambda_{1212}}\right|^{2}

- 44 \lambda_{1221} \left|{\lambda_{1212}}\right|^{2}

- 2 \lambda_{1212} \lambda_{2323} \lambda_{1313}^{*}

- 12 \lambda_{2233} \left|{\lambda_{1313}}\right|^{2}

- 2 \lambda_{2332} \left|{\lambda_{1313}}\right|^{2}

- 2 \lambda_{1313} \lambda_{1212}^{*} \lambda_{2323}^{*}

+ 24 g_1^{2} \lambda_{11} \lambda_{1122}

+ 72 g_2^{2} \lambda_{11} \lambda_{1122}

+ 8 g_1^{2} \lambda_{11} \lambda_{1221}

+ 36 g_2^{2} \lambda_{11} \lambda_{1221}

+ 24 g_1^{2} \lambda_{1122} \lambda_{22}

+ 72 g_2^{2} \lambda_{1122} \lambda_{22}

+ 2 g_1^{2} \lambda_{1122}^{2}

+ 6 g_2^{2} \lambda_{1122}^{2}

- 12 g_2^{2} \lambda_{1122} \lambda_{1221}

+ 8 g_1^{2} \lambda_{1221} \lambda_{22}

+ 36 g_2^{2} \lambda_{1221} \lambda_{22}

- 2 g_1^{2} \lambda_{1221}^{2}

+ 6 g_2^{2} \lambda_{1221}^{2}

+ 8 g_1^{2} \lambda_{1133} \lambda_{2233}

+ 24 g_2^{2} \lambda_{1133} \lambda_{2233}

+ 4 g_1^{2} \lambda_{1133} \lambda_{2332}

+ 12 g_2^{2} \lambda_{1133} \lambda_{2332}

+ 4 g_1^{2} \lambda_{1331} \lambda_{2233}

+ 12 g_2^{2} \lambda_{1331} \lambda_{2233}

+ 6 g_2^{2} \lambda_{1331} \lambda_{2332}

+ 4 g_1^{2} \left|{\lambda_{1212}}\right|^{2}

+ \frac{15}{2} g_1^{4} \lambda_{11}

- 5 g_1^{2} g_2^{2} \lambda_{11}

+ \frac{45}{2} g_2^{4} \lambda_{11}

+ \frac{15}{2} g_1^{4} \lambda_{22}

- 5 g_1^{2} g_2^{2} \lambda_{22}

+ \frac{45}{2} g_2^{4} \lambda_{22}

+ \frac{613}{24} g_1^{4} \lambda_{1122}

+ \frac{11}{4} g_1^{2} g_2^{2} \lambda_{1122}

-  \frac{89}{8} g_2^{4} \lambda_{1122}

+ \frac{5}{2} g_1^{4} \lambda_{1221}

- 3 g_1^{2} g_2^{2} \lambda_{1221}

+ \frac{15}{2} g_2^{4} \lambda_{1221}

+ \frac{5}{2} g_1^{4} \lambda_{1133}

+ \frac{15}{2} g_2^{4} \lambda_{1133}

+ \frac{5}{4} g_1^{4} \lambda_{1331}

-  \frac{5}{2} g_1^{2} g_2^{2} \lambda_{1331}

+ \frac{15}{4} g_2^{4} \lambda_{1331}

+ \frac{5}{2} g_1^{4} \lambda_{2233}

+ \frac{15}{2} g_2^{4} \lambda_{2233}

+ \frac{5}{4} g_1^{4} \lambda_{2332}

-  \frac{5}{2} g_1^{2} g_2^{2} \lambda_{2332}

+ \frac{15}{4} g_2^{4} \lambda_{2332}

-  \frac{407}{24} g_1^{6}

+ \frac{317}{24} g_1^{4} g_2^{2}

+ \frac{47}{24} g_1^{2} g_2^{4}

+ \frac{277}{8} g_2^{6}

- 72 \lambda_{11} \lambda_{1122} \tr\left(y_d y_d^{\dagger} \right)

- 24 \lambda_{11} \lambda_{1221} \tr\left(y_d y_d^{\dagger} \right)

- 72 \lambda_{1122} \lambda_{22} \tr\left(y_u y_u^{\dagger} \right)

- 12 \lambda_{1122}^{2} \tr\left(y_d y_d^{\dagger} \right)

- 12 \lambda_{1122}^{2} \tr\left(y_u y_u^{\dagger} \right)

- 24 \lambda_{1221} \lambda_{22} \tr\left(y_u y_u^{\dagger} \right)

- 6 \lambda_{1221}^{2} \tr\left(y_d y_d^{\dagger} \right)

- 6 \lambda_{1221}^{2} \tr\left(y_u y_u^{\dagger} \right)

- 8 \lambda_{1133} \lambda_{2233} \tr\left(y_e y_e^{\dagger} \right)

- 4 \lambda_{1133} \lambda_{2332} \tr\left(y_e y_e^{\dagger} \right)

- 4 \lambda_{1331} \lambda_{2233} \tr\left(y_e y_e^{\dagger} \right)

- 6 \left|{\lambda_{1212}}\right|^{2} \tr\left(y_d y_d^{\dagger} \right)

- 6 \left|{\lambda_{1212}}\right|^{2} \tr\left(y_u y_u^{\dagger} \right)

+ \frac{25}{12} g_1^{2} \lambda_{1122} \tr\left(y_d y_d^{\dagger} \right)

+ \frac{85}{12} g_1^{2} \lambda_{1122} \tr\left(y_u y_u^{\dagger} \right)

+ \frac{45}{4} g_2^{2} \lambda_{1122} \tr\left(y_d y_d^{\dagger} \right)

+ \frac{45}{4} g_2^{2} \lambda_{1122} \tr\left(y_u y_u^{\dagger} \right)

+ 40 g_3^{2} \lambda_{1122} \tr\left(y_d y_d^{\dagger} \right)

+ 40 g_3^{2} \lambda_{1122} \tr\left(y_u y_u^{\dagger} \right)

+ \frac{5}{4} g_1^{4} \tr\left(y_d y_d^{\dagger} \right)

-  \frac{19}{4} g_1^{4} \tr\left(y_u y_u^{\dagger} \right)

-  \frac{9}{2} g_1^{2} g_2^{2} \tr\left(y_d y_d^{\dagger} \right)

-  \frac{21}{2} g_1^{2} g_2^{2} \tr\left(y_u y_u^{\dagger} \right)

-  \frac{9}{4} g_2^{4} \tr\left(y_d y_d^{\dagger} \right)

-  \frac{9}{4} g_2^{4} \tr\left(y_u y_u^{\dagger} \right)

-  \frac{27}{2} \lambda_{1122} \tr\left(y_d y_d^{\dagger} y_d y_d^{\dagger} \right)

+ 15 \lambda_{1122} \tr\left(y_d y_d^{\dagger} y_u y_u^{\dagger} \right)

-  \frac{27}{2} \lambda_{1122} \tr\left(y_u y_u^{\dagger} y_u y_u^{\dagger} \right)

-  \frac{4}{3} g_1^{2} \tr\left(y_d y_d^{\dagger} y_u y_u^{\dagger} \right)

- 64 g_3^{2} \tr\left(y_d y_d^{\dagger} y_u y_u^{\dagger} \right)

+ 36 \tr\left(y_d y_d^{\dagger} y_d y_d^{\dagger} y_u y_u^{\dagger} \right)

+ 36 \tr\left(y_d y_d^{\dagger} y_u y_u^{\dagger} y_u y_u^{\dagger} \right)
\,.\end{autobreak}
\end{align}
\begin{align}
\begin{autobreak}
\beta^{(1)}(\lambda_{1221}) =

 4 \lambda_{11} \lambda_{1221}

+ 8 \lambda_{1122} \lambda_{1221}

+ 4 \lambda_{1221} \lambda_{22}

+ 4 \lambda_{1221}^{2}

+ 2 \lambda_{1331} \lambda_{2332}

+ 8 \left|{\lambda_{1212}}\right|^{2}

- 3 g_1^{2} \lambda_{1221}

- 9 g_2^{2} \lambda_{1221}

+ 3 g_1^{2} g_2^{2}

+ 6 \lambda_{1221} \tr\left(y_d y_d^{\dagger} \right)

+ 6 \lambda_{1221} \tr\left(y_u y_u^{\dagger} \right)

+ 12 \tr\left(y_d y_d^{\dagger} y_u y_u^{\dagger} \right)
\,.\end{autobreak}
\end{align}
\begin{align}
\begin{autobreak}
\beta^{(2)}(\lambda_{1221}) =

- 80 \lambda_{11} \lambda_{1122} \lambda_{1221}

- 40 \lambda_{11} \lambda_{1221}^{2}

- 48 \lambda_{11} \left|{\lambda_{1212}}\right|^{2}

- 40 \lambda_{1221}^{2} \lambda_{22}

- 80 \lambda_{1122} \lambda_{1221} \lambda_{22}

- 28 \lambda_{1122} \lambda_{1221}^{2}

- 8 \lambda_{1122} \lambda_{1331} \lambda_{2332}

- 48 \lambda_{1122} \left|{\lambda_{1212}}\right|^{2}

- 28 \lambda_{11}^{2} \lambda_{1221}

- 28 \lambda_{1221} \lambda_{22}^{2}

- 28 \lambda_{1122}^{2} \lambda_{1221}

- 2 \lambda_{1133}^{2} \lambda_{1221}

+ 2 \lambda_{1221} \lambda_{1331}^{2}

- 8 \lambda_{1221} \lambda_{1331} \lambda_{2233}

- 8 \lambda_{1221} \lambda_{1331} \lambda_{2332}

- 2 \lambda_{1221} \lambda_{2233}^{2}

- 2 \lambda_{1221} \lambda_{2233} \lambda_{2332}

+ 2 \lambda_{1221} \lambda_{2332}^{2}

+ \lambda_{1221} \left|{\lambda_{1313}}\right|^{2}

+ \lambda_{1221} \left|{\lambda_{2323}}\right|^{2}

- 2 \lambda_{1133} \lambda_{1221} \lambda_{1331}

- 16 \lambda_{1133} \lambda_{1221} \lambda_{2233}

- 8 \lambda_{1133} \lambda_{1221} \lambda_{2332}

- 8 \lambda_{1133} \lambda_{1331} \lambda_{2332}

- 8 \lambda_{1331} \lambda_{2233} \lambda_{2332}

- 4 \lambda_{1331} \lambda_{2332}^{2}

- 8 \lambda_{1331} \left|{\lambda_{2323}}\right|^{2}

- 4 \lambda_{1331}^{2} \lambda_{2332}

- 48 \lambda_{22} \left|{\lambda_{1212}}\right|^{2}

- 26 \lambda_{1221} \left|{\lambda_{1212}}\right|^{2}

- 8 \lambda_{1212} \lambda_{2323} \lambda_{1313}^{*}

- 8 \lambda_{2332} \left|{\lambda_{1313}}\right|^{2}

- 8 \lambda_{1313} \lambda_{1212}^{*} \lambda_{2323}^{*}

+ 8 g_1^{2} \lambda_{11} \lambda_{1221}

+ 4 g_1^{2} \lambda_{1122} \lambda_{1221}

+ 36 g_2^{2} \lambda_{1122} \lambda_{1221}

+ 8 g_1^{2} \lambda_{1221} \lambda_{22}

+ 8 g_1^{2} \lambda_{1221}^{2}

+ 18 g_2^{2} \lambda_{1221}^{2}

+ 4 g_1^{2} \lambda_{1331} \lambda_{2332}

+ 16 g_1^{2} \left|{\lambda_{1212}}\right|^{2}

+ 54 g_2^{2} \left|{\lambda_{1212}}\right|^{2}

+ 10 g_1^{2} g_2^{2} \lambda_{11}

+ 10 g_1^{2} g_2^{2} \lambda_{22}

+ 2 g_1^{2} g_2^{2} \lambda_{1122}

+ \frac{493}{24} g_1^{4} \lambda_{1221}

+ \frac{51}{4} g_1^{2} g_2^{2} \lambda_{1221}

-  \frac{209}{8} g_2^{4} \lambda_{1221}

+ 5 g_1^{2} g_2^{2} \lambda_{1331}

+ 5 g_1^{2} g_2^{2} \lambda_{2332}

-  \frac{113}{3} g_1^{4} g_2^{2}

-  \frac{91}{6} g_1^{2} g_2^{4}

- 24 \lambda_{11} \lambda_{1221} \tr\left(y_d y_d^{\dagger} \right)

- 24 \lambda_{1122} \lambda_{1221} \tr\left(y_d y_d^{\dagger} \right)

- 24 \lambda_{1122} \lambda_{1221} \tr\left(y_u y_u^{\dagger} \right)

- 24 \lambda_{1221} \lambda_{22} \tr\left(y_u y_u^{\dagger} \right)

- 12 \lambda_{1221}^{2} \tr\left(y_d y_d^{\dagger} \right)

- 12 \lambda_{1221}^{2} \tr\left(y_u y_u^{\dagger} \right)

- 4 \lambda_{1331} \lambda_{2332} \tr\left(y_e y_e^{\dagger} \right)

- 24 \left|{\lambda_{1212}}\right|^{2} \tr\left(y_d y_d^{\dagger} \right)

- 24 \left|{\lambda_{1212}}\right|^{2} \tr\left(y_u y_u^{\dagger} \right)

+ \frac{25}{12} g_1^{2} \lambda_{1221} \tr\left(y_d y_d^{\dagger} \right)

+ \frac{85}{12} g_1^{2} \lambda_{1221} \tr\left(y_u y_u^{\dagger} \right)

+ \frac{45}{4} g_2^{2} \lambda_{1221} \tr\left(y_d y_d^{\dagger} \right)

+ \frac{45}{4} g_2^{2} \lambda_{1221} \tr\left(y_u y_u^{\dagger} \right)

+ 40 g_3^{2} \lambda_{1221} \tr\left(y_d y_d^{\dagger} \right)

+ 40 g_3^{2} \lambda_{1221} \tr\left(y_u y_u^{\dagger} \right)

+ 9 g_1^{2} g_2^{2} \tr\left(y_d y_d^{\dagger} \right)

+ 21 g_1^{2} g_2^{2} \tr\left(y_u y_u^{\dagger} \right)

- 24 \lambda_{1122} \tr\left(y_d y_d^{\dagger} y_u y_u^{\dagger} \right)

-  \frac{27}{2} \lambda_{1221} \tr\left(y_d y_d^{\dagger} y_d y_d^{\dagger} \right)

- 33 \lambda_{1221} \tr\left(y_d y_d^{\dagger} y_u y_u^{\dagger} \right)

-  \frac{27}{2} \lambda_{1221} \tr\left(y_u y_u^{\dagger} y_u y_u^{\dagger} \right)

+ \frac{4}{3} g_1^{2} \tr\left(y_d y_d^{\dagger} y_u y_u^{\dagger} \right)

+ 64 g_3^{2} \tr\left(y_d y_d^{\dagger} y_u y_u^{\dagger} \right)

- 24 \tr\left(y_d y_d^{\dagger} y_d y_d^{\dagger} y_u y_u^{\dagger} \right)

- 24 \tr\left(y_d y_d^{\dagger} y_u y_u^{\dagger} y_u y_u^{\dagger} \right)
\,.\end{autobreak}
\end{align}
\begin{align}
\begin{autobreak}
\beta^{(1)}(\lambda_{1133}) =

 12 \lambda_{11} \lambda_{1133}

+ 4 \lambda_{11} \lambda_{1331}

+ 4 \lambda_{1122} \lambda_{2233}

+ 2 \lambda_{1122} \lambda_{2332}

+ 2 \lambda_{1221} \lambda_{2233}

+ 12 \lambda_{1133} \lambda_{33}

+ 4 \lambda_{1133}^{2}

+ 4 \lambda_{1331} \lambda_{33}

+ 2 \lambda_{1331}^{2}

+ 2 \left|{\lambda_{1313}}\right|^{2}

- 3 g_1^{2} \lambda_{1133}

- 9 g_2^{2} \lambda_{1133}

+ \frac{3}{4} g_1^{4}

-  \frac{3}{2} g_1^{2} g_2^{2}

+ \frac{9}{4} g_2^{4}

+ 6 \lambda_{1133} \tr\left(y_d y_d^{\dagger} \right)

+ 2 \lambda_{1133} \tr\left(y_e y_e^{\dagger} \right)
\,.\end{autobreak}
\end{align}
\begin{align}
\begin{autobreak}
\beta^{(2)}(\lambda_{1133}) =

- 72 \lambda_{11} \lambda_{1133}^{2}

- 32 \lambda_{11} \lambda_{1133} \lambda_{1331}

- 28 \lambda_{11} \lambda_{1331}^{2}

- 36 \lambda_{11} \left|{\lambda_{1313}}\right|^{2}

- 72 \lambda_{1133}^{2} \lambda_{33}

- 28 \lambda_{1331}^{2} \lambda_{33}

- 8 \lambda_{1122} \lambda_{1221} \lambda_{2233}

- 2 \lambda_{1122} \lambda_{1133} \lambda_{1221}

- 16 \lambda_{1122} \lambda_{1133} \lambda_{2233}

- 8 \lambda_{1122} \lambda_{1133} \lambda_{2332}

- 8 \lambda_{1122} \lambda_{2233}^{2}

- 8 \lambda_{1122} \lambda_{2233} \lambda_{2332}

- 8 \lambda_{1122} \lambda_{2332}^{2}

- 12 \lambda_{1122} \left|{\lambda_{2323}}\right|^{2}

- 4 \lambda_{1221} \lambda_{1331} \lambda_{2332}

- 4 \lambda_{1221} \lambda_{2233}^{2}

- 2 \lambda_{1221} \lambda_{2332}^{2}

- 2 \lambda_{1221} \left|{\lambda_{2323}}\right|^{2}

- 60 \lambda_{11}^{2} \lambda_{1133}

- 60 \lambda_{1133} \lambda_{33}^{2}

- 2 \lambda_{1122}^{2} \lambda_{1133}

- 2 \lambda_{1133} \lambda_{1221}^{2}

- 8 \lambda_{1133} \lambda_{1221} \lambda_{2233}

- 12 \lambda_{1133}^{3}

- 32 \lambda_{1133} \lambda_{1331} \lambda_{33}

- 16 \lambda_{1133} \lambda_{1331}^{2}

- 2 \lambda_{1133} \lambda_{2233}^{2}

- 2 \lambda_{1133} \lambda_{2233} \lambda_{2332}

- 2 \lambda_{1133} \lambda_{2332}^{2}

- 3 \lambda_{1133} \left|{\lambda_{1212}}\right|^{2}

- 18 \lambda_{1133} \left|{\lambda_{1313}}\right|^{2}

- 3 \lambda_{1133} \left|{\lambda_{2323}}\right|^{2}

- 16 \lambda_{11}^{2} \lambda_{1331}

- 16 \lambda_{1331} \lambda_{33}^{2}

- 2 \lambda_{1221}^{2} \lambda_{1331}

- 4 \lambda_{1133}^{2} \lambda_{1331}

- 12 \lambda_{1331}^{3}

- 2 \lambda_{1331} \lambda_{2332}^{2}

- 2 \lambda_{1331} \left|{\lambda_{2323}}\right|^{2}

- 8 \lambda_{1122}^{2} \lambda_{2233}

- 8 \lambda_{1221}^{2} \lambda_{2233}

- 4 \lambda_{1122}^{2} \lambda_{2332}

- 2 \lambda_{1221}^{2} \lambda_{2332}

- 2 \lambda_{1331} \left|{\lambda_{1212}}\right|^{2}

- 12 \lambda_{2233} \left|{\lambda_{1212}}\right|^{2}

- 2 \lambda_{2332} \left|{\lambda_{1212}}\right|^{2}

- 2 \lambda_{1212} \lambda_{2323} \lambda_{1313}^{*}

- 36 \lambda_{33} \left|{\lambda_{1313}}\right|^{2}

- 44 \lambda_{1331} \left|{\lambda_{1313}}\right|^{2}

- 2 \lambda_{1313} \lambda_{1212}^{*} \lambda_{2323}^{*}

+ 24 g_1^{2} \lambda_{11} \lambda_{1133}

+ 72 g_2^{2} \lambda_{11} \lambda_{1133}

+ 8 g_1^{2} \lambda_{11} \lambda_{1331}

+ 36 g_2^{2} \lambda_{11} \lambda_{1331}

+ 8 g_1^{2} \lambda_{1122} \lambda_{2233}

+ 24 g_2^{2} \lambda_{1122} \lambda_{2233}

+ 4 g_1^{2} \lambda_{1122} \lambda_{2332}

+ 12 g_2^{2} \lambda_{1122} \lambda_{2332}

+ 4 g_1^{2} \lambda_{1221} \lambda_{2233}

+ 12 g_2^{2} \lambda_{1221} \lambda_{2233}

+ 6 g_2^{2} \lambda_{1221} \lambda_{2332}

+ 24 g_1^{2} \lambda_{1133} \lambda_{33}

+ 72 g_2^{2} \lambda_{1133} \lambda_{33}

+ 2 g_1^{2} \lambda_{1133}^{2}

+ 6 g_2^{2} \lambda_{1133}^{2}

- 12 g_2^{2} \lambda_{1133} \lambda_{1331}

+ 8 g_1^{2} \lambda_{1331} \lambda_{33}

+ 36 g_2^{2} \lambda_{1331} \lambda_{33}

- 2 g_1^{2} \lambda_{1331}^{2}

+ 6 g_2^{2} \lambda_{1331}^{2}

+ 4 g_1^{2} \left|{\lambda_{1313}}\right|^{2}

+ \frac{15}{2} g_1^{4} \lambda_{11}

- 5 g_1^{2} g_2^{2} \lambda_{11}

+ \frac{45}{2} g_2^{4} \lambda_{11}

+ \frac{15}{2} g_1^{4} \lambda_{33}

- 5 g_1^{2} g_2^{2} \lambda_{33}

+ \frac{45}{2} g_2^{4} \lambda_{33}

+ \frac{5}{2} g_1^{4} \lambda_{1122}

+ \frac{15}{2} g_2^{4} \lambda_{1122}

+ \frac{5}{4} g_1^{4} \lambda_{1221}

-  \frac{5}{2} g_1^{2} g_2^{2} \lambda_{1221}

+ \frac{15}{4} g_2^{4} \lambda_{1221}

+ \frac{613}{24} g_1^{4} \lambda_{1133}

+ \frac{11}{4} g_1^{2} g_2^{2} \lambda_{1133}

-  \frac{89}{8} g_2^{4} \lambda_{1133}

+ \frac{5}{2} g_1^{4} \lambda_{1331}

- 3 g_1^{2} g_2^{2} \lambda_{1331}

+ \frac{15}{2} g_2^{4} \lambda_{1331}

+ \frac{5}{2} g_1^{4} \lambda_{2233}

+ \frac{15}{2} g_2^{4} \lambda_{2233}

+ \frac{5}{4} g_1^{4} \lambda_{2332}

-  \frac{5}{2} g_1^{2} g_2^{2} \lambda_{2332}

+ \frac{15}{4} g_2^{4} \lambda_{2332}

-  \frac{407}{24} g_1^{6}

+ \frac{317}{24} g_1^{4} g_2^{2}

+ \frac{47}{24} g_1^{2} g_2^{4}

+ \frac{277}{8} g_2^{6}

- 72 \lambda_{11} \lambda_{1133} \tr\left(y_d y_d^{\dagger} \right)

- 24 \lambda_{11} \lambda_{1331} \tr\left(y_d y_d^{\dagger} \right)

- 24 \lambda_{1122} \lambda_{2233} \tr\left(y_u y_u^{\dagger} \right)

- 12 \lambda_{1122} \lambda_{2332} \tr\left(y_u y_u^{\dagger} \right)

- 12 \lambda_{1221} \lambda_{2233} \tr\left(y_u y_u^{\dagger} \right)

- 24 \lambda_{1133} \lambda_{33} \tr\left(y_e y_e^{\dagger} \right)

- 12 \lambda_{1133}^{2} \tr\left(y_d y_d^{\dagger} \right)

- 4 \lambda_{1133}^{2} \tr\left(y_e y_e^{\dagger} \right)

- 8 \lambda_{1331} \lambda_{33} \tr\left(y_e y_e^{\dagger} \right)

- 6 \lambda_{1331}^{2} \tr\left(y_d y_d^{\dagger} \right)

- 2 \lambda_{1331}^{2} \tr\left(y_e y_e^{\dagger} \right)

- 6 \left|{\lambda_{1313}}\right|^{2} \tr\left(y_d y_d^{\dagger} \right)

- 2 \left|{\lambda_{1313}}\right|^{2} \tr\left(y_e y_e^{\dagger} \right)

+ \frac{25}{12} g_1^{2} \lambda_{1133} \tr\left(y_d y_d^{\dagger} \right)

+ \frac{25}{4} g_1^{2} \lambda_{1133} \tr\left(y_e y_e^{\dagger} \right)

+ \frac{45}{4} g_2^{2} \lambda_{1133} \tr\left(y_d y_d^{\dagger} \right)

+ \frac{15}{4} g_2^{2} \lambda_{1133} \tr\left(y_e y_e^{\dagger} \right)

+ 40 g_3^{2} \lambda_{1133} \tr\left(y_d y_d^{\dagger} \right)

+ \frac{5}{4} g_1^{4} \tr\left(y_d y_d^{\dagger} \right)

-  \frac{25}{4} g_1^{4} \tr\left(y_e y_e^{\dagger} \right)

-  \frac{9}{2} g_1^{2} g_2^{2} \tr\left(y_d y_d^{\dagger} \right)

-  \frac{11}{2} g_1^{2} g_2^{2} \tr\left(y_e y_e^{\dagger} \right)

-  \frac{9}{4} g_2^{4} \tr\left(y_d y_d^{\dagger} \right)

-  \frac{3}{4} g_2^{4} \tr\left(y_e y_e^{\dagger} \right)

-  \frac{27}{2} \lambda_{1133} \tr\left(y_d y_d^{\dagger} y_d y_d^{\dagger} \right)

-  \frac{9}{2} \lambda_{1133} \tr\left(y_d y_d^{\dagger} y_u y_u^{\dagger} \right)

-  \frac{9}{2} \lambda_{1133} \tr\left(y_e y_e^{\dagger} y_e y_e^{\dagger} \right)
\,.\end{autobreak}
\end{align}
\begin{align}
\begin{autobreak}
\beta^{(1)}(\lambda_{1331}) =

+ 4 \lambda_{11} \lambda_{1331}

+ 2 \lambda_{1221} \lambda_{2332}

+ 8 \lambda_{1133} \lambda_{1331}

+ 4 \lambda_{1331} \lambda_{33}

+ 4 \lambda_{1331}^{2}

+ 8 \left|{\lambda_{1313}}\right|^{2}

- 3 g_1^{2} \lambda_{1331}

- 9 g_2^{2} \lambda_{1331}

+ 3 g_1^{2} g_2^{2}

+ 6 \lambda_{1331} \tr\left(y_d y_d^{\dagger} \right)

+ 2 \lambda_{1331} \tr\left(y_e y_e^{\dagger} \right)
\,.\end{autobreak}
\end{align}
\begin{align}
\begin{autobreak}
\beta^{(2)}(\lambda_{1331}) =

- 80 \lambda_{11} \lambda_{1133} \lambda_{1331}

- 40 \lambda_{11} \lambda_{1331}^{2}

- 48 \lambda_{11} \left|{\lambda_{1313}}\right|^{2}

- 40 \lambda_{1331}^{2} \lambda_{33}

- 2 \lambda_{1122} \lambda_{1221} \lambda_{1331}

- 8 \lambda_{1122} \lambda_{1221} \lambda_{2332}

- 16 \lambda_{1122} \lambda_{1331} \lambda_{2233}

- 8 \lambda_{1122} \lambda_{1331} \lambda_{2332}

- 8 \lambda_{1221} \lambda_{1331} \lambda_{2233}

- 8 \lambda_{1221} \lambda_{1331} \lambda_{2332}

- 8 \lambda_{1221} \lambda_{2233} \lambda_{2332}

- 4 \lambda_{1221} \lambda_{2332}^{2}

- 8 \lambda_{1221} \left|{\lambda_{2323}}\right|^{2}

- 8 \lambda_{1133} \lambda_{1221} \lambda_{2332}

- 80 \lambda_{1133} \lambda_{1331} \lambda_{33}

- 28 \lambda_{1133} \lambda_{1331}^{2}

- 48 \lambda_{1133} \left|{\lambda_{1313}}\right|^{2}

- 28 \lambda_{11}^{2} \lambda_{1331}

- 28 \lambda_{1331} \lambda_{33}^{2}

- 2 \lambda_{1122}^{2} \lambda_{1331}

+ 2 \lambda_{1221}^{2} \lambda_{1331}

- 28 \lambda_{1133}^{2} \lambda_{1331}

- 2 \lambda_{1331} \lambda_{2233}^{2}

- 2 \lambda_{1331} \lambda_{2233} \lambda_{2332}

+ 2 \lambda_{1331} \lambda_{2332}^{2}

+ \lambda_{1331} \left|{\lambda_{2323}}\right|^{2}

- 4 \lambda_{1221}^{2} \lambda_{2332}

+ \lambda_{1331} \left|{\lambda_{1212}}\right|^{2}

- 8 \lambda_{2332} \left|{\lambda_{1212}}\right|^{2}

- 8 \lambda_{1212} \lambda_{2323} \lambda_{1313}^{*}

- 48 \lambda_{33} \left|{\lambda_{1313}}\right|^{2}

- 26 \lambda_{1331} \left|{\lambda_{1313}}\right|^{2}

- 8 \lambda_{1313} \lambda_{1212}^{*} \lambda_{2323}^{*}

+ 8 g_1^{2} \lambda_{11} \lambda_{1331}

+ 4 g_1^{2} \lambda_{1221} \lambda_{2332}

+ 4 g_1^{2} \lambda_{1133} \lambda_{1331}

+ 36 g_2^{2} \lambda_{1133} \lambda_{1331}

+ 8 g_1^{2} \lambda_{1331} \lambda_{33}

+ 8 g_1^{2} \lambda_{1331}^{2}

+ 18 g_2^{2} \lambda_{1331}^{2}

+ 16 g_1^{2} \left|{\lambda_{1313}}\right|^{2}

+ 54 g_2^{2} \left|{\lambda_{1313}}\right|^{2}

+ 10 g_1^{2} g_2^{2} \lambda_{11}

+ 10 g_1^{2} g_2^{2} \lambda_{33}

+ 5 g_1^{2} g_2^{2} \lambda_{1221}

+ 2 g_1^{2} g_2^{2} \lambda_{1133}

+ \frac{493}{24} g_1^{4} \lambda_{1331}

+ \frac{51}{4} g_1^{2} g_2^{2} \lambda_{1331}

-  \frac{209}{8} g_2^{4} \lambda_{1331}

+ 5 g_1^{2} g_2^{2} \lambda_{2332}

-  \frac{113}{3} g_1^{4} g_2^{2}

-  \frac{91}{6} g_1^{2} g_2^{4}

- 24 \lambda_{11} \lambda_{1331} \tr\left(y_d y_d^{\dagger} \right)

- 12 \lambda_{1221} \lambda_{2332} \tr\left(y_u y_u^{\dagger} \right)

- 24 \lambda_{1133} \lambda_{1331} \tr\left(y_d y_d^{\dagger} \right)

- 8 \lambda_{1133} \lambda_{1331} \tr\left(y_e y_e^{\dagger} \right)

- 8 \lambda_{1331} \lambda_{33} \tr\left(y_e y_e^{\dagger} \right)

- 12 \lambda_{1331}^{2} \tr\left(y_d y_d^{\dagger} \right)

- 4 \lambda_{1331}^{2} \tr\left(y_e y_e^{\dagger} \right)

- 24 \left|{\lambda_{1313}}\right|^{2} \tr\left(y_d y_d^{\dagger} \right)

- 8 \left|{\lambda_{1313}}\right|^{2} \tr\left(y_e y_e^{\dagger} \right)

+ \frac{25}{12} g_1^{2} \lambda_{1331} \tr\left(y_d y_d^{\dagger} \right)

+ \frac{25}{4} g_1^{2} \lambda_{1331} \tr\left(y_e y_e^{\dagger} \right)

+ \frac{45}{4} g_2^{2} \lambda_{1331} \tr\left(y_d y_d^{\dagger} \right)

+ \frac{15}{4} g_2^{2} \lambda_{1331} \tr\left(y_e y_e^{\dagger} \right)

+ 40 g_3^{2} \lambda_{1331} \tr\left(y_d y_d^{\dagger} \right)

+ 9 g_1^{2} g_2^{2} \tr\left(y_d y_d^{\dagger} \right)

+ 11 g_1^{2} g_2^{2} \tr\left(y_e y_e^{\dagger} \right)

-  \frac{27}{2} \lambda_{1331} \tr\left(y_d y_d^{\dagger} y_d y_d^{\dagger} \right)

-  \frac{9}{2} \lambda_{1331} \tr\left(y_d y_d^{\dagger} y_u y_u^{\dagger} \right)

-  \frac{9}{2} \lambda_{1331} \tr\left(y_e y_e^{\dagger} y_e y_e^{\dagger} \right)
\,.\end{autobreak}
\end{align}
\begin{align}
\begin{autobreak}
\beta^{(1)}(\lambda_{2233}) =

 12 \lambda_{22} \lambda_{2233}

+ 4 \lambda_{22} \lambda_{2332}

+ 4 \lambda_{1122} \lambda_{1133}

+ 2 \lambda_{1122} \lambda_{1331}

+ 2 \lambda_{1133} \lambda_{1221}

+ 12 \lambda_{2233} \lambda_{33}

+ 4 \lambda_{2233}^{2}

+ 4 \lambda_{2332} \lambda_{33}

+ 2 \lambda_{2332}^{2}

+ 2 \left|{\lambda_{2323}}\right|^{2}

- 3 g_1^{2} \lambda_{2233}

- 9 g_2^{2} \lambda_{2233}

+ \frac{3}{4} g_1^{4}

-  \frac{3}{2} g_1^{2} g_2^{2}

+ \frac{9}{4} g_2^{4}

+ 6 \lambda_{2233} \tr\left(y_u y_u^{\dagger} \right)

+ 2 \lambda_{2233} \tr\left(y_e y_e^{\dagger} \right)
\,.\end{autobreak}
\end{align}
\begin{align}
\begin{autobreak}
\beta^{(2)}(\lambda_{2233}) =

- 72 \lambda_{22} \lambda_{2233}^{2}

- 32 \lambda_{22} \lambda_{2233} \lambda_{2332}

- 28 \lambda_{22} \lambda_{2332}^{2}

- 36 \lambda_{22} \left|{\lambda_{2323}}\right|^{2}

- 72 \lambda_{2233}^{2} \lambda_{33}

- 28 \lambda_{2332}^{2} \lambda_{33}

- 2 \lambda_{1122} \lambda_{1221} \lambda_{2233}

- 8 \lambda_{1122} \lambda_{1133} \lambda_{1221}

- 8 \lambda_{1122} \lambda_{1133}^{2}

- 8 \lambda_{1122} \lambda_{1133} \lambda_{1331}

- 16 \lambda_{1122} \lambda_{1133} \lambda_{2233}

- 8 \lambda_{1122} \lambda_{1331}^{2}

- 8 \lambda_{1122} \lambda_{1331} \lambda_{2233}

- 12 \lambda_{1122} \left|{\lambda_{1313}}\right|^{2}

- 4 \lambda_{1133}^{2} \lambda_{1221}

- 2 \lambda_{1221} \lambda_{1331}^{2}

- 4 \lambda_{1221} \lambda_{1331} \lambda_{2332}

- 2 \lambda_{1221} \left|{\lambda_{1313}}\right|^{2}

- 8 \lambda_{1122}^{2} \lambda_{1133}

- 8 \lambda_{1133} \lambda_{1221}^{2}

- 8 \lambda_{1133} \lambda_{1221} \lambda_{2233}

- 2 \lambda_{1133} \lambda_{1331} \lambda_{2233}

- 12 \lambda_{1133} \left|{\lambda_{1212}}\right|^{2}

- 4 \lambda_{1122}^{2} \lambda_{1331}

- 2 \lambda_{1221}^{2} \lambda_{1331}

- 60 \lambda_{22}^{2} \lambda_{2233}

- 60 \lambda_{2233} \lambda_{33}^{2}

- 2 \lambda_{1122}^{2} \lambda_{2233}

- 2 \lambda_{1221}^{2} \lambda_{2233}

- 2 \lambda_{1133}^{2} \lambda_{2233}

- 2 \lambda_{1331}^{2} \lambda_{2233}

- 12 \lambda_{2233}^{3}

- 32 \lambda_{2233} \lambda_{2332} \lambda_{33}

- 16 \lambda_{2233} \lambda_{2332}^{2}

- 18 \lambda_{2233} \left|{\lambda_{2323}}\right|^{2}

- 16 \lambda_{22}^{2} \lambda_{2332}

- 16 \lambda_{2332} \lambda_{33}^{2}

- 2 \lambda_{1221}^{2} \lambda_{2332}

- 2 \lambda_{1331}^{2} \lambda_{2332}

- 4 \lambda_{2233}^{2} \lambda_{2332}

- 12 \lambda_{2332}^{3}

- 2 \lambda_{1331} \left|{\lambda_{1212}}\right|^{2}

- 3 \lambda_{2233} \left|{\lambda_{1212}}\right|^{2}

- 2 \lambda_{2332} \left|{\lambda_{1212}}\right|^{2}

- 2 \lambda_{1212} \lambda_{2323} \lambda_{1313}^{*}

- 3 \lambda_{2233} \left|{\lambda_{1313}}\right|^{2}

- 2 \lambda_{2332} \left|{\lambda_{1313}}\right|^{2}

- 2 \lambda_{1313} \lambda_{1212}^{*} \lambda_{2323}^{*}

- 36 \lambda_{33} \left|{\lambda_{2323}}\right|^{2}

- 44 \lambda_{2332} \left|{\lambda_{2323}}\right|^{2}

+ 24 g_1^{2} \lambda_{22} \lambda_{2233}

+ 72 g_2^{2} \lambda_{22} \lambda_{2233}

+ 8 g_1^{2} \lambda_{22} \lambda_{2332}

+ 36 g_2^{2} \lambda_{22} \lambda_{2332}

+ 8 g_1^{2} \lambda_{1122} \lambda_{1133}

+ 24 g_2^{2} \lambda_{1122} \lambda_{1133}

+ 4 g_1^{2} \lambda_{1122} \lambda_{1331}

+ 12 g_2^{2} \lambda_{1122} \lambda_{1331}

+ 6 g_2^{2} \lambda_{1221} \lambda_{1331}

+ 4 g_1^{2} \lambda_{1133} \lambda_{1221}

+ 12 g_2^{2} \lambda_{1133} \lambda_{1221}

+ 24 g_1^{2} \lambda_{2233} \lambda_{33}

+ 72 g_2^{2} \lambda_{2233} \lambda_{33}

+ 2 g_1^{2} \lambda_{2233}^{2}

+ 6 g_2^{2} \lambda_{2233}^{2}

- 12 g_2^{2} \lambda_{2233} \lambda_{2332}

+ 8 g_1^{2} \lambda_{2332} \lambda_{33}

+ 36 g_2^{2} \lambda_{2332} \lambda_{33}

- 2 g_1^{2} \lambda_{2332}^{2}

+ 6 g_2^{2} \lambda_{2332}^{2}

+ 4 g_1^{2} \left|{\lambda_{2323}}\right|^{2}

+ \frac{15}{2} g_1^{4} \lambda_{22}

- 5 g_1^{2} g_2^{2} \lambda_{22}

+ \frac{45}{2} g_2^{4} \lambda_{22}

+ \frac{15}{2} g_1^{4} \lambda_{33}

- 5 g_1^{2} g_2^{2} \lambda_{33}

+ \frac{45}{2} g_2^{4} \lambda_{33}

+ \frac{5}{2} g_1^{4} \lambda_{1122}

+ \frac{15}{2} g_2^{4} \lambda_{1122}

+ \frac{5}{4} g_1^{4} \lambda_{1221}

-  \frac{5}{2} g_1^{2} g_2^{2} \lambda_{1221}

+ \frac{15}{4} g_2^{4} \lambda_{1221}

+ \frac{5}{2} g_1^{4} \lambda_{1133}

+ \frac{15}{2} g_2^{4} \lambda_{1133}

+ \frac{5}{4} g_1^{4} \lambda_{1331}

-  \frac{5}{2} g_1^{2} g_2^{2} \lambda_{1331}

+ \frac{15}{4} g_2^{4} \lambda_{1331}

+ \frac{613}{24} g_1^{4} \lambda_{2233}

+ \frac{11}{4} g_1^{2} g_2^{2} \lambda_{2233}

-  \frac{89}{8} g_2^{4} \lambda_{2233}

+ \frac{5}{2} g_1^{4} \lambda_{2332}

- 3 g_1^{2} g_2^{2} \lambda_{2332}

+ \frac{15}{2} g_2^{4} \lambda_{2332}

-  \frac{407}{24} g_1^{6}

+ \frac{317}{24} g_1^{4} g_2^{2}

+ \frac{47}{24} g_1^{2} g_2^{4}

+ \frac{277}{8} g_2^{6}

- 72 \lambda_{22} \lambda_{2233} \tr\left(y_u y_u^{\dagger} \right)

- 24 \lambda_{22} \lambda_{2332} \tr\left(y_u y_u^{\dagger} \right)

- 24 \lambda_{1122} \lambda_{1133} \tr\left(y_d y_d^{\dagger} \right)

- 12 \lambda_{1122} \lambda_{1331} \tr\left(y_d y_d^{\dagger} \right)

- 12 \lambda_{1133} \lambda_{1221} \tr\left(y_d y_d^{\dagger} \right)

- 24 \lambda_{2233} \lambda_{33} \tr\left(y_e y_e^{\dagger} \right)

- 12 \lambda_{2233}^{2} \tr\left(y_u y_u^{\dagger} \right)

- 4 \lambda_{2233}^{2} \tr\left(y_e y_e^{\dagger} \right)

- 8 \lambda_{2332} \lambda_{33} \tr\left(y_e y_e^{\dagger} \right)

- 6 \lambda_{2332}^{2} \tr\left(y_u y_u^{\dagger} \right)

- 2 \lambda_{2332}^{2} \tr\left(y_e y_e^{\dagger} \right)

- 6 \left|{\lambda_{2323}}\right|^{2} \tr\left(y_u y_u^{\dagger} \right)

- 2 \left|{\lambda_{2323}}\right|^{2} \tr\left(y_e y_e^{\dagger} \right)

+ \frac{85}{12} g_1^{2} \lambda_{2233} \tr\left(y_u y_u^{\dagger} \right)

+ \frac{25}{4} g_1^{2} \lambda_{2233} \tr\left(y_e y_e^{\dagger} \right)

+ \frac{45}{4} g_2^{2} \lambda_{2233} \tr\left(y_u y_u^{\dagger} \right)

+ \frac{15}{4} g_2^{2} \lambda_{2233} \tr\left(y_e y_e^{\dagger} \right)

+ 40 g_3^{2} \lambda_{2233} \tr\left(y_u y_u^{\dagger} \right)

-  \frac{19}{4} g_1^{4} \tr\left(y_u y_u^{\dagger} \right)

-  \frac{25}{4} g_1^{4} \tr\left(y_e y_e^{\dagger} \right)

-  \frac{21}{2} g_1^{2} g_2^{2} \tr\left(y_u y_u^{\dagger} \right)

-  \frac{11}{2} g_1^{2} g_2^{2} \tr\left(y_e y_e^{\dagger} \right)

-  \frac{9}{4} g_2^{4} \tr\left(y_u y_u^{\dagger} \right)

-  \frac{3}{4} g_2^{4} \tr\left(y_e y_e^{\dagger} \right)

-  \frac{9}{2} \lambda_{2233} \tr\left(y_d y_d^{\dagger} y_u y_u^{\dagger} \right)

-  \frac{27}{2} \lambda_{2233} \tr\left(y_u y_u^{\dagger} y_u y_u^{\dagger} \right)

-  \frac{9}{2} \lambda_{2233} \tr\left(y_e y_e^{\dagger} y_e y_e^{\dagger} \right)
\,.\end{autobreak}
\end{align}
\begin{align}
\begin{autobreak}
\beta^{(1)}(\lambda_{2332}) =

 4 \lambda_{22} \lambda_{2332}

+ 2 \lambda_{1221} \lambda_{1331}

+ 8 \lambda_{2233} \lambda_{2332}

+ 4 \lambda_{2332} \lambda_{33}

+ 4 \lambda_{2332}^{2}

+ 8 \left|{\lambda_{2323}}\right|^{2}

- 3 g_1^{2} \lambda_{2332}

- 9 g_2^{2} \lambda_{2332}

+ 3 g_1^{2} g_2^{2}

+ 6 \lambda_{2332} \tr\left(y_u y_u^{\dagger} \right)

+ 2 \lambda_{2332} \tr\left(y_e y_e^{\dagger} \right)
\,.\end{autobreak}
\end{align}
\begin{align}
\begin{autobreak}
\beta^{(2)}(\lambda_{2332}) =

- 80 \lambda_{22} \lambda_{2233} \lambda_{2332}

- 40 \lambda_{22} \lambda_{2332}^{2}

- 48 \lambda_{22} \left|{\lambda_{2323}}\right|^{2}

- 40 \lambda_{2332}^{2} \lambda_{33}

- 8 \lambda_{1122} \lambda_{1221} \lambda_{1331}

- 2 \lambda_{1122} \lambda_{1221} \lambda_{2332}

- 16 \lambda_{1122} \lambda_{1133} \lambda_{2332}

- 8 \lambda_{1122} \lambda_{1331} \lambda_{2332}

- 4 \lambda_{1221} \lambda_{1331}^{2}

- 8 \lambda_{1221} \lambda_{1331} \lambda_{2233}

- 8 \lambda_{1221} \lambda_{1331} \lambda_{2332}

- 8 \lambda_{1221} \left|{\lambda_{1313}}\right|^{2}

- 8 \lambda_{1133} \lambda_{1221} \lambda_{1331}

- 8 \lambda_{1133} \lambda_{1221} \lambda_{2332}

- 2 \lambda_{1133} \lambda_{1331} \lambda_{2332}

- 4 \lambda_{1221}^{2} \lambda_{1331}

- 80 \lambda_{2233} \lambda_{2332} \lambda_{33}

- 28 \lambda_{2233} \lambda_{2332}^{2}

- 48 \lambda_{2233} \left|{\lambda_{2323}}\right|^{2}

- 28 \lambda_{22}^{2} \lambda_{2332}

- 28 \lambda_{2332} \lambda_{33}^{2}

- 2 \lambda_{1122}^{2} \lambda_{2332}

+ 2 \lambda_{1221}^{2} \lambda_{2332}

- 2 \lambda_{1133}^{2} \lambda_{2332}

+ 2 \lambda_{1331}^{2} \lambda_{2332}

- 28 \lambda_{2233}^{2} \lambda_{2332}

- 8 \lambda_{1331} \left|{\lambda_{1212}}\right|^{2}

+ \lambda_{2332} \left|{\lambda_{1212}}\right|^{2}

- 8 \lambda_{1212} \lambda_{2323} \lambda_{1313}^{*}

+ \lambda_{2332} \left|{\lambda_{1313}}\right|^{2}

- 8 \lambda_{1313} \lambda_{1212}^{*} \lambda_{2323}^{*}

- 48 \lambda_{33} \left|{\lambda_{2323}}\right|^{2}

- 26 \lambda_{2332} \left|{\lambda_{2323}}\right|^{2}

+ 8 g_1^{2} \lambda_{22} \lambda_{2332}

+ 4 g_1^{2} \lambda_{1221} \lambda_{1331}

+ 4 g_1^{2} \lambda_{2233} \lambda_{2332}

+ 36 g_2^{2} \lambda_{2233} \lambda_{2332}

+ 8 g_1^{2} \lambda_{2332} \lambda_{33}

+ 8 g_1^{2} \lambda_{2332}^{2}

+ 18 g_2^{2} \lambda_{2332}^{2}

+ 16 g_1^{2} \left|{\lambda_{2323}}\right|^{2}

+ 54 g_2^{2} \left|{\lambda_{2323}}\right|^{2}

+ 10 g_1^{2} g_2^{2} \lambda_{22}

+ 10 g_1^{2} g_2^{2} \lambda_{33}

+ 5 g_1^{2} g_2^{2} \lambda_{1221}

+ 5 g_1^{2} g_2^{2} \lambda_{1331}

+ 2 g_1^{2} g_2^{2} \lambda_{2233}

+ \frac{493}{24} g_1^{4} \lambda_{2332}

+ \frac{51}{4} g_1^{2} g_2^{2} \lambda_{2332}

-  \frac{209}{8} g_2^{4} \lambda_{2332}

-  \frac{113}{3} g_1^{4} g_2^{2}

-  \frac{91}{6} g_1^{2} g_2^{4}

- 24 \lambda_{22} \lambda_{2332} \tr\left(y_u y_u^{\dagger} \right)

- 12 \lambda_{1221} \lambda_{1331} \tr\left(y_d y_d^{\dagger} \right)

- 24 \lambda_{2233} \lambda_{2332} \tr\left(y_u y_u^{\dagger} \right)

- 8 \lambda_{2233} \lambda_{2332} \tr\left(y_e y_e^{\dagger} \right)

- 8 \lambda_{2332} \lambda_{33} \tr\left(y_e y_e^{\dagger} \right)

- 12 \lambda_{2332}^{2} \tr\left(y_u y_u^{\dagger} \right)

- 4 \lambda_{2332}^{2} \tr\left(y_e y_e^{\dagger} \right)

- 24 \left|{\lambda_{2323}}\right|^{2} \tr\left(y_u y_u^{\dagger} \right)

- 8 \left|{\lambda_{2323}}\right|^{2} \tr\left(y_e y_e^{\dagger} \right)

+ \frac{85}{12} g_1^{2} \lambda_{2332} \tr\left(y_u y_u^{\dagger} \right)

+ \frac{25}{4} g_1^{2} \lambda_{2332} \tr\left(y_e y_e^{\dagger} \right)

+ \frac{45}{4} g_2^{2} \lambda_{2332} \tr\left(y_u y_u^{\dagger} \right)

+ \frac{15}{4} g_2^{2} \lambda_{2332} \tr\left(y_e y_e^{\dagger} \right)

+ 40 g_3^{2} \lambda_{2332} \tr\left(y_u y_u^{\dagger} \right)

+ 21 g_1^{2} g_2^{2} \tr\left(y_u y_u^{\dagger} \right)

+ 11 g_1^{2} g_2^{2} \tr\left(y_e y_e^{\dagger} \right)

-  \frac{9}{2} \lambda_{2332} \tr\left(y_d y_d^{\dagger} y_u y_u^{\dagger} \right)

-  \frac{27}{2} \lambda_{2332} \tr\left(y_u y_u^{\dagger} y_u y_u^{\dagger} \right)

-  \frac{9}{2} \lambda_{2332} \tr\left(y_e y_e^{\dagger} y_e y_e^{\dagger} \right)
\,.\end{autobreak}
\end{align}
\begin{align}
\begin{autobreak}
\beta^{(1)}(\lambda_{1212}) =

 4 \lambda_{11} \lambda_{1212}

+ 8 \lambda_{1122} \lambda_{1212}

+ 4 \lambda_{1212} \lambda_{22}

+ 12 \lambda_{1212} \lambda_{1221}

+ 2 \lambda_{1313} \lambda_{2323}^{*}

- 3 g_1^{2} \lambda_{1212}

- 9 g_2^{2} \lambda_{1212}

+ 6 \lambda_{1212} \tr\left(y_d y_d^{\dagger} \right)

+ 6 \lambda_{1212} \tr\left(y_u y_u^{\dagger} \right)
\,.\end{autobreak}
\end{align}
\begin{align}
\begin{autobreak}
\beta^{(2)}(\lambda_{1212}) =

- 80 \lambda_{11} \lambda_{1122} \lambda_{1212}

- 88 \lambda_{11} \lambda_{1212} \lambda_{1221}

- 80 \lambda_{1122} \lambda_{1212} \lambda_{22}

- 76 \lambda_{1122} \lambda_{1212} \lambda_{1221}

- 8 \lambda_{1122} \lambda_{1313} \lambda_{2323}^{*}

- 12 \lambda_{1221} \lambda_{1313} \lambda_{2323}^{*}

- 2 \lambda_{1133} \lambda_{1212} \lambda_{1331}

- 16 \lambda_{1133} \lambda_{1212} \lambda_{2233}

- 8 \lambda_{1133} \lambda_{1212} \lambda_{2332}

- 8 \lambda_{1133} \lambda_{1313} \lambda_{2323}^{*}

- 28 \lambda_{11}^{2} \lambda_{1212}

- 28 \lambda_{1212} \lambda_{22}^{2}

- 28 \lambda_{1122}^{2} \lambda_{1212}

- 88 \lambda_{1212} \lambda_{1221} \lambda_{22}

- 32 \lambda_{1212} \lambda_{1221}^{2}

- 2 \lambda_{1133}^{2} \lambda_{1212}

- 8 \lambda_{1212} \lambda_{1331} \lambda_{2233}

- 12 \lambda_{1212} \lambda_{1331} \lambda_{2332}

- 2 \lambda_{1212} \lambda_{2233}^{2}

- 2 \lambda_{1212} \lambda_{2233} \lambda_{2332}

+ 6 \lambda_{1212} \left|{\lambda_{1212}}\right|^{2}

+ 3 \lambda_{1212} \left|{\lambda_{1313}}\right|^{2}

+ 3 \lambda_{1212} \left|{\lambda_{2323}}\right|^{2}

- 12 \lambda_{1313} \lambda_{1331} \lambda_{2323}^{*}

- 8 \lambda_{1313} \lambda_{2233} \lambda_{2323}^{*}

- 12 \lambda_{1313} \lambda_{2332} \lambda_{2323}^{*}

- 4 g_1^{2} \lambda_{11} \lambda_{1212}

+ 16 g_1^{2} \lambda_{1122} \lambda_{1212}

+ 36 g_2^{2} \lambda_{1122} \lambda_{1212}

- 4 g_1^{2} \lambda_{1212} \lambda_{22}

+ 24 g_1^{2} \lambda_{1212} \lambda_{1221}

+ 72 g_2^{2} \lambda_{1212} \lambda_{1221}

- 2 g_1^{2} \lambda_{1313} \lambda_{2323}^{*}

+ \frac{493}{24} g_1^{4} \lambda_{1212}

+ \frac{19}{4} g_1^{2} g_2^{2} \lambda_{1212}

-  \frac{209}{8} g_2^{4} \lambda_{1212}

- 24 \lambda_{11} \lambda_{1212} \tr\left(y_d y_d^{\dagger} \right)

- 24 \lambda_{1122} \lambda_{1212} \tr\left(y_d y_d^{\dagger} \right)

- 24 \lambda_{1122} \lambda_{1212} \tr\left(y_u y_u^{\dagger} \right)

- 24 \lambda_{1212} \lambda_{22} \tr\left(y_u y_u^{\dagger} \right)

- 36 \lambda_{1212} \lambda_{1221} \tr\left(y_d y_d^{\dagger} \right)

- 36 \lambda_{1212} \lambda_{1221} \tr\left(y_u y_u^{\dagger} \right)

- 4 \lambda_{1313} \lambda_{2323}^{*} \tr\left(y_e y_e^{\dagger} \right)

+ \frac{25}{12} g_1^{2} \lambda_{1212} \tr\left(y_d y_d^{\dagger} \right)

+ \frac{85}{12} g_1^{2} \lambda_{1212} \tr\left(y_u y_u^{\dagger} \right)

+ \frac{45}{4} g_2^{2} \lambda_{1212} \tr\left(y_d y_d^{\dagger} \right)

+ \frac{45}{4} g_2^{2} \lambda_{1212} \tr\left(y_u y_u^{\dagger} \right)

+ 40 g_3^{2} \lambda_{1212} \tr\left(y_d y_d^{\dagger} \right)

+ 40 g_3^{2} \lambda_{1212} \tr\left(y_u y_u^{\dagger} \right)

-  \frac{3}{2} \lambda_{1212} \tr\left(y_d y_d^{\dagger} y_d y_d^{\dagger} \right)

- 33 \lambda_{1212} \tr\left(y_d y_d^{\dagger} y_u y_u^{\dagger} \right)

-  \frac{3}{2} \lambda_{1212} \tr\left(y_u y_u^{\dagger} y_u y_u^{\dagger} \right)
\,.\end{autobreak}
\end{align}
\begin{align}
\begin{autobreak}
\beta^{(1)}(\lambda_{1212}^{*}) =

 4 \lambda_{11} \lambda_{1212}^{*}

+ 4 \lambda_{22} \lambda_{1212}^{*}

+ 8 \lambda_{1122} \lambda_{1212}^{*}

+ 12 \lambda_{1221} \lambda_{1212}^{*}

+ 2 \lambda_{2323} \lambda_{1313}^{*}

- 3 g_1^{2} \lambda_{1212}^{*}

- 9 g_2^{2} \lambda_{1212}^{*}

+ 6 \lambda_{1212}^{*} \tr\left(y_d y_d^{\dagger} \right)

+ 6 \lambda_{1212}^{*} \tr\left(y_u y_u^{\dagger} \right)
\,.\end{autobreak}
\end{align}
\begin{align}
\begin{autobreak}
\beta^{(2)}(\lambda_{1212}^{*}) =

- 28 \lambda_{11}^{2} \lambda_{1212}^{*}

- 80 \lambda_{11} \lambda_{1122} \lambda_{1212}^{*}

- 88 \lambda_{11} \lambda_{1221} \lambda_{1212}^{*}

- 28 \lambda_{22}^{2} \lambda_{1212}^{*}

- 80 \lambda_{1122} \lambda_{22} \lambda_{1212}^{*}

- 28 \lambda_{1122}^{2} \lambda_{1212}^{*}

- 76 \lambda_{1122} \lambda_{1221} \lambda_{1212}^{*}

- 8 \lambda_{1122} \lambda_{2323} \lambda_{1313}^{*}

- 88 \lambda_{1221} \lambda_{22} \lambda_{1212}^{*}

- 32 \lambda_{1221}^{2} \lambda_{1212}^{*}

- 12 \lambda_{1221} \lambda_{2323} \lambda_{1313}^{*}

- 2 \lambda_{1133}^{2} \lambda_{1212}^{*}

- 2 \lambda_{1133} \lambda_{1331} \lambda_{1212}^{*}

- 16 \lambda_{1133} \lambda_{2233} \lambda_{1212}^{*}

- 8 \lambda_{1133} \lambda_{2332} \lambda_{1212}^{*}

- 8 \lambda_{1133} \lambda_{2323} \lambda_{1313}^{*}

- 8 \lambda_{1331} \lambda_{2233} \lambda_{1212}^{*}

- 12 \lambda_{1331} \lambda_{2332} \lambda_{1212}^{*}

- 12 \lambda_{1331} \lambda_{2323} \lambda_{1313}^{*}

- 2 \lambda_{2233}^{2} \lambda_{1212}^{*}

- 2 \lambda_{2233} \lambda_{2332} \lambda_{1212}^{*}

- 8 \lambda_{2233} \lambda_{2323} \lambda_{1313}^{*}

+ 6 \lambda_{1212}^{*} \left|{\lambda_{1212}}\right|^{2}

+ 3 \lambda_{1212}^{*} \left|{\lambda_{1313}}\right|^{2}

- 12 \lambda_{2323} \lambda_{2332} \lambda_{1313}^{*}

+ 3 \lambda_{1212}^{*} \left|{\lambda_{2323}}\right|^{2}

- 4 g_1^{2} \lambda_{11} \lambda_{1212}^{*}

- 4 g_1^{2} \lambda_{22} \lambda_{1212}^{*}

+ 16 g_1^{2} \lambda_{1122} \lambda_{1212}^{*}

+ 36 g_2^{2} \lambda_{1122} \lambda_{1212}^{*}

+ 24 g_1^{2} \lambda_{1221} \lambda_{1212}^{*}

+ 72 g_2^{2} \lambda_{1221} \lambda_{1212}^{*}

- 2 g_1^{2} \lambda_{2323} \lambda_{1313}^{*}

+ \frac{493}{24} g_1^{4} \lambda_{1212}^{*}

+ \frac{19}{4} g_1^{2} g_2^{2} \lambda_{1212}^{*}

-  \frac{209}{8} g_2^{4} \lambda_{1212}^{*}

- 24 \lambda_{11} \lambda_{1212}^{*} \tr\left(y_d y_d^{\dagger} \right)

- 24 \lambda_{22} \lambda_{1212}^{*} \tr\left(y_u y_u^{\dagger} \right)

- 24 \lambda_{1122} \lambda_{1212}^{*} \tr\left(y_d y_d^{\dagger} \right)

- 24 \lambda_{1122} \lambda_{1212}^{*} \tr\left(y_u y_u^{\dagger} \right)

- 36 \lambda_{1221} \lambda_{1212}^{*} \tr\left(y_d y_d^{\dagger} \right)

- 36 \lambda_{1221} \lambda_{1212}^{*} \tr\left(y_u y_u^{\dagger} \right)

- 4 \lambda_{2323} \lambda_{1313}^{*} \tr\left(y_e y_e^{\dagger} \right)

+ \frac{25}{12} g_1^{2} \lambda_{1212}^{*} \tr\left(y_d y_d^{\dagger} \right)

+ \frac{85}{12} g_1^{2} \lambda_{1212}^{*} \tr\left(y_u y_u^{\dagger} \right)

+ \frac{45}{4} g_2^{2} \lambda_{1212}^{*} \tr\left(y_d y_d^{\dagger} \right)

+ \frac{45}{4} g_2^{2} \lambda_{1212}^{*} \tr\left(y_u y_u^{\dagger} \right)

+ 40 g_3^{2} \lambda_{1212}^{*} \tr\left(y_d y_d^{\dagger} \right)

+ 40 g_3^{2} \lambda_{1212}^{*} \tr\left(y_u y_u^{\dagger} \right)

-  \frac{3}{2} \lambda_{1212}^{*} \tr\left(y_d y_d^{\dagger} y_d y_d^{\dagger} \right)

- 33 \lambda_{1212}^{*} \tr\left(y_d y_d^{\dagger} y_u y_u^{\dagger} \right)

-  \frac{3}{2} \lambda_{1212}^{*} \tr\left(y_u y_u^{\dagger} y_u y_u^{\dagger} \right)
\,.\end{autobreak}
\end{align}
\begin{align}
\begin{autobreak}
\beta^{(1)}(\lambda_{1313}) =

 4 \lambda_{11} \lambda_{1313}

+ 8 \lambda_{1133} \lambda_{1313}

+ 2 \lambda_{1212} \lambda_{2323}

+ 4 \lambda_{1313} \lambda_{33}

+ 12 \lambda_{1313} \lambda_{1331}

- 3 g_1^{2} \lambda_{1313}

- 9 g_2^{2} \lambda_{1313}

+ 6 \lambda_{1313} \tr\left(y_d y_d^{\dagger} \right)

+ 2 \lambda_{1313} \tr\left(y_e y_e^{\dagger} \right)
\,.\end{autobreak}
\end{align}
\begin{align}
\begin{autobreak}
\beta^{(2)}(\lambda_{1313}) =

- 80 \lambda_{11} \lambda_{1133} \lambda_{1313}

- 88 \lambda_{11} \lambda_{1313} \lambda_{1331}

- 2 \lambda_{1122} \lambda_{1221} \lambda_{1313}

- 8 \lambda_{1122} \lambda_{1212} \lambda_{2323}

- 16 \lambda_{1122} \lambda_{1313} \lambda_{2233}

- 8 \lambda_{1122} \lambda_{1313} \lambda_{2332}

- 8 \lambda_{1221} \lambda_{1313} \lambda_{2233}

- 12 \lambda_{1221} \lambda_{1313} \lambda_{2332}

- 8 \lambda_{1133} \lambda_{1212} \lambda_{2323}

- 80 \lambda_{1133} \lambda_{1313} \lambda_{33}

- 76 \lambda_{1133} \lambda_{1313} \lambda_{1331}

- 12 \lambda_{1212} \lambda_{1221} \lambda_{2323}

- 12 \lambda_{1212} \lambda_{1331} \lambda_{2323}

- 8 \lambda_{1212} \lambda_{2233} \lambda_{2323}

+ 3 \lambda_{1313} \left|{\lambda_{1212}}\right|^{2}

- 12 \lambda_{1212} \lambda_{2323} \lambda_{2332}

- 28 \lambda_{11}^{2} \lambda_{1313}

- 28 \lambda_{1313} \lambda_{33}^{2}

- 2 \lambda_{1122}^{2} \lambda_{1313}

- 28 \lambda_{1133}^{2} \lambda_{1313}

- 88 \lambda_{1313} \lambda_{1331} \lambda_{33}

- 32 \lambda_{1313} \lambda_{1331}^{2}

- 2 \lambda_{1313} \lambda_{2233}^{2}

- 2 \lambda_{1313} \lambda_{2233} \lambda_{2332}

+ 6 \lambda_{1313} \left|{\lambda_{1313}}\right|^{2}

+ 3 \lambda_{1313} \left|{\lambda_{2323}}\right|^{2}

- 4 g_1^{2} \lambda_{11} \lambda_{1313}

+ 16 g_1^{2} \lambda_{1133} \lambda_{1313}

+ 36 g_2^{2} \lambda_{1133} \lambda_{1313}

- 2 g_1^{2} \lambda_{1212} \lambda_{2323}

- 4 g_1^{2} \lambda_{1313} \lambda_{33}

+ 24 g_1^{2} \lambda_{1313} \lambda_{1331}

+ 72 g_2^{2} \lambda_{1313} \lambda_{1331}

+ \frac{493}{24} g_1^{4} \lambda_{1313}

+ \frac{19}{4} g_1^{2} g_2^{2} \lambda_{1313}

-  \frac{209}{8} g_2^{4} \lambda_{1313}

- 24 \lambda_{11} \lambda_{1313} \tr\left(y_d y_d^{\dagger} \right)

- 24 \lambda_{1133} \lambda_{1313} \tr\left(y_d y_d^{\dagger} \right)

- 8 \lambda_{1133} \lambda_{1313} \tr\left(y_e y_e^{\dagger} \right)

- 12 \lambda_{1212} \lambda_{2323} \tr\left(y_u y_u^{\dagger} \right)

- 8 \lambda_{1313} \lambda_{33} \tr\left(y_e y_e^{\dagger} \right)

- 36 \lambda_{1313} \lambda_{1331} \tr\left(y_d y_d^{\dagger} \right)

- 12 \lambda_{1313} \lambda_{1331} \tr\left(y_e y_e^{\dagger} \right)

+ \frac{25}{12} g_1^{2} \lambda_{1313} \tr\left(y_d y_d^{\dagger} \right)

+ \frac{25}{4} g_1^{2} \lambda_{1313} \tr\left(y_e y_e^{\dagger} \right)

+ \frac{45}{4} g_2^{2} \lambda_{1313} \tr\left(y_d y_d^{\dagger} \right)

+ \frac{15}{4} g_2^{2} \lambda_{1313} \tr\left(y_e y_e^{\dagger} \right)

+ 40 g_3^{2} \lambda_{1313} \tr\left(y_d y_d^{\dagger} \right)

-  \frac{3}{2} \lambda_{1313} \tr\left(y_d y_d^{\dagger} y_d y_d^{\dagger} \right)

-  \frac{9}{2} \lambda_{1313} \tr\left(y_d y_d^{\dagger} y_u y_u^{\dagger} \right)

-  \frac{1}{2} \lambda_{1313} \tr\left(y_e y_e^{\dagger} y_e y_e^{\dagger} \right)
\,.\end{autobreak}
\end{align}
\begin{align}
\begin{autobreak}
\beta^{(1)}(\lambda_{1313}^{*}) =

 4 \lambda_{11} \lambda_{1313}^{*}

+ 4 \lambda_{33} \lambda_{1313}^{*}

+ 8 \lambda_{1133} \lambda_{1313}^{*}

+ 12 \lambda_{1331} \lambda_{1313}^{*}

+ 2 \lambda_{1212}^{*} \lambda_{2323}^{*}

- 3 g_1^{2} \lambda_{1313}^{*}

- 9 g_2^{2} \lambda_{1313}^{*}

+ 6 \lambda_{1313}^{*} \tr\left(y_d y_d^{\dagger} \right)

+ 2 \lambda_{1313}^{*} \tr\left(y_e y_e^{\dagger} \right)
\,.\end{autobreak}
\end{align}
\begin{align}
\begin{autobreak}
\beta^{(2)}(\lambda_{1313}^{*}) =

- 28 \lambda_{11}^{2} \lambda_{1313}^{*}

- 80 \lambda_{11} \lambda_{1133} \lambda_{1313}^{*}

- 88 \lambda_{11} \lambda_{1331} \lambda_{1313}^{*}

- 28 \lambda_{33}^{2} \lambda_{1313}^{*}

- 2 \lambda_{1122}^{2} \lambda_{1313}^{*}

- 2 \lambda_{1122} \lambda_{1221} \lambda_{1313}^{*}

- 16 \lambda_{1122} \lambda_{2233} \lambda_{1313}^{*}

- 8 \lambda_{1122} \lambda_{2332} \lambda_{1313}^{*}

- 8 \lambda_{1122} \lambda_{1212}^{*} \lambda_{2323}^{*}

- 8 \lambda_{1221} \lambda_{2233} \lambda_{1313}^{*}

- 12 \lambda_{1221} \lambda_{2332} \lambda_{1313}^{*}

- 12 \lambda_{1221} \lambda_{1212}^{*} \lambda_{2323}^{*}

- 80 \lambda_{1133} \lambda_{33} \lambda_{1313}^{*}

- 28 \lambda_{1133}^{2} \lambda_{1313}^{*}

- 76 \lambda_{1133} \lambda_{1331} \lambda_{1313}^{*}

- 8 \lambda_{1133} \lambda_{1212}^{*} \lambda_{2323}^{*}

- 88 \lambda_{1331} \lambda_{33} \lambda_{1313}^{*}

- 32 \lambda_{1331}^{2} \lambda_{1313}^{*}

- 12 \lambda_{1331} \lambda_{1212}^{*} \lambda_{2323}^{*}

- 2 \lambda_{2233}^{2} \lambda_{1313}^{*}

- 2 \lambda_{2233} \lambda_{2332} \lambda_{1313}^{*}

- 8 \lambda_{2233} \lambda_{1212}^{*} \lambda_{2323}^{*}

- 12 \lambda_{2332} \lambda_{1212}^{*} \lambda_{2323}^{*}

+ 3 \lambda_{1313}^{*} \left|{\lambda_{1212}}\right|^{2}

+ 6 \lambda_{1313}^{*} \left|{\lambda_{1313}}\right|^{2}

+ 3 \lambda_{1313}^{*} \left|{\lambda_{2323}}\right|^{2}

- 4 g_1^{2} \lambda_{11} \lambda_{1313}^{*}

- 4 g_1^{2} \lambda_{33} \lambda_{1313}^{*}

+ 16 g_1^{2} \lambda_{1133} \lambda_{1313}^{*}

+ 36 g_2^{2} \lambda_{1133} \lambda_{1313}^{*}

+ 24 g_1^{2} \lambda_{1331} \lambda_{1313}^{*}

+ 72 g_2^{2} \lambda_{1331} \lambda_{1313}^{*}

- 2 g_1^{2} \lambda_{1212}^{*} \lambda_{2323}^{*}

+ \frac{493}{24} g_1^{4} \lambda_{1313}^{*}

+ \frac{19}{4} g_1^{2} g_2^{2} \lambda_{1313}^{*}

-  \frac{209}{8} g_2^{4} \lambda_{1313}^{*}

- 24 \lambda_{11} \lambda_{1313}^{*} \tr\left(y_d y_d^{\dagger} \right)

- 8 \lambda_{33} \lambda_{1313}^{*} \tr\left(y_e y_e^{\dagger} \right)

- 24 \lambda_{1133} \lambda_{1313}^{*} \tr\left(y_d y_d^{\dagger} \right)

- 8 \lambda_{1133} \lambda_{1313}^{*} \tr\left(y_e y_e^{\dagger} \right)

- 36 \lambda_{1331} \lambda_{1313}^{*} \tr\left(y_d y_d^{\dagger} \right)

- 12 \lambda_{1331} \lambda_{1313}^{*} \tr\left(y_e y_e^{\dagger} \right)

- 12 \lambda_{1212}^{*} \lambda_{2323}^{*} \tr\left(y_u y_u^{\dagger} \right)

+ \frac{25}{12} g_1^{2} \lambda_{1313}^{*} \tr\left(y_d y_d^{\dagger} \right)

+ \frac{25}{4} g_1^{2} \lambda_{1313}^{*} \tr\left(y_e y_e^{\dagger} \right)

+ \frac{45}{4} g_2^{2} \lambda_{1313}^{*} \tr\left(y_d y_d^{\dagger} \right)

+ \frac{15}{4} g_2^{2} \lambda_{1313}^{*} \tr\left(y_e y_e^{\dagger} \right)

+ 40 g_3^{2} \lambda_{1313}^{*} \tr\left(y_d y_d^{\dagger} \right)

-  \frac{3}{2} \lambda_{1313}^{*} \tr\left(y_d y_d^{\dagger} y_d y_d^{\dagger} \right)

-  \frac{9}{2} \lambda_{1313}^{*} \tr\left(y_d y_d^{\dagger} y_u y_u^{\dagger} \right)

-  \frac{1}{2} \lambda_{1313}^{*} \tr\left(y_e y_e^{\dagger} y_e y_e^{\dagger} \right)
\,.\end{autobreak}
\end{align}
\begin{align}
\begin{autobreak}
\beta^{(1)}(\lambda_{2323}) =

 4 \lambda_{22} \lambda_{2323}

+ 8 \lambda_{2233} \lambda_{2323}

+ 2 \lambda_{1313} \lambda_{1212}^{*}

+ 4 \lambda_{2323} \lambda_{33}

+ 12 \lambda_{2323} \lambda_{2332}

- 3 g_1^{2} \lambda_{2323}

- 9 g_2^{2} \lambda_{2323}

+ 6 \lambda_{2323} \tr\left(y_u y_u^{\dagger} \right)

+ 2 \lambda_{2323} \tr\left(y_e y_e^{\dagger} \right)
\,.\end{autobreak}
\end{align}
\begin{align}
\begin{autobreak}
\beta^{(2)}(\lambda_{2323}) =

- 80 \lambda_{22} \lambda_{2233} \lambda_{2323}

- 88 \lambda_{22} \lambda_{2323} \lambda_{2332}

- 2 \lambda_{1122} \lambda_{1221} \lambda_{2323}

- 16 \lambda_{1122} \lambda_{1133} \lambda_{2323}

- 8 \lambda_{1122} \lambda_{1331} \lambda_{2323}

- 8 \lambda_{1122} \lambda_{1313} \lambda_{1212}^{*}

- 12 \lambda_{1221} \lambda_{1331} \lambda_{2323}

- 12 \lambda_{1221} \lambda_{1313} \lambda_{1212}^{*}

- 8 \lambda_{1133} \lambda_{1221} \lambda_{2323}

- 2 \lambda_{1133} \lambda_{1331} \lambda_{2323}

- 8 \lambda_{1133} \lambda_{1313} \lambda_{1212}^{*}

- 80 \lambda_{2233} \lambda_{2323} \lambda_{33}

- 76 \lambda_{2233} \lambda_{2323} \lambda_{2332}

+ 3 \lambda_{2323} \left|{\lambda_{1212}}\right|^{2}

- 12 \lambda_{1313} \lambda_{1331} \lambda_{1212}^{*}

- 8 \lambda_{1313} \lambda_{2233} \lambda_{1212}^{*}

- 12 \lambda_{1313} \lambda_{2332} \lambda_{1212}^{*}

+ 3 \lambda_{2323} \left|{\lambda_{1313}}\right|^{2}

- 28 \lambda_{22}^{2} \lambda_{2323}

- 28 \lambda_{2323} \lambda_{33}^{2}

- 2 \lambda_{1122}^{2} \lambda_{2323}

- 2 \lambda_{1133}^{2} \lambda_{2323}

- 28 \lambda_{2233}^{2} \lambda_{2323}

- 88 \lambda_{2323} \lambda_{2332} \lambda_{33}

- 32 \lambda_{2323} \lambda_{2332}^{2}

+ 6 \lambda_{2323} \left|{\lambda_{2323}}\right|^{2}

- 4 g_1^{2} \lambda_{22} \lambda_{2323}

+ 16 g_1^{2} \lambda_{2233} \lambda_{2323}

+ 36 g_2^{2} \lambda_{2233} \lambda_{2323}

- 2 g_1^{2} \lambda_{1313} \lambda_{1212}^{*}

- 4 g_1^{2} \lambda_{2323} \lambda_{33}

+ 24 g_1^{2} \lambda_{2323} \lambda_{2332}

+ 72 g_2^{2} \lambda_{2323} \lambda_{2332}

+ \frac{493}{24} g_1^{4} \lambda_{2323}

+ \frac{19}{4} g_1^{2} g_2^{2} \lambda_{2323}

-  \frac{209}{8} g_2^{4} \lambda_{2323}

- 24 \lambda_{22} \lambda_{2323} \tr\left(y_u y_u^{\dagger} \right)

- 24 \lambda_{2233} \lambda_{2323} \tr\left(y_u y_u^{\dagger} \right)

- 8 \lambda_{2233} \lambda_{2323} \tr\left(y_e y_e^{\dagger} \right)

- 12 \lambda_{1313} \lambda_{1212}^{*} \tr\left(y_d y_d^{\dagger} \right)

- 8 \lambda_{2323} \lambda_{33} \tr\left(y_e y_e^{\dagger} \right)

- 36 \lambda_{2323} \lambda_{2332} \tr\left(y_u y_u^{\dagger} \right)

- 12 \lambda_{2323} \lambda_{2332} \tr\left(y_e y_e^{\dagger} \right)

+ \frac{85}{12} g_1^{2} \lambda_{2323} \tr\left(y_u y_u^{\dagger} \right)

+ \frac{25}{4} g_1^{2} \lambda_{2323} \tr\left(y_e y_e^{\dagger} \right)

+ \frac{45}{4} g_2^{2} \lambda_{2323} \tr\left(y_u y_u^{\dagger} \right)

+ \frac{15}{4} g_2^{2} \lambda_{2323} \tr\left(y_e y_e^{\dagger} \right)

+ 40 g_3^{2} \lambda_{2323} \tr\left(y_u y_u^{\dagger} \right)

-  \frac{9}{2} \lambda_{2323} \tr\left(y_d y_d^{\dagger} y_u y_u^{\dagger} \right)

-  \frac{3}{2} \lambda_{2323} \tr\left(y_u y_u^{\dagger} y_u y_u^{\dagger} \right)

-  \frac{1}{2} \lambda_{2323} \tr\left(y_e y_e^{\dagger} y_e y_e^{\dagger} \right)
\,.\end{autobreak}
\end{align}
\begin{align}
\begin{autobreak}
\beta^{(1)}(\lambda_{2323}^{*}) =

 4 \lambda_{22} \lambda_{2323}^{*}

+ 4 \lambda_{33} \lambda_{2323}^{*}

+ 8 \lambda_{2233} \lambda_{2323}^{*}

+ 12 \lambda_{2332} \lambda_{2323}^{*}

+ 2 \lambda_{1212} \lambda_{1313}^{*}

- 3 g_1^{2} \lambda_{2323}^{*}

- 9 g_2^{2} \lambda_{2323}^{*}

+ 6 \lambda_{2323}^{*} \tr\left(y_u y_u^{\dagger} \right)

+ 2 \lambda_{2323}^{*} \tr\left(y_e y_e^{\dagger} \right)
\,.\end{autobreak}
\end{align}
\begin{align}
\begin{autobreak}
\beta^{(2)}(\lambda_{2323}^{*}) =

- 28 \lambda_{22}^{2} \lambda_{2323}^{*}

- 80 \lambda_{22} \lambda_{2233} \lambda_{2323}^{*}

- 88 \lambda_{22} \lambda_{2332} \lambda_{2323}^{*}

- 28 \lambda_{33}^{2} \lambda_{2323}^{*}

- 2 \lambda_{1122}^{2} \lambda_{2323}^{*}

- 2 \lambda_{1122} \lambda_{1221} \lambda_{2323}^{*}

- 16 \lambda_{1122} \lambda_{1133} \lambda_{2323}^{*}

- 8 \lambda_{1122} \lambda_{1331} \lambda_{2323}^{*}

- 8 \lambda_{1122} \lambda_{1212} \lambda_{1313}^{*}

- 12 \lambda_{1221} \lambda_{1331} \lambda_{2323}^{*}

- 8 \lambda_{1133} \lambda_{1221} \lambda_{2323}^{*}

- 2 \lambda_{1133}^{2} \lambda_{2323}^{*}

- 2 \lambda_{1133} \lambda_{1331} \lambda_{2323}^{*}

- 8 \lambda_{1133} \lambda_{1212} \lambda_{1313}^{*}

- 80 \lambda_{2233} \lambda_{33} \lambda_{2323}^{*}

- 28 \lambda_{2233}^{2} \lambda_{2323}^{*}

- 76 \lambda_{2233} \lambda_{2332} \lambda_{2323}^{*}

- 88 \lambda_{2332} \lambda_{33} \lambda_{2323}^{*}

- 32 \lambda_{2332}^{2} \lambda_{2323}^{*}

- 12 \lambda_{1212} \lambda_{1221} \lambda_{1313}^{*}

- 12 \lambda_{1212} \lambda_{1331} \lambda_{1313}^{*}

- 8 \lambda_{1212} \lambda_{2233} \lambda_{1313}^{*}

- 12 \lambda_{1212} \lambda_{2332} \lambda_{1313}^{*}

+ 3 \lambda_{2323}^{*} \left|{\lambda_{1212}}\right|^{2}

+ 3 \lambda_{2323}^{*} \left|{\lambda_{1313}}\right|^{2}

+ 6 \lambda_{2323}^{*} \left|{\lambda_{2323}}\right|^{2}

- 4 g_1^{2} \lambda_{22} \lambda_{2323}^{*}

- 4 g_1^{2} \lambda_{33} \lambda_{2323}^{*}

+ 16 g_1^{2} \lambda_{2233} \lambda_{2323}^{*}

+ 36 g_2^{2} \lambda_{2233} \lambda_{2323}^{*}

+ 24 g_1^{2} \lambda_{2332} \lambda_{2323}^{*}

+ 72 g_2^{2} \lambda_{2332} \lambda_{2323}^{*}

- 2 g_1^{2} \lambda_{1212} \lambda_{1313}^{*}

+ \frac{493}{24} g_1^{4} \lambda_{2323}^{*}

+ \frac{19}{4} g_1^{2} g_2^{2} \lambda_{2323}^{*}

-  \frac{209}{8} g_2^{4} \lambda_{2323}^{*}

- 24 \lambda_{22} \lambda_{2323}^{*} \tr\left(y_u y_u^{\dagger} \right)

- 8 \lambda_{33} \lambda_{2323}^{*} \tr\left(y_e y_e^{\dagger} \right)

- 24 \lambda_{2233} \lambda_{2323}^{*} \tr\left(y_u y_u^{\dagger} \right)

- 8 \lambda_{2233} \lambda_{2323}^{*} \tr\left(y_e y_e^{\dagger} \right)

- 36 \lambda_{2332} \lambda_{2323}^{*} \tr\left(y_u y_u^{\dagger} \right)

- 12 \lambda_{2332} \lambda_{2323}^{*} \tr\left(y_e y_e^{\dagger} \right)

- 12 \lambda_{1212} \lambda_{1313}^{*} \tr\left(y_d y_d^{\dagger} \right)

+ \frac{85}{12} g_1^{2} \lambda_{2323}^{*} \tr\left(y_u y_u^{\dagger} \right)

+ \frac{25}{4} g_1^{2} \lambda_{2323}^{*} \tr\left(y_e y_e^{\dagger} \right)

+ \frac{45}{4} g_2^{2} \lambda_{2323}^{*} \tr\left(y_u y_u^{\dagger} \right)

+ \frac{15}{4} g_2^{2} \lambda_{2323}^{*} \tr\left(y_e y_e^{\dagger} \right)

+ 40 g_3^{2} \lambda_{2323}^{*} \tr\left(y_u y_u^{\dagger} \right)

-  \frac{9}{2} \lambda_{2323}^{*} \tr\left(y_d y_d^{\dagger} y_u y_u^{\dagger} \right)

-  \frac{3}{2} \lambda_{2323}^{*} \tr\left(y_u y_u^{\dagger} y_u y_u^{\dagger} \right)

-  \frac{1}{2} \lambda_{2323}^{*} \tr\left(y_e y_e^{\dagger} y_e y_e^{\dagger} \right)
\,.\end{autobreak}
\end{align}
}

\subsection{One- and Two-Loop RGEs of Mass Parameters}\label{RGE:d}
The one- and two-loop RGEs of the mass parameters in the 3HDM Type-V may be obtained as:
{\allowdisplaybreaks \small
\begin{align}
\begin{autobreak}
\beta^{(1)}(m_{11}^2) =

-  \frac{3}{2} g_1^{2} m_{11}^2

-  \frac{9}{2} g_2^{2} m_{11}^2

+ 12 \lambda_{11} m_{11}^2

+ 4 \lambda_{1122} m_{22}^2

+ 2 \lambda_{1221} m_{22}^2

+ 4 \lambda_{1133} m_{33}^2

+ 2 \lambda_{1331} m_{33}^2

+ 6 m_{11}^2 \tr\left(y_d y_d^{\dagger} \right)
\,.\end{autobreak}
\end{align}
\begin{align}
\begin{autobreak}
\beta^{(2)}(m_{11}^2) =

 \frac{601}{48} g_1^{4} m_{11}^2

+ \frac{15}{8} g_1^{2} g_2^{2} m_{11}^2

-  \frac{101}{16} g_2^{4} m_{11}^2

+ \frac{5}{2} g_1^{4} m_{22}^2

+ \frac{15}{2} g_2^{4} m_{22}^2

+ \frac{5}{2} g_1^{4} m_{33}^2

+ \frac{15}{2} g_2^{4} m_{33}^2

+ 24 g_1^{2} \lambda_{11} m_{11}^2

+ 72 g_2^{2} \lambda_{11} m_{11}^2

+ 8 g_1^{2} \lambda_{1122} m_{22}^2

+ 4 g_1^{2} \lambda_{1221} m_{22}^2

+ 24 g_2^{2} \lambda_{1122} m_{22}^2

+ 12 g_2^{2} \lambda_{1221} m_{22}^2

+ 8 g_1^{2} \lambda_{1133} m_{33}^2

+ 4 g_1^{2} \lambda_{1331} m_{33}^2

+ 24 g_2^{2} \lambda_{1133} m_{33}^2

+ 12 g_2^{2} \lambda_{1331} m_{33}^2

- 60 \lambda_{11}^{2} m_{11}^2

- 2 \lambda_{1122}^{2} m_{11}^2

- 2 \lambda_{1122} \lambda_{1221} m_{11}^2

- 2 \lambda_{1221}^{2} m_{11}^2

- 2 \lambda_{1133}^{2} m_{11}^2

- 2 \lambda_{1133} \lambda_{1331} m_{11}^2

- 2 \lambda_{1331}^{2} m_{11}^2

- 3 m_{11}^2 \left|{\lambda_{1212}}\right|^{2}

- 3 m_{11}^2 \left|{\lambda_{1313}}\right|^{2}

- 8 \lambda_{1122}^{2} m_{22}^2

- 8 \lambda_{1122} \lambda_{1221} m_{22}^2

- 8 \lambda_{1221}^{2} m_{22}^2

- 12 m_{22}^2 \left|{\lambda_{1212}}\right|^{2}

- 8 \lambda_{1133}^{2} m_{33}^2

- 8 \lambda_{1133} \lambda_{1331} m_{33}^2

- 8 \lambda_{1331}^{2} m_{33}^2

- 12 m_{33}^2 \left|{\lambda_{1313}}\right|^{2}

+ \frac{25}{12} g_1^{2} m_{11}^2 \tr\left(y_d y_d^{\dagger} \right)

+ \frac{45}{4} g_2^{2} m_{11}^2 \tr\left(y_d y_d^{\dagger} \right)

+ 40 g_3^{2} m_{11}^2 \tr\left(y_d y_d^{\dagger} \right)

- 72 \lambda_{11} m_{11}^2 \tr\left(y_d y_d^{\dagger} \right)

- 24 \lambda_{1122} m_{22}^2 \tr\left(y_u y_u^{\dagger} \right)

- 12 \lambda_{1221} m_{22}^2 \tr\left(y_u y_u^{\dagger} \right)

- 8 \lambda_{1133} m_{33}^2 \tr\left(y_e y_e^{\dagger} \right)

- 4 \lambda_{1331} m_{33}^2 \tr\left(y_e y_e^{\dagger} \right)

-  \frac{27}{2} m_{11}^2 \tr\left(y_d y_d^{\dagger} y_d y_d^{\dagger} \right)

-  \frac{9}{2} m_{11}^2 \tr\left(y_d y_d^{\dagger} y_u y_u^{\dagger} \right)
\,.\end{autobreak}
\end{align}
\begin{align}
\begin{autobreak}
\beta^{(1)}(m_{22}^2) =

-  \frac{3}{2} g_1^{2} m_{22}^2

-  \frac{9}{2} g_2^{2} m_{22}^2

+ 4 \lambda_{1122} m_{11}^2

+ 2 \lambda_{1221} m_{11}^2

+ 12 \lambda_{22} m_{22}^2

+ 4 \lambda_{2233} m_{33}^2

+ 2 \lambda_{2332} m_{33}^2

+ 6 m_{22}^2 \tr\left(y_u y_u^{\dagger} \right)
\,.\end{autobreak}
\end{align}
\begin{align}
\begin{autobreak}
\beta^{(2)}(m_{22}^2) =

 \frac{5}{2} g_1^{4} m_{11}^2

+ \frac{15}{2} g_2^{4} m_{11}^2

+ \frac{601}{48} g_1^{4} m_{22}^2

+ \frac{15}{8} g_1^{2} g_2^{2} m_{22}^2

-  \frac{101}{16} g_2^{4} m_{22}^2

+ \frac{5}{2} g_1^{4} m_{33}^2

+ \frac{15}{2} g_2^{4} m_{33}^2

+ 8 g_1^{2} \lambda_{1122} m_{11}^2

+ 4 g_1^{2} \lambda_{1221} m_{11}^2

+ 24 g_2^{2} \lambda_{1122} m_{11}^2

+ 12 g_2^{2} \lambda_{1221} m_{11}^2

+ 24 g_1^{2} \lambda_{22} m_{22}^2

+ 72 g_2^{2} \lambda_{22} m_{22}^2

+ 8 g_1^{2} \lambda_{2233} m_{33}^2

+ 4 g_1^{2} \lambda_{2332} m_{33}^2

+ 24 g_2^{2} \lambda_{2233} m_{33}^2

+ 12 g_2^{2} \lambda_{2332} m_{33}^2

- 8 \lambda_{1122}^{2} m_{11}^2

- 8 \lambda_{1122} \lambda_{1221} m_{11}^2

- 8 \lambda_{1221}^{2} m_{11}^2

- 12 m_{11}^2 \left|{\lambda_{1212}}\right|^{2}

- 60 \lambda_{22}^{2} m_{22}^2

- 2 \lambda_{1122}^{2} m_{22}^2

- 2 \lambda_{1122} \lambda_{1221} m_{22}^2

- 2 \lambda_{1221}^{2} m_{22}^2

- 2 \lambda_{2233}^{2} m_{22}^2

- 2 \lambda_{2233} \lambda_{2332} m_{22}^2

- 2 \lambda_{2332}^{2} m_{22}^2

- 3 m_{22}^2 \left|{\lambda_{1212}}\right|^{2}

- 3 m_{22}^2 \left|{\lambda_{2323}}\right|^{2}

- 8 \lambda_{2233}^{2} m_{33}^2

- 8 \lambda_{2233} \lambda_{2332} m_{33}^2

- 8 \lambda_{2332}^{2} m_{33}^2

- 12 m_{33}^2 \left|{\lambda_{2323}}\right|^{2}

+ \frac{85}{12} g_1^{2} m_{22}^2 \tr\left(y_u y_u^{\dagger} \right)

+ \frac{45}{4} g_2^{2} m_{22}^2 \tr\left(y_u y_u^{\dagger} \right)

+ 40 g_3^{2} m_{22}^2 \tr\left(y_u y_u^{\dagger} \right)

- 24 \lambda_{1122} m_{11}^2 \tr\left(y_d y_d^{\dagger} \right)

- 12 \lambda_{1221} m_{11}^2 \tr\left(y_d y_d^{\dagger} \right)

- 72 \lambda_{22} m_{22}^2 \tr\left(y_u y_u^{\dagger} \right)

- 8 \lambda_{2233} m_{33}^2 \tr\left(y_e y_e^{\dagger} \right)

- 4 \lambda_{2332} m_{33}^2 \tr\left(y_e y_e^{\dagger} \right)

-  \frac{9}{2} m_{22}^2 \tr\left(y_d y_d^{\dagger} y_u y_u^{\dagger} \right)

-  \frac{27}{2} m_{22}^2 \tr\left(y_u y_u^{\dagger} y_u y_u^{\dagger} \right)
\,.\end{autobreak}
\end{align}
\begin{align}
\begin{autobreak}
\beta^{(1)}(m_{33}^2) =

-  \frac{3}{2} g_1^{2} m_{33}^2

-  \frac{9}{2} g_2^{2} m_{33}^2

+ 4 \lambda_{1133} m_{11}^2

+ 2 \lambda_{1331} m_{11}^2

+ 4 \lambda_{2233} m_{22}^2

+ 2 \lambda_{2332} m_{22}^2

+ 12 \lambda_{33} m_{33}^2

+ 2 m_{33}^2 \tr\left(y_e y_e^{\dagger} \right)
\,.\end{autobreak}
\end{align}
\begin{align}
\begin{autobreak}
\beta^{(2)}(m_{33}^2) =

 \frac{5}{2} g_1^{4} m_{11}^2

+ \frac{15}{2} g_2^{4} m_{11}^2

+ \frac{5}{2} g_1^{4} m_{22}^2

+ \frac{15}{2} g_2^{4} m_{22}^2

+ \frac{601}{48} g_1^{4} m_{33}^2

+ \frac{15}{8} g_1^{2} g_2^{2} m_{33}^2

-  \frac{101}{16} g_2^{4} m_{33}^2

+ 8 g_1^{2} \lambda_{1133} m_{11}^2

+ 4 g_1^{2} \lambda_{1331} m_{11}^2

+ 24 g_2^{2} \lambda_{1133} m_{11}^2

+ 12 g_2^{2} \lambda_{1331} m_{11}^2

+ 8 g_1^{2} \lambda_{2233} m_{22}^2

+ 4 g_1^{2} \lambda_{2332} m_{22}^2

+ 24 g_2^{2} \lambda_{2233} m_{22}^2

+ 12 g_2^{2} \lambda_{2332} m_{22}^2

+ 24 g_1^{2} \lambda_{33} m_{33}^2

+ 72 g_2^{2} \lambda_{33} m_{33}^2

- 8 \lambda_{1133}^{2} m_{11}^2

- 8 \lambda_{1133} \lambda_{1331} m_{11}^2

- 8 \lambda_{1331}^{2} m_{11}^2

- 12 m_{11}^2 \left|{\lambda_{1313}}\right|^{2}

- 8 \lambda_{2233}^{2} m_{22}^2

- 8 \lambda_{2233} \lambda_{2332} m_{22}^2

- 8 \lambda_{2332}^{2} m_{22}^2

- 12 m_{22}^2 \left|{\lambda_{2323}}\right|^{2}

- 60 \lambda_{33}^{2} m_{33}^2

- 2 \lambda_{1133}^{2} m_{33}^2

- 2 \lambda_{1133} \lambda_{1331} m_{33}^2

- 2 \lambda_{1331}^{2} m_{33}^2

- 2 \lambda_{2233}^{2} m_{33}^2

- 2 \lambda_{2233} \lambda_{2332} m_{33}^2

- 2 \lambda_{2332}^{2} m_{33}^2

- 3 m_{33}^2 \left|{\lambda_{1313}}\right|^{2}

- 3 m_{33}^2 \left|{\lambda_{2323}}\right|^{2}

+ \frac{25}{4} g_1^{2} m_{33}^2 \tr\left(y_e y_e^{\dagger} \right)

+ \frac{15}{4} g_2^{2} m_{33}^2 \tr\left(y_e y_e^{\dagger} \right)

- 24 \lambda_{1133} m_{11}^2 \tr\left(y_d y_d^{\dagger} \right)

- 12 \lambda_{1331} m_{11}^2 \tr\left(y_d y_d^{\dagger} \right)

- 24 \lambda_{2233} m_{22}^2 \tr\left(y_u y_u^{\dagger} \right)

- 12 \lambda_{2332} m_{22}^2 \tr\left(y_u y_u^{\dagger} \right)

- 24 \lambda_{33} m_{33}^2 \tr\left(y_e y_e^{\dagger} \right)

-  \frac{9}{2} m_{33}^2 \tr\left(y_e y_e^{\dagger} y_e y_e^{\dagger} \right)
\,.\end{autobreak}
\end{align}
\begin{align}
\begin{autobreak}
\beta^{(1)}(m_{12}^2) =

-  \frac{3}{2} g_1^{2} m_{12}^2

-  \frac{9}{2} g_2^{2} m_{12}^2

+ 2 \lambda_{1122} m_{12}^2

+ 4 \lambda_{1221} m_{12}^2

+ 6 \lambda_{1212} m_{12}^{2\,*}

+ 3 m_{12}^2 \tr\left(y_d y_d^{\dagger} \right)

+ 3 m_{12}^2 \tr\left(y_u y_u^{\dagger} \right)
\,.\end{autobreak}
\end{align}
\begin{align}
\begin{autobreak}
\beta^{(2)}(m_{12}^2) =

\frac{481}{48} g_1^{4} m_{12}^2

+ \frac{15}{8} g_1^{2} g_2^{2} m_{12}^2

-  \frac{221}{16} g_2^{4} m_{12}^2

+ 4 g_1^{2} \lambda_{1122} m_{12}^2

+ 8 g_1^{2} \lambda_{1221} m_{12}^2

+ 12 g_1^{2} \lambda_{1212} m_{12}^{2\,*}

+ 12 g_2^{2} \lambda_{1122} m_{12}^2

+ 24 g_2^{2} \lambda_{1221} m_{12}^2

+ 36 g_2^{2} \lambda_{1212} m_{12}^{2\,*}

+ 6 \lambda_{11}^{2} m_{12}^2

- 12 \lambda_{11} \lambda_{1122} m_{12}^2

- 12 \lambda_{11} \lambda_{1221} m_{12}^2

- 12 \lambda_{11} \lambda_{1212} m_{12}^{2\,*}

+ 6 \lambda_{22}^{2} m_{12}^2

- 12 \lambda_{1122} \lambda_{22} m_{12}^2

- 6 \lambda_{1122} \lambda_{1221} m_{12}^2

- 12 \lambda_{1122} \lambda_{1212} m_{12}^{2\,*}

- 12 \lambda_{1221} \lambda_{22} m_{12}^2

+ \lambda_{1133}^{2} m_{12}^2

+ \lambda_{1133} \lambda_{1331} m_{12}^2

- 4 \lambda_{1133} \lambda_{2233} m_{12}^2

- 2 \lambda_{1133} \lambda_{2332} m_{12}^2

+ \lambda_{1331}^{2} m_{12}^2

- 2 \lambda_{1331} \lambda_{2233} m_{12}^2

- 4 \lambda_{1331} \lambda_{2332} m_{12}^2

+ \lambda_{2233}^{2} m_{12}^2

+ \lambda_{2233} \lambda_{2332} m_{12}^2

+ \lambda_{2332}^{2} m_{12}^2

- 12 \lambda_{1212} \lambda_{22} m_{12}^{2\,*}

- 12 \lambda_{1212} \lambda_{1221} m_{12}^{2\,*}

+ 3 m_{12}^2 \left|{\lambda_{1212}}\right|^{2}

+ \frac{3}{2} m_{12}^2 \left|{\lambda_{1313}}\right|^{2}

- 6 \lambda_{1313} \lambda_{2323}^{*} m_{12}^{2\,*}

+ \frac{3}{2} m_{12}^2 \left|{\lambda_{2323}}\right|^{2}

+ \frac{25}{24} g_1^{2} m_{12}^2 \tr\left(y_d y_d^{\dagger} \right)

+ \frac{85}{24} g_1^{2} m_{12}^2 \tr\left(y_u y_u^{\dagger} \right)

+ \frac{45}{8} g_2^{2} m_{12}^2 \tr\left(y_d y_d^{\dagger} \right)

+ \frac{45}{8} g_2^{2} m_{12}^2 \tr\left(y_u y_u^{\dagger} \right)

+ 20 g_3^{2} m_{12}^2 \tr\left(y_d y_d^{\dagger} \right)

+ 20 g_3^{2} m_{12}^2 \tr\left(y_u y_u^{\dagger} \right)

- 6 \lambda_{1122} m_{12}^2 \tr\left(y_d y_d^{\dagger} \right)

- 6 \lambda_{1122} m_{12}^2 \tr\left(y_u y_u^{\dagger} \right)

- 12 \lambda_{1221} m_{12}^2 \tr\left(y_d y_d^{\dagger} \right)

- 12 \lambda_{1221} m_{12}^2 \tr\left(y_u y_u^{\dagger} \right)

- 18 \lambda_{1212} m_{12}^{2\,*} \tr\left(y_d y_d^{\dagger} \right)

- 18 \lambda_{1212} m_{12}^{2\,*} \tr\left(y_u y_u^{\dagger} \right)

-  \frac{27}{4} m_{12}^2 \tr\left(y_d y_d^{\dagger} y_d y_d^{\dagger} \right)

-  \frac{33}{2} m_{12}^2 \tr\left(y_d y_d^{\dagger} y_u y_u^{\dagger} \right)

-  \frac{27}{4} m_{12}^2 \tr\left(y_u y_u^{\dagger} y_u y_u^{\dagger} \right)
\,.\end{autobreak}
\end{align}
\begin{align}
\begin{autobreak}
\beta^{(1)}(m_{12}^{2\,*}) =

-  \frac{3}{2} g_1^{2} m_{12}^{2\,*}

-  \frac{9}{2} g_2^{2} m_{12}^{2\,*}

+ 2 \lambda_{1122} m_{12}^{2\,*}

+ 4 \lambda_{1221} m_{12}^{2\,*}

+ 6 m_{12}^2 \lambda_{1212}^{*}

+ 3 m_{12}^{2\,*} \tr\left(y_d y_d^{\dagger} \right)

+ 3 m_{12}^{2\,*} \tr\left(y_u y_u^{\dagger} \right)
\,.\end{autobreak}
\end{align}
\begin{align}
\begin{autobreak}
\beta^{(2)}(m_{12}^{2\,*}) =

\frac{481}{48} g_1^{4} m_{12}^{2\,*}

+ \frac{15}{8} g_1^{2} g_2^{2} m_{12}^{2\,*}

-  \frac{221}{16} g_2^{4} m_{12}^{2\,*}

+ 4 g_1^{2} \lambda_{1122} m_{12}^{2\,*}

+ 8 g_1^{2} \lambda_{1221} m_{12}^{2\,*}

+ 12 g_1^{2} m_{12}^2 \lambda_{1212}^{*}

+ 12 g_2^{2} \lambda_{1122} m_{12}^{2\,*}

+ 24 g_2^{2} \lambda_{1221} m_{12}^{2\,*}

+ 36 g_2^{2} m_{12}^2 \lambda_{1212}^{*}

+ 6 \lambda_{11}^{2} m_{12}^{2\,*}

- 12 \lambda_{11} \lambda_{1122} m_{12}^{2\,*}

- 12 \lambda_{11} \lambda_{1221} m_{12}^{2\,*}

- 12 \lambda_{11} m_{12}^2 \lambda_{1212}^{*}

+ 6 \lambda_{22}^{2} m_{12}^{2\,*}

- 12 \lambda_{22} m_{12}^2 \lambda_{1212}^{*}

- 12 \lambda_{1122} \lambda_{22} m_{12}^{2\,*}

- 6 \lambda_{1122} \lambda_{1221} m_{12}^{2\,*}

- 12 \lambda_{1122} m_{12}^2 \lambda_{1212}^{*}

- 12 \lambda_{1221} \lambda_{22} m_{12}^{2\,*}

- 12 \lambda_{1221} m_{12}^2 \lambda_{1212}^{*}

+ \lambda_{1133}^{2} m_{12}^{2\,*}

+ \lambda_{1133} \lambda_{1331} m_{12}^{2\,*}

- 4 \lambda_{1133} \lambda_{2233} m_{12}^{2\,*}

- 2 \lambda_{1133} \lambda_{2332} m_{12}^{2\,*}

+ \lambda_{1331}^{2} m_{12}^{2\,*}

- 2 \lambda_{1331} \lambda_{2233} m_{12}^{2\,*}

- 4 \lambda_{1331} \lambda_{2332} m_{12}^{2\,*}

+ \lambda_{2233}^{2} m_{12}^{2\,*}

+ \lambda_{2233} \lambda_{2332} m_{12}^{2\,*}

+ \lambda_{2332}^{2} m_{12}^{2\,*}

+ 3 m_{12}^{2\,*} \left|{\lambda_{1212}}\right|^{2}

+ \frac{3}{2} m_{12}^{2\,*} \left|{\lambda_{1313}}\right|^{2}

- 6 \lambda_{2323} m_{12}^2 \lambda_{1313}^{*}

+ \frac{3}{2} m_{12}^{2\,*} \left|{\lambda_{2323}}\right|^{2}

+ \frac{25}{24} g_1^{2} m_{12}^{2\,*} \tr\left(y_d y_d^{\dagger} \right)

+ \frac{85}{24} g_1^{2} m_{12}^{2\,*} \tr\left(y_u y_u^{\dagger} \right)

+ \frac{45}{8} g_2^{2} m_{12}^{2\,*} \tr\left(y_d y_d^{\dagger} \right)

+ \frac{45}{8} g_2^{2} m_{12}^{2\,*} \tr\left(y_u y_u^{\dagger} \right)

+ 20 g_3^{2} m_{12}^{2\,*} \tr\left(y_d y_d^{\dagger} \right)

+ 20 g_3^{2} m_{12}^{2\,*} \tr\left(y_u y_u^{\dagger} \right)

- 6 \lambda_{1122} m_{12}^{2\,*} \tr\left(y_d y_d^{\dagger} \right)

- 6 \lambda_{1122} m_{12}^{2\,*} \tr\left(y_u y_u^{\dagger} \right)

- 12 \lambda_{1221} m_{12}^{2\,*} \tr\left(y_d y_d^{\dagger} \right)

- 12 \lambda_{1221} m_{12}^{2\,*} \tr\left(y_u y_u^{\dagger} \right)

- 18 m_{12}^2 \lambda_{1212}^{*} \tr\left(y_d y_d^{\dagger} \right)

- 18 m_{12}^2 \lambda_{1212}^{*} \tr\left(y_u y_u^{\dagger} \right)

-  \frac{27}{4} m_{12}^{2\,*} \tr\left(y_d y_d^{\dagger} y_d y_d^{\dagger} \right)

-  \frac{33}{2} m_{12}^{2\,*} \tr\left(y_d y_d^{\dagger} y_u y_u^{\dagger} \right)

-  \frac{27}{4} m_{12}^{2\,*} \tr\left(y_u y_u^{\dagger} y_u y_u^{\dagger} \right)
\,.\end{autobreak}
\end{align}
\begin{align}
\begin{autobreak}
\beta^{(1)}(m_{13}^2) =

-  \frac{3}{2} g_1^{2} m_{13}^2

-  \frac{9}{2} g_2^{2} m_{13}^2

+ 2 \lambda_{1133} m_{13}^2

+ 4 \lambda_{1331} m_{13}^2

+ 6 \lambda_{1313} m_{13}^{2\,*}

+ 3 m_{13}^2 \tr\left(y_d y_d^{\dagger} \right)

+ m_{13}^2 \tr\left(y_e y_e^{\dagger} \right)
\,.\end{autobreak}
\end{align}
\begin{align}
\begin{autobreak}
\beta^{(2)}(m_{13}^2) =

 \frac{481}{48} g_1^{4} m_{13}^2

+ \frac{15}{8} g_1^{2} g_2^{2} m_{13}^2

-  \frac{221}{16} g_2^{4} m_{13}^2

+ 4 g_1^{2} \lambda_{1133} m_{13}^2

+ 8 g_1^{2} \lambda_{1331} m_{13}^2

+ 12 g_1^{2} \lambda_{1313} m_{13}^{2\,*}

+ 12 g_2^{2} \lambda_{1133} m_{13}^2

+ 24 g_2^{2} \lambda_{1331} m_{13}^2

+ 36 g_2^{2} \lambda_{1313} m_{13}^{2\,*}

+ 6 \lambda_{11}^{2} m_{13}^2

- 12 \lambda_{11} \lambda_{1133} m_{13}^2

- 12 \lambda_{11} \lambda_{1331} m_{13}^2

- 12 \lambda_{11} \lambda_{1313} m_{13}^{2\,*}

+ 6 \lambda_{33}^{2} m_{13}^2

+ \lambda_{1122}^{2} m_{13}^2

+ \lambda_{1122} \lambda_{1221} m_{13}^2

- 4 \lambda_{1122} \lambda_{2233} m_{13}^2

- 2 \lambda_{1122} \lambda_{2332} m_{13}^2

+ \lambda_{1221}^{2} m_{13}^2

- 2 \lambda_{1221} \lambda_{2233} m_{13}^2

- 4 \lambda_{1221} \lambda_{2332} m_{13}^2

- 12 \lambda_{1133} \lambda_{33} m_{13}^2

- 6 \lambda_{1133} \lambda_{1331} m_{13}^2

- 12 \lambda_{1133} \lambda_{1313} m_{13}^{2\,*}

- 12 \lambda_{1331} \lambda_{33} m_{13}^2

+ \lambda_{2233}^{2} m_{13}^2

+ \lambda_{2233} \lambda_{2332} m_{13}^2

+ \lambda_{2332}^{2} m_{13}^2

+ \frac{3}{2} m_{13}^2 \left|{\lambda_{1212}}\right|^{2}

- 6 \lambda_{1212} \lambda_{2323} m_{13}^{2\,*}

- 12 \lambda_{1313} \lambda_{33} m_{13}^{2\,*}

- 12 \lambda_{1313} \lambda_{1331} m_{13}^{2\,*}

+ 3 m_{13}^2 \left|{\lambda_{1313}}\right|^{2}

+ \frac{3}{2} m_{13}^2 \left|{\lambda_{2323}}\right|^{2}

+ \frac{25}{24} g_1^{2} m_{13}^2 \tr\left(y_d y_d^{\dagger} \right)

+ \frac{25}{8} g_1^{2} m_{13}^2 \tr\left(y_e y_e^{\dagger} \right)

+ \frac{45}{8} g_2^{2} m_{13}^2 \tr\left(y_d y_d^{\dagger} \right)

+ \frac{15}{8} g_2^{2} m_{13}^2 \tr\left(y_e y_e^{\dagger} \right)

+ 20 g_3^{2} m_{13}^2 \tr\left(y_d y_d^{\dagger} \right)

- 6 \lambda_{1133} m_{13}^2 \tr\left(y_d y_d^{\dagger} \right)

- 2 \lambda_{1133} m_{13}^2 \tr\left(y_e y_e^{\dagger} \right)

- 12 \lambda_{1331} m_{13}^2 \tr\left(y_d y_d^{\dagger} \right)

- 4 \lambda_{1331} m_{13}^2 \tr\left(y_e y_e^{\dagger} \right)

- 18 \lambda_{1313} m_{13}^{2\,*} \tr\left(y_d y_d^{\dagger} \right)

- 6 \lambda_{1313} m_{13}^{2\,*} \tr\left(y_e y_e^{\dagger} \right)

-  \frac{27}{4} m_{13}^2 \tr\left(y_d y_d^{\dagger} y_d y_d^{\dagger} \right)

-  \frac{9}{4} m_{13}^2 \tr\left(y_d y_d^{\dagger} y_u y_u^{\dagger} \right)

-  \frac{9}{4} m_{13}^2 \tr\left(y_e y_e^{\dagger} y_e y_e^{\dagger} \right)
\,.\end{autobreak}
\end{align}
\begin{align}
\begin{autobreak}
\beta^{(1)}(m_{13}^{2\,*}) =

-  \frac{3}{2} g_1^{2} m_{13}^{2\,*}

-  \frac{9}{2} g_2^{2} m_{13}^{2\,*}

+ 2 \lambda_{1133} m_{13}^{2\,*}

+ 4 \lambda_{1331} m_{13}^{2\,*}

+ 6 m_{13}^2 \lambda_{1313}^{*}

+ 3 m_{13}^{2\,*} \tr\left(y_d y_d^{\dagger} \right)

+ m_{13}^{2\,*} \tr\left(y_e y_e^{\dagger} \right)
\,.\end{autobreak}
\end{align}
\begin{align}
\begin{autobreak}
\beta^{(2)}(m_{13}^{2\,*}) =

 \frac{481}{48} g_1^{4} m_{13}^{2\,*}

+ \frac{15}{8} g_1^{2} g_2^{2} m_{13}^{2\,*}

-  \frac{221}{16} g_2^{4} m_{13}^{2\,*}

+ 4 g_1^{2} \lambda_{1133} m_{13}^{2\,*}

+ 8 g_1^{2} \lambda_{1331} m_{13}^{2\,*}

+ 12 g_1^{2} m_{13}^2 \lambda_{1313}^{*}

+ 12 g_2^{2} \lambda_{1133} m_{13}^{2\,*}

+ 24 g_2^{2} \lambda_{1331} m_{13}^{2\,*}

+ 36 g_2^{2} m_{13}^2 \lambda_{1313}^{*}

+ 6 \lambda_{11}^{2} m_{13}^{2\,*}

- 12 \lambda_{11} \lambda_{1133} m_{13}^{2\,*}

- 12 \lambda_{11} \lambda_{1331} m_{13}^{2\,*}

- 12 \lambda_{11} m_{13}^2 \lambda_{1313}^{*}

+ 6 \lambda_{33}^{2} m_{13}^{2\,*}

- 12 \lambda_{33} m_{13}^2 \lambda_{1313}^{*}

+ \lambda_{1122}^{2} m_{13}^{2\,*}

+ \lambda_{1122} \lambda_{1221} m_{13}^{2\,*}

- 4 \lambda_{1122} \lambda_{2233} m_{13}^{2\,*}

- 2 \lambda_{1122} \lambda_{2332} m_{13}^{2\,*}

+ \lambda_{1221}^{2} m_{13}^{2\,*}

- 2 \lambda_{1221} \lambda_{2233} m_{13}^{2\,*}

- 4 \lambda_{1221} \lambda_{2332} m_{13}^{2\,*}

- 12 \lambda_{1133} \lambda_{33} m_{13}^{2\,*}

- 6 \lambda_{1133} \lambda_{1331} m_{13}^{2\,*}

- 12 \lambda_{1133} m_{13}^2 \lambda_{1313}^{*}

- 12 \lambda_{1331} \lambda_{33} m_{13}^{2\,*}

- 12 \lambda_{1331} m_{13}^2 \lambda_{1313}^{*}

+ \lambda_{2233}^{2} m_{13}^{2\,*}

+ \lambda_{2233} \lambda_{2332} m_{13}^{2\,*}

+ \lambda_{2332}^{2} m_{13}^{2\,*}

+ \frac{3}{2} m_{13}^{2\,*} \left|{\lambda_{1212}}\right|^{2}

- 6 m_{13}^2 \lambda_{1212}^{*} \lambda_{2323}^{*}

+ 3 m_{13}^{2\,*} \left|{\lambda_{1313}}\right|^{2}

+ \frac{3}{2} m_{13}^{2\,*} \left|{\lambda_{2323}}\right|^{2}

+ \frac{25}{24} g_1^{2} m_{13}^{2\,*} \tr\left(y_d y_d^{\dagger} \right)

+ \frac{25}{8} g_1^{2} m_{13}^{2\,*} \tr\left(y_e y_e^{\dagger} \right)

+ \frac{45}{8} g_2^{2} m_{13}^{2\,*} \tr\left(y_d y_d^{\dagger} \right)

+ \frac{15}{8} g_2^{2} m_{13}^{2\,*} \tr\left(y_e y_e^{\dagger} \right)

+ 20 g_3^{2} m_{13}^{2\,*} \tr\left(y_d y_d^{\dagger} \right)

- 6 \lambda_{1133} m_{13}^{2\,*} \tr\left(y_d y_d^{\dagger} \right)

- 2 \lambda_{1133} m_{13}^{2\,*} \tr\left(y_e y_e^{\dagger} \right)

- 12 \lambda_{1331} m_{13}^{2\,*} \tr\left(y_d y_d^{\dagger} \right)

- 4 \lambda_{1331} m_{13}^{2\,*} \tr\left(y_e y_e^{\dagger} \right)

- 18 m_{13}^2 \lambda_{1313}^{*} \tr\left(y_d y_d^{\dagger} \right)

- 6 m_{13}^2 \lambda_{1313}^{*} \tr\left(y_e y_e^{\dagger} \right)

-  \frac{27}{4} m_{13}^{2\,*} \tr\left(y_d y_d^{\dagger} y_d y_d^{\dagger} \right)

-  \frac{9}{4} m_{13}^{2\,*} \tr\left(y_d y_d^{\dagger} y_u y_u^{\dagger} \right)

-  \frac{9}{4} m_{13}^{2\,*} \tr\left(y_e y_e^{\dagger} y_e y_e^{\dagger} \right)
\,.\end{autobreak}
\end{align}
\begin{align}
\begin{autobreak}
\beta^{(1)}(m_{23}^2) =

-  \frac{3}{2} g_1^{2} m_{23}^2

-  \frac{9}{2} g_2^{2} m_{23}^2

+ 2 \lambda_{2233} m_{23}^2

+ 4 \lambda_{2332} m_{23}^2

+ 6 \lambda_{2323} m_{23}^{2\,*}

+ 3 m_{23}^2 \tr\left(y_u y_u^{\dagger} \right)

+ m_{23}^2 \tr\left(y_e y_e^{\dagger} \right)
\,.\end{autobreak}
\end{align}
\begin{align}
\begin{autobreak}
\beta^{(2)}(m_{23}^2) =

\frac{481}{48} g_1^{4} m_{23}^2

+ \frac{15}{8} g_1^{2} g_2^{2} m_{23}^2

-  \frac{221}{16} g_2^{4} m_{23}^2

+ 4 g_1^{2} \lambda_{2233} m_{23}^2

+ 8 g_1^{2} \lambda_{2332} m_{23}^2

+ 12 g_1^{2} \lambda_{2323} m_{23}^{2\,*}

+ 12 g_2^{2} \lambda_{2233} m_{23}^2

+ 24 g_2^{2} \lambda_{2332} m_{23}^2

+ 36 g_2^{2} \lambda_{2323} m_{23}^{2\,*}

+ 6 \lambda_{22}^{2} m_{23}^2

- 12 \lambda_{22} \lambda_{2233} m_{23}^2

- 12 \lambda_{22} \lambda_{2332} m_{23}^2

- 12 \lambda_{22} \lambda_{2323} m_{23}^{2\,*}

+ 6 \lambda_{33}^{2} m_{23}^2

+ \lambda_{1122}^{2} m_{23}^2

+ \lambda_{1122} \lambda_{1221} m_{23}^2

- 4 \lambda_{1122} \lambda_{1133} m_{23}^2

- 2 \lambda_{1122} \lambda_{1331} m_{23}^2

+ \lambda_{1221}^{2} m_{23}^2

- 4 \lambda_{1221} \lambda_{1331} m_{23}^2

- 2 \lambda_{1133} \lambda_{1221} m_{23}^2

+ \lambda_{1133}^{2} m_{23}^2

+ \lambda_{1133} \lambda_{1331} m_{23}^2

+ \lambda_{1331}^{2} m_{23}^2

- 12 \lambda_{2233} \lambda_{33} m_{23}^2

- 6 \lambda_{2233} \lambda_{2332} m_{23}^2

- 12 \lambda_{2233} \lambda_{2323} m_{23}^{2\,*}

- 12 \lambda_{2332} \lambda_{33} m_{23}^2

+ \frac{3}{2} m_{23}^2 \left|{\lambda_{1212}}\right|^{2}

- 6 \lambda_{1313} \lambda_{1212}^{*} m_{23}^{2\,*}

+ \frac{3}{2} m_{23}^2 \left|{\lambda_{1313}}\right|^{2}

- 12 \lambda_{2323} \lambda_{33} m_{23}^{2\,*}

- 12 \lambda_{2323} \lambda_{2332} m_{23}^{2\,*}

+ 3 m_{23}^2 \left|{\lambda_{2323}}\right|^{2}

+ \frac{85}{24} g_1^{2} m_{23}^2 \tr\left(y_u y_u^{\dagger} \right)

+ \frac{25}{8} g_1^{2} m_{23}^2 \tr\left(y_e y_e^{\dagger} \right)

+ \frac{45}{8} g_2^{2} m_{23}^2 \tr\left(y_u y_u^{\dagger} \right)

+ \frac{15}{8} g_2^{2} m_{23}^2 \tr\left(y_e y_e^{\dagger} \right)

+ 20 g_3^{2} m_{23}^2 \tr\left(y_u y_u^{\dagger} \right)

- 6 \lambda_{2233} m_{23}^2 \tr\left(y_u y_u^{\dagger} \right)

- 2 \lambda_{2233} m_{23}^2 \tr\left(y_e y_e^{\dagger} \right)

- 12 \lambda_{2332} m_{23}^2 \tr\left(y_u y_u^{\dagger} \right)

- 4 \lambda_{2332} m_{23}^2 \tr\left(y_e y_e^{\dagger} \right)

- 18 \lambda_{2323} m_{23}^{2\,*} \tr\left(y_u y_u^{\dagger} \right)

- 6 \lambda_{2323} m_{23}^{2\,*} \tr\left(y_e y_e^{\dagger} \right)

-  \frac{9}{4} m_{23}^2 \tr\left(y_d y_d^{\dagger} y_u y_u^{\dagger} \right)

-  \frac{27}{4} m_{23}^2 \tr\left(y_u y_u^{\dagger} y_u y_u^{\dagger} \right)

-  \frac{9}{4} m_{23}^2 \tr\left(y_e y_e^{\dagger} y_e y_e^{\dagger} \right)
\,.\end{autobreak}
\end{align}
\begin{align}
\begin{autobreak}
\beta^{(1)}(m_{23}^{2\,*}) =

-  \frac{3}{2} g_1^{2} m_{23}^{2\,*}

-  \frac{9}{2} g_2^{2} m_{23}^{2\,*}

+ 2 \lambda_{2233} m_{23}^{2\,*}

+ 4 \lambda_{2332} m_{23}^{2\,*}

+ 6 m_{23}^2 \lambda_{2323}^{*}

+ 3 m_{23}^{2\,*} \tr\left(y_u y_u^{\dagger} \right)

+ m_{23}^{2\,*} \tr\left(y_e y_e^{\dagger} \right)
\,.\end{autobreak}
\end{align}
\begin{align}
\begin{autobreak}
\beta^{(2)}(m_{23}^{2\,*}) =

 \frac{481}{48} g_1^{4} m_{23}^{2\,*}

+ \frac{15}{8} g_1^{2} g_2^{2} m_{23}^{2\,*}

-  \frac{221}{16} g_2^{4} m_{23}^{2\,*}

+ 4 g_1^{2} \lambda_{2233} m_{23}^{2\,*}

+ 8 g_1^{2} \lambda_{2332} m_{23}^{2\,*}

+ 12 g_1^{2} m_{23}^2 \lambda_{2323}^{*}

+ 12 g_2^{2} \lambda_{2233} m_{23}^{2\,*}

+ 24 g_2^{2} \lambda_{2332} m_{23}^{2\,*}

+ 36 g_2^{2} m_{23}^2 \lambda_{2323}^{*}

+ 6 \lambda_{22}^{2} m_{23}^{2\,*}

- 12 \lambda_{22} \lambda_{2233} m_{23}^{2\,*}

- 12 \lambda_{22} \lambda_{2332} m_{23}^{2\,*}

- 12 \lambda_{22} m_{23}^2 \lambda_{2323}^{*}

+ 6 \lambda_{33}^{2} m_{23}^{2\,*}

- 12 \lambda_{33} m_{23}^2 \lambda_{2323}^{*}

+ \lambda_{1122}^{2} m_{23}^{2\,*}

+ \lambda_{1122} \lambda_{1221} m_{23}^{2\,*}

- 4 \lambda_{1122} \lambda_{1133} m_{23}^{2\,*}

- 2 \lambda_{1122} \lambda_{1331} m_{23}^{2\,*}

+ \lambda_{1221}^{2} m_{23}^{2\,*}

- 4 \lambda_{1221} \lambda_{1331} m_{23}^{2\,*}

- 2 \lambda_{1133} \lambda_{1221} m_{23}^{2\,*}

+ \lambda_{1133}^{2} m_{23}^{2\,*}

+ \lambda_{1133} \lambda_{1331} m_{23}^{2\,*}

+ \lambda_{1331}^{2} m_{23}^{2\,*}

- 12 \lambda_{2233} \lambda_{33} m_{23}^{2\,*}

- 6 \lambda_{2233} \lambda_{2332} m_{23}^{2\,*}

- 12 \lambda_{2233} m_{23}^2 \lambda_{2323}^{*}

- 12 \lambda_{2332} \lambda_{33} m_{23}^{2\,*}

- 12 \lambda_{2332} m_{23}^2 \lambda_{2323}^{*}

+ \frac{3}{2} m_{23}^{2\,*} \left|{\lambda_{1212}}\right|^{2}

- 6 \lambda_{1212} m_{23}^2 \lambda_{1313}^{*}

+ \frac{3}{2} m_{23}^{2\,*} \left|{\lambda_{1313}}\right|^{2}

+ 3 m_{23}^{2\,*} \left|{\lambda_{2323}}\right|^{2}

+ \frac{85}{24} g_1^{2} m_{23}^{2\,*} \tr\left(y_u y_u^{\dagger} \right)

+ \frac{25}{8} g_1^{2} m_{23}^{2\,*} \tr\left(y_e y_e^{\dagger} \right)

+ \frac{45}{8} g_2^{2} m_{23}^{2\,*} \tr\left(y_u y_u^{\dagger} \right)

+ \frac{15}{8} g_2^{2} m_{23}^{2\,*} \tr\left(y_e y_e^{\dagger} \right)

+ 20 g_3^{2} m_{23}^{2\,*} \tr\left(y_u y_u^{\dagger} \right)

- 6 \lambda_{2233} m_{23}^{2\,*} \tr\left(y_u y_u^{\dagger} \right)

- 2 \lambda_{2233} m_{23}^{2\,*} \tr\left(y_e y_e^{\dagger} \right)

- 12 \lambda_{2332} m_{23}^{2\,*} \tr\left(y_u y_u^{\dagger} \right)

- 4 \lambda_{2332} m_{23}^{2\,*} \tr\left(y_e y_e^{\dagger} \right)

- 18 m_{23}^2 \lambda_{2323}^{*} \tr\left(y_u y_u^{\dagger} \right)

- 6 m_{23}^2 \lambda_{2323}^{*} \tr\left(y_e y_e^{\dagger} \right)

-  \frac{9}{4} m_{23}^{2\,*} \tr\left(y_d y_d^{\dagger} y_u y_u^{\dagger} \right)

-  \frac{27}{4} m_{23}^{2\,*} \tr\left(y_u y_u^{\dagger} y_u y_u^{\dagger} \right)

-  \frac{9}{4} m_{23}^{2\,*} \tr\left(y_e y_e^{\dagger} y_e y_e^{\dagger} \right)
\,.\end{autobreak}
\end{align}
}

\subsection{One- and Two-Loop RGEs of VEVs}\label{RGE:e}

The one- and two-loop RGEs of $v_1$, $v_2$ and $v_3$ take on the following forms:
{\allowdisplaybreaks \small
\begin{align}
\begin{autobreak}
\beta^{(1)}(v_{1}) =

 \frac{3}{4} g_1^{2} v_{1}

+ \frac{1}{4} \xi g_1^{2} v_{1}

+ \frac{9}{4} g_2^{2} v_{1}

+ \frac{3}{4} \xi g_2^{2} v_{1}

- 3 v_{1} \tr\left(y_d y_d^{\dagger} \right)
\,.\end{autobreak}
\end{align}
\begin{align}
\begin{autobreak}
\beta^{(2)}(v_{1}) =

-  \frac{475}{96} g_1^{4} v_{1}

+ \frac{1}{8} \xi g_1^{4} v_{1}

+ \frac{1}{8} \xi^{2} g_1^{4} v_{1}

-  \frac{9}{16} g_1^{2} g_2^{2} v_{1}

+ \frac{3}{4} \xi g_1^{2} g_2^{2} v_{1}

+ \frac{3}{4} \xi^{2} g_1^{2} g_2^{2} v_{1}

+ \frac{227}{32} g_2^{4} v_{1}

-  \frac{9}{8} \xi g_2^{4} v_{1}

-  \frac{9}{8} \xi^{2} g_2^{4} v_{1}

-  \frac{25}{24} g_1^{2} v_{1} \tr\left(y_d y_d^{\dagger} \right)

-  \frac{3}{4} \xi g_1^{2} v_{1} \tr\left(y_d y_d^{\dagger} \right)

-  \frac{45}{8} g_2^{2} v_{1} \tr\left(y_d y_d^{\dagger} \right)

-  \frac{9}{4} \xi g_2^{2} v_{1} \tr\left(y_d y_d^{\dagger} \right)

- 20 g_3^{2} v_{1} \tr\left(y_d y_d^{\dagger} \right)

+ \frac{27}{4} v_{1} \tr\left(y_d y_d^{\dagger} y_d y_d^{\dagger} \right)

+ \frac{9}{4} v_{1} \tr\left(y_d y_d^{\dagger} y_u y_u^{\dagger} \right)

- 6 \lambda_{11}^{2} v_{1}

-  \lambda_{1122}^{2} v_{1}

-  \lambda_{1122} \lambda_{1221} v_{1}

-  \lambda_{1221}^{2} v_{1}

-  \lambda_{1133}^{2} v_{1}

-  \lambda_{1133} \lambda_{1331} v_{1}

-  \lambda_{1331}^{2} v_{1}

-  \frac{3}{2} v_{1} \left|{\lambda_{1212}}\right|^{2}

-  \frac{3}{2} v_{1} \left|{\lambda_{1313}}\right|^{2}
\,.\end{autobreak}
\end{align}
\begin{align}
\begin{autobreak}
\beta^{(1)}(v_{2}) =

 \frac{3}{4} g_1^{2} v_{2}

+ \frac{1}{4} \xi g_1^{2} v_{2}

+ \frac{9}{4} g_2^{2} v_{2}

+ \frac{3}{4} \xi g_2^{2} v_{2}

- 3 v_{2} \tr\left(y_u y_u^{\dagger} \right)
\,.\end{autobreak}
\end{align}
\begin{align}
\begin{autobreak}
\beta^{(2)}(v_{2}) =

-  \frac{475}{96} g_1^{4} v_{2}

+ \frac{1}{8} \xi g_1^{4} v_{2}

+ \frac{1}{8} \xi^{2} g_1^{4} v_{2}

-  \frac{9}{16} g_1^{2} g_2^{2} v_{2}

+ \frac{3}{4} \xi g_1^{2} g_2^{2} v_{2}

+ \frac{3}{4} \xi^{2} g_1^{2} g_2^{2} v_{2}

+ \frac{227}{32} g_2^{4} v_{2}

-  \frac{9}{8} \xi g_2^{4} v_{2}

-  \frac{9}{8} \xi^{2} g_2^{4} v_{2}

-  \frac{85}{24} g_1^{2} v_{2} \tr\left(y_u y_u^{\dagger} \right)

-  \frac{3}{4} \xi g_1^{2} v_{2} \tr\left(y_u y_u^{\dagger} \right)

-  \frac{45}{8} g_2^{2} v_{2} \tr\left(y_u y_u^{\dagger} \right)

-  \frac{9}{4} \xi g_2^{2} v_{2} \tr\left(y_u y_u^{\dagger} \right)

- 20 g_3^{2} v_{2} \tr\left(y_u y_u^{\dagger} \right)

+ \frac{9}{4} v_{2} \tr\left(y_d y_d^{\dagger} y_u y_u^{\dagger} \right)

+ \frac{27}{4} v_{2} \tr\left(y_u y_u^{\dagger} y_u y_u^{\dagger} \right)

- 6 \lambda_{22}^{2} v_{2}

-  \lambda_{1122}^{2} v_{2}

-  \lambda_{1122} \lambda_{1221} v_{2}

-  \lambda_{1221}^{2} v_{2}

-  \lambda_{2233}^{2} v_{2}

-  \lambda_{2233} \lambda_{2332} v_{2}

-  \lambda_{2332}^{2} v_{2}

-  \frac{3}{2} v_{2} \left|{\lambda_{1212}}\right|^{2}

-  \frac{3}{2} v_{2} \left|{\lambda_{2323}}\right|^{2}
\,.\end{autobreak}
\end{align}
\begin{align}
\begin{autobreak}
\beta^{(1)}(v_{3}) =

\frac{3}{4} g_1^{2} v_{3}

+ \frac{1}{4} \xi g_1^{2} v_{3}

+ \frac{9}{4} g_2^{2} v_{3}

+ \frac{3}{4} \xi g_2^{2} v_{3}

-  v_{3} \tr\left(y_e y_e^{\dagger} \right)
\,.\end{autobreak}
\end{align}
\begin{align}
\begin{autobreak}
\beta^{(2)}(v_{3}) =

-  \frac{475}{96} g_1^{4} v_{3}

+ \frac{1}{8} \xi g_1^{4} v_{3}

+ \frac{1}{8} \xi^{2} g_1^{4} v_{3}

-  \frac{9}{16} g_1^{2} g_2^{2} v_{3}

+ \frac{3}{4} \xi g_1^{2} g_2^{2} v_{3}

+ \frac{3}{4} \xi^{2} g_1^{2} g_2^{2} v_{3}

+ \frac{227}{32} g_2^{4} v_{3}

-  \frac{9}{8} \xi g_2^{4} v_{3}

-  \frac{9}{8} \xi^{2} g_2^{4} v_{3}

-  \frac{25}{8} g_1^{2} v_{3} \tr\left(y_e y_e^{\dagger} \right)

-  \frac{1}{4} \xi g_1^{2} v_{3} \tr\left(y_e y_e^{\dagger} \right)

-  \frac{15}{8} g_2^{2} v_{3} \tr\left(y_e y_e^{\dagger} \right)

-  \frac{3}{4} \xi g_2^{2} v_{3} \tr\left(y_e y_e^{\dagger} \right)

+ \frac{9}{4} v_{3} \tr\left(y_e y_e^{\dagger} y_e y_e^{\dagger} \right)

- 6 \lambda_{33}^{2} v_{3}

-  \lambda_{1133}^{2} v_{3}

-  \lambda_{1133} \lambda_{1331} v_{3}

-  \lambda_{1331}^{2} v_{3}

-  \lambda_{2233}^{2} v_{3}

-  \lambda_{2233} \lambda_{2332} v_{3}

-  \lambda_{2332}^{2} v_{3}

-  \frac{3}{2} v_{3} \left|{\lambda_{1313}}\right|^{2}

-  \frac{3}{2} v_{3} \left|{\lambda_{2323}}\right|^{2}
\,.\end{autobreak}
\end{align}
}
In the above relations, $\xi$ is the gauge fixing parameter.
\vfill\eject

\end{document}